\PassOptionsToPackage{unicode}{hyperref}
\PassOptionsToPackage{hyphens}{url}
\PassOptionsToPackage{dvipsnames,svgnames,x11names}{xcolor}
\documentclass[12pt]{article}

\usepackage{amsmath,amssymb}
\usepackage{iftex}
\ifPDFTeX
  \usepackage[T1]{fontenc}
  \usepackage[utf8]{inputenc}
  \usepackage{textcomp} 
\else 
  \usepackage{unicode-math}
  \defaultfontfeatures{Scale=MatchLowercase}
  \defaultfontfeatures[\rmfamily]{Ligatures=TeX,Scale=1}
\fi

\usepackage{inconsolata}
\usepackage{fourier}

\ifPDFTeX\else  
\fi
\IfFileExists{upquote.sty}{\usepackage{upquote}}{}
\IfFileExists{microtype.sty}{
  \usepackage[]{microtype}
  \UseMicrotypeSet[protrusion]{basicmath} 
}{}
\makeatletter
\@ifundefined{KOMAClassName}{
  \IfFileExists{parskip.sty}{%
    \usepackage{parskip}
  }{
    \setlength{\parindent}{0pt}
    \setlength{\parskip}{6pt plus 2pt minus 1pt}}
}{
  \KOMAoptions{parskip=half}}
\makeatother
\usepackage{xcolor}
\ifx\paragraph\undefined\else
  \let\oldparagraph\paragraph
  \renewcommand{\paragraph}[1]{\oldparagraph{#1}\mbox{}}
\fi
\ifx\subparagraph\undefined\else
  \let\oldsubparagraph\subparagraph
  \renewcommand{\subparagraph}[1]{\oldsubparagraph{#1}\mbox{}}
\fi

\usepackage{color}
\usepackage{fancyvrb}

\DefineVerbatimEnvironment{Highlighting}{Verbatim}{commandchars=\\\{\}}
\usepackage{framed}
\definecolor{shadecolor}{RGB}{241,243,245}
\newenvironment{Shaded}{\begin{snugshade}}{\end{snugshade}}
\newcommand{\AlertTok}[1]{\textcolor[rgb]{0.68,0.00,0.00}{#1}}

\newcommand{\AttributeTok}[1]{\textcolor[rgb]{0.40,0.45,0.13}{#1}}

\newcommand{\CommentTok}[1]{\textcolor[rgb]{0.37,0.37,0.37}{#1}}

\newcommand{\ConstantTok}[1]{\textcolor[rgb]{0.56,0.35,0.01}{#1}}
\newcommand{\ControlFlowTok}[1]{\textcolor[rgb]{0.00,0.23,0.31}{#1}}

\newcommand{\DecValTok}[1]{\textcolor[rgb]{0.68,0.00,0.00}{#1}}

\newcommand{\FloatTok}[1]{\textcolor[rgb]{0.68,0.00,0.00}{#1}}
\newcommand{\FunctionTok}[1]{\textcolor[rgb]{0.28,0.35,0.67}{#1}}

\newcommand{\NormalTok}[1]{\textcolor[rgb]{0.00,0.23,0.31}{#1}}

\newcommand{\OtherTok}[1]{\textcolor[rgb]{0.00,0.23,0.31}{#1}}

\newcommand{\SpecialCharTok}[1]{\textcolor[rgb]{0.37,0.37,0.37}{#1}}

\newcommand{\StringTok}[1]{\textcolor[rgb]{0.13,0.47,0.30}{#1}}

\providecommand{\tightlist}{%
  \setlength{\itemsep}{0pt}\setlength{\parskip}{0pt}}\usepackage{longtable,booktabs,array}
\usepackage{calc} 
\usepackage{etoolbox}
\makeatletter
\patchcmd\longtable{\par}{\if@noskipsec\mbox{}\fi\par}{}{}
\makeatother
\IfFileExists{footnotehyper.sty}{\usepackage{footnotehyper}}{\usepackage{footnote}}
\makesavenoteenv{longtable}
\usepackage{graphicx}
\makeatletter
\def\maxwidth{\ifdim\Gin@nat@width>\linewidth\linewidth\else\Gin@nat@width\fi}
\def\maxheight{\ifdim\Gin@nat@height>\textheight\textheight\else\Gin@nat@height\fi}
\makeatother
\setkeys{Gin}{width=\maxwidth,height=\maxheight,keepaspectratio}
\makeatletter
\def\fps@figure{htbp}
\makeatother

\usepackage{graphicx}
\usepackage{amsmath}
\usepackage{amsfonts}
\usepackage{amssymb}
\usepackage{booktabs}
\usepackage{threeparttable}
\usepackage{adjustbox}
\usepackage{pdflscape} 
\usepackage{rotating}
\usepackage{array}
\usepackage{makecell}
\newcommand{\tabitem}{\hspace{1em}~~\llap{\textbullet}~~}

\newcommand{\proglang}[1]{\textsc{#1}}
\newcommand{\pkg}[1]{\textbf{\texttt{#1}}}

\usepackage[top=2cm, bottom=2cm, left=2cm, right=2cm]{geometry}

\makeatletter
\let\input\@@input
\makeatother

\makeatletter
\@ifpackageloaded{caption}{}{\usepackage{caption}}
\AtBeginDocument{%
\ifdefined\contentsname
  \renewcommand*\contentsname{Table of contents}
\else
  \newcommand\contentsname{Table of contents}
\fi
\ifdefined\listfigurename
  \renewcommand*\listfigurename{List of Figures}
\else
  \newcommand\listfigurename{List of Figures}
\fi
\ifdefined\listtablename
  \renewcommand*\listtablename{List of Tables}
\else
  \newcommand\listtablename{List of Tables}
\fi
\ifdefined\figurename
  \renewcommand*\figurename{Figure}
\else
  \newcommand\figurename{Figure}
\fi
\ifdefined\tablename
  \renewcommand*\tablename{Table}
\else
  \newcommand\tablename{Table}
\fi
}
\@ifpackageloaded{float}{}{\usepackage{float}}
\floatstyle{ruled}
\@ifundefined{c@chapter}{\newfloat{codelisting}{h}{lop}}{\newfloat{codelisting}{h}{lop}[chapter]}
\floatname{codelisting}{Listing}

\makeatother
\makeatletter
\@ifpackageloaded{caption}{}{\usepackage{caption}}
\@ifpackageloaded{subcaption}{}{\usepackage{subcaption}}
\makeatother
\makeatletter
\@ifpackageloaded{tcolorbox}{}{\usepackage[skins,breakable]{tcolorbox}}
\makeatother
\makeatletter
\@ifundefined{shadecolor}{\definecolor{shadecolor}{rgb}{.97, .97, .97}}
\makeatother
\ifLuaTeX
  \usepackage{selnolig}  
\fi
\usepackage[round]{natbib}
\IfFileExists{bookmark.sty}{\usepackage{bookmark}}{\usepackage{hyperref}}
\IfFileExists{xurl.sty}{\usepackage{xurl}}{} 
\hypersetup{
  pdftitle={fixest: A fast and feature-rich framework for econometric estimations in R},
  pdfauthor={Laurent R. Bergé; Kyle Butts; Grant McDermott},
  colorlinks=true,
  linkcolor={blue},
  filecolor={Maroon},
  citecolor={Blue},
  urlcolor={Blue},
  pdfcreator={LaTeX via pandoc}}

\newcommand\smallaffil[1]{\small{\textsc{#1}}}
\newcommand\email[1]{\href{mailto:#1}{#1}}

\title{\texttt{fixest}: A fast and feature-rich framework for
econometric estimations in R}
\author{
  Laurent R. Bergé\thanks{\email{laurent.berge@u-bordeaux.fr}}\\\smallaffil{BxSE -- UMR CNRS 6060}\\\smallaffil{University of Bordeaux}
  \and Kyle Butts\thanks{\email{kbutts@uark.edu}}\\\smallaffil{Department of Economics}\\\smallaffil{University of Arkansas}
  \and Grant McDermott\thanks{\email{contact@grantmcdermott.com}}\\\smallaffil{Department of Economics}\\\smallaffil{University of Oregon}
}
\date{2026-04-02}
\begin{document}
\maketitle
\begin{abstract}
\pkg{fixest} is an \proglang{R} package for fast and flexible
econometric estimation. It provides a unified framework for applied
research, with comprehensive support for a diverse class of models:
ordinary least squares, instrumental variables, generalized linear
models, maximum likelihood, and difference-in-differences. The package
particularly excels at fixed-effects estimation, supported by a novel
fixed-point acceleration algorithm implemented in \proglang{C++}. This
algorithm achieves rapid convergence across a variety of data contexts
and enables efficient estimation of complex models, including those with
varying slopes. An expressive formula interface facilitates multiple
estimations, stepwise regressions, and variable interpolation in a
single call. Users can adjust inference strategies on the fly, choosing
from an array of built-in robust standard errors. The package also
provides methods for publication-ready regression tables and coefficient
plots. Benchmarks demonstrate that \pkg{fixest} offers best-in-class
performance against leading alternatives in \proglang{R},
\proglang{Python}, and \proglang{Julia}.
\end{abstract}

\section{Introduction}\label{introduction}

\label{intro}

Fixed-effects estimation is a cornerstone of modern applied
econometrics. By absorbing unobserved heterogeneity through categorical
controls, fixed-effects models allow researchers to isolate causal
relationships in panel data, firm-employee records, trade flows, and
many other settings. The computational challenge is that fixed-effects
can have thousands or millions of levels, making standard estimation
routines impractical. This has motivated a body of work on efficient
algorithms for ``demeaning'' the fixed-effects out of the data before
estimating the coefficients of interest, grounded in the canonical
Frisch-Waugh-Lovell (FWL) theorem
\citep{frisch1933PartialTimeRegressions, lovell1963SeasonalAdjustmentEconomic}.

A variety of software packages have been developed to tackle the
demeaning problem. In \proglang{R} \citep{rproglang2025}, \pkg{lfe}
\citep{gaure2013lfe} introduced efficient fixed-effects estimation using
the method of alternating projections \citep{gaure2013ols}, with support
for ordinary least squares (OLS), instrumental varaibles (IV), and
clustered standard errors. \pkg{alpaca}
\citep{stammann2018FastFeasibleEstimation} extended the same approach to
generalized linear models (GLMs). In \proglang{Julia}
\citep{bezanson2017julia}, \pkg{FixedEffectModels}
\citep{gomez2024FixedEffectModelsJlFast} uses the LSMR algorithm
\citep{fong2011LSMRIterativeAlgorithm} for OLS, while
\pkg{GLFixedEffectModels} \citep{boehm2025GLFixedEffectModelsJl} handles
GLMs. In \proglang{Python} \citep{python2025}, \pkg{PyFixest}
\citep{fischer2024Pyfixest} offers similar functionality.\footnote{In
  this paper, our comparison focuses on free and open source software.
  Note that in the not free and open source language \proglang{Stata},
  \citet{correia2017_reghdfe} and \citet{correia2020fast} use a variant
  of the method of alternating projections to handle fixed-effects in
  OLS and GLM.}

These packages have generally made fixed-effects estimation accessible
and fast. However, they each address only part of the typical workflow
for applied researchers. For example, \pkg{lfe} does not support GLMs
and is incompatible with the \pkg{sandwich}
\citep{zeileis2020VariousVersatileVariances} ecosystem for flexible
variance-covariance (VCOV) estimation. \pkg{alpaca} handles GLMs but not
OLS or IV. The core alternating projections algorithm of these two
packages is also not well suited to ``sparse'' network structures, which
is common in many real-life datasets (e.g., employee-firm settings).
Similarly, the \pkg{PyFixest} library, developed by independent authors,
implements the same algorithm and hence suffers from the same
limitations. It does, however, provide a set of features close to
\pkg{fixest}, since its original objective was to port \pkg{fixest}
syntax and features to \proglang{Python}. In contrast, the two Julia
packages offer strong performance for their respective use-cases, but
with limited post-estimation methods and options (e.g., conducting
inference on different VCOVs). In sum, none of the existing packages
provides a unified interface that spans estimation, inference, and
presentation within a single, highly performant framework.

It is against this background that we introduce the \proglang{R} package
\pkg{fixest}. The package name reflects the fact that \pkg{fixest}
originated as a specialized library for fixed-effects regression.
However, it has since evolved into a general purpose tool that covers a
wide range of econometric methods and use cases. Its defining ambition
is to be both fast and coherent, and thus to fill a gap left by existing
software options. For many applied research projects and researchers,
\pkg{fixest} should suffice as a standalone package able to cover all of
their econometric needs.

\pkg{fixest} offers best-in-class performance for many frequentist
estimators, supported by a unified framework that enables estimation,
inference, and reporting in a single package. It supports OLS, IV, GLM,
and maximum likelihood (MLE) estimation through a common syntax that is
fully compatible with standard \proglang{R} \pkg{stats} methods. For
fixed-effects estimation, specifically, it introduces a novel algorithm
based on fixed-point acceleration using the method of
\citet{irons1969version}, which achieves fast convergence across many
different data contexts. The performance-critical parts of the codebase
are implemented in parallel \proglang{C++} via \pkg{Rcpp}
\citep{eddelbuettel2011jss_rcpp}, further ensuring optimal efficiency.

We illustrate \pkg{fixest} performance in a wide array of benchmarks.
Using simulated data, we generate scenarios with simple or difficult
fixed-effects convergence and compare \pkg{fixest} to leading
alternatives in \proglang{R}, \proglang{Python} and \proglang{Julia}.
\pkg{fixest} is consistently the fastest in all situations, closely
followed by the \proglang{Julia} software. A second set of benchmarks
uses a large real-life data set to illustrate \pkg{fixest} specific
features useful in applied econometrics (detailed below). In those
cases, \pkg{fixest} speed is unmatched.

In addition to speed, \pkg{fixest} contributes several features that
distinguish it from existing software. For example, one of the core
design principles is that inference is separate from estimation. This
means that users can swap out a variety of VCOVs---including
heteroskedasticity-robust, clustered, Newey-West
\citep{newey1987simple}, Driscoll-Kraay \citep{driscoll1998consistent},
Conley \citep{conley1999gmm}, and other standard errors---at virtually
zero computational cost, since the underlying model does not have to be
re-estimated. An expressive formula interface supports programmatic
variable interpolation, factor manipulation, and multiple estimations in
a single call, where shared computations are pooled across models for
additional speed gains. \pkg{fixest} also provides built-in tools for
publication-ready regression tables and coefficient plots, and invests
heavily in informative error messages that help users diagnose problems
quickly.

This article describes \pkg{fixest} version 0.14.0, available from the
Comprehensive R Archive Network at
\url{https://cran.r-project.org/package=fixest}. We begin with a
self-contained motivating example in Section \ref{sec_example} that
introduces the core syntax through an end-to-end analysis. Section
\ref{sec_theory} presents the econometric theory and associated
computational considerations for estimating fixed-effects models, as
well detailing \pkg{fixest}'s novel fixed-point acceleration solution.
Sections \ref{sec_estimation} through \ref{sec_presentation} describe
the package's estimation functions, formula interface, VCOV options, and
presentation tools. Section \ref{sec_benchmarks} reports benchmarks
against leading alternatives. Section \ref{sec_conclu} discusses
\pkg{fixest}'s integration with the broader \proglang{R} ecosystem and
pedagogical context, before concluding with the package's development
philosophy.

\section{Motivating example: Estimating the returns to science
funding}\label{motivating-example-estimating-the-returns-to-science-funding}

\label{sec_example}

Assume that we have data on scientific researchers in the EU and US. We
are able to track their individual productivity levels (as measured by
the number of articles produced every year), as well as their access to
funding (in \$US'000). Our goal is to estimate the effect that funding
has on scientific production, whilst controlling for differences across
individuals, years and regions.

For demonstration purposes, let us create a dummy dataset called
\texttt{scipubs} with six variables: number of articles
(\texttt{articles}), funding received (\texttt{funding}), region
(\texttt{eu\_us}), a dummy representing a policy shock
(\texttt{policy}), scientist identifier (\texttt{indiv}), and the year
(\texttt{year}).

\begin{Shaded}
\begin{Highlighting}[]
\FunctionTok{library}\NormalTok{(}\StringTok{"fixest"}\NormalTok{)}
\FunctionTok{data}\NormalTok{(base\_did)}
\NormalTok{scipubs }\OtherTok{=} \FunctionTok{with}\NormalTok{(}
\NormalTok{  base\_did, }
  \FunctionTok{data.frame}\NormalTok{(}
    \AttributeTok{articles =} \FunctionTok{round}\NormalTok{(y }\SpecialCharTok{{-}} \FunctionTok{min}\NormalTok{(y)), }
    \AttributeTok{funding  =} \FunctionTok{round}\NormalTok{(x1 }\SpecialCharTok{{-}} \FunctionTok{min}\NormalTok{(x1)) }\SpecialCharTok{*} \DecValTok{10}\NormalTok{,}
    \AttributeTok{eu\_us    =} \FunctionTok{ifelse}\NormalTok{(treat, }\StringTok{"EU"}\NormalTok{, }\StringTok{"US"}\NormalTok{), }
    \AttributeTok{policy   =} \SpecialCharTok{+}\NormalTok{(period }\SpecialCharTok{{-}}\NormalTok{ id }\SpecialCharTok{\textgreater{}} \DecValTok{0}\NormalTok{),}
    \AttributeTok{indiv    =}\NormalTok{ id, }
    \AttributeTok{year     =}\NormalTok{ period}
\NormalTok{  )}
\NormalTok{)}
\end{Highlighting}
\end{Shaded}

\subsection{Ordinary least squares
estimation}\label{ordinary-least-squares-estimation}

\label{sec_example_ols}

We start with a simple ordinary least squares (OLS) model, where we
regress science production (\texttt{articles}) on funding
(\texttt{funding}) and a set of control variables. OLS estimations in
\pkg{fixest} are performed with the function \texttt{feols}. In its
simplest guise, \texttt{feols} can be seen as a drop-in replacement for
the base \proglang{R} \texttt{stats::lm} function. For example, we can
use simple indicator (factor) variables to control for individual and
year fixed-effects:

\begin{Shaded}
\begin{Highlighting}[]
\FunctionTok{feols}\NormalTok{(articles }\SpecialCharTok{\textasciitilde{}}\NormalTok{ funding }\SpecialCharTok{+} \FunctionTok{as.factor}\NormalTok{(indiv) }\SpecialCharTok{+} \FunctionTok{as.factor}\NormalTok{(year), scipubs)}
\CommentTok{\#\textgreater{} OLS estimation, Dep. Var.: articles}
\CommentTok{\#\textgreater{} Observations: 1,080}
\CommentTok{\#\textgreater{} Standard{-}errors: IID }
\CommentTok{\#\textgreater{}                    Estimate Std. Error   t value   Pr(\textgreater{}|t|)    }
\CommentTok{\#\textgreater{} (Intercept)        5.777038   1.508346  3.830048 0.00013642 ***}
\CommentTok{\#\textgreater{} funding            0.096365   0.004702 20.492577  \textless{} 2.2e{-}16 ***}
\CommentTok{\#\textgreater{} as.factor(indiv)2  2.634578   1.943905  1.355302 0.17563956    }
\CommentTok{\#\textgreater{} as.factor(indiv)3 {-}2.925442   1.942340 {-}1.506143 0.13235874    }
\CommentTok{\#\textgreater{} as.factor(indiv)4  1.114538   1.942152  0.573868 0.56619154    }
\CommentTok{\#\textgreater{} as.factor(indiv)5 {-}0.192731   1.942084 {-}0.099239 0.92096903    }
\CommentTok{\#\textgreater{} as.factor(indiv)6  1.950884   1.943177  1.003966 0.31564731    }
\CommentTok{\#\textgreater{} as.factor(indiv)7  2.347249   1.943023  1.208040 0.22732881    }
\CommentTok{\#\textgreater{} ... 110 coefficients remaining (display them with summary() or use}
\CommentTok{\#\textgreater{} argument n)}
\CommentTok{\#\textgreater{} {-}{-}{-}}
\CommentTok{\#\textgreater{} Signif. codes:  0 \textquotesingle{}***\textquotesingle{} 0.001 \textquotesingle{}**\textquotesingle{} 0.01 \textquotesingle{}*\textquotesingle{} 0.05 \textquotesingle{}.\textquotesingle{} 0.1 \textquotesingle{} \textquotesingle{} 1}
\CommentTok{\#\textgreater{} RMSE: 4.09849   Adj. R2: 0.420879}
\end{Highlighting}
\end{Shaded}

The above \texttt{feols} results are identical to those from an
equivalent \texttt{stats::lm} call.\footnote{Specifically,
  \texttt{lm(articles\ \textasciitilde{}\ funding\ +\ as.factor(indiv)\ +\ as.factor(year),\ scipubs)}.}
However, the output is slightly different insofar as \texttt{feols}
automatically displays the results in summary form (including standard
errors, R-squared, etc.) and avoids displaying all the coefficients
unless explicitly requested by the user. In contrast, the default
\texttt{lm} print method would only display the point estimates and
would do so for all 118 coefficients.

This interchangeability between \texttt{feols} and \texttt{lm} is
convenient, since only the function call changes. However, it is not the
preferred way to estimate a fixed-effects model with \texttt{feols}.
Rather, we can take advantage of \pkg{fixest}'s specialized architecture
for fixed-effects by invoking the syntax:

\[
\mathtt{dep\_var} \sim \mathtt{indep\_vars} \mathtt{\ |\ } \mathtt{fixed\_effects} 
\]

In a \pkg{fixest} formula, any variable after the pipe
(\texttt{\textbar{}}) is treated as a fixed-effect and passed to a
highly optimized internal routine. Applying this principle to our
example, we have:

\begin{Shaded}
\begin{Highlighting}[]
\FunctionTok{feols}\NormalTok{(articles }\SpecialCharTok{\textasciitilde{}}\NormalTok{ funding }\SpecialCharTok{|}\NormalTok{ indiv }\SpecialCharTok{+}\NormalTok{ year, scipubs)}
\CommentTok{\#\textgreater{} OLS estimation, Dep. Var.: articles}
\CommentTok{\#\textgreater{} Observations: 1,080}
\CommentTok{\#\textgreater{} Fixed{-}effects: indiv: 108,  year: 10}
\CommentTok{\#\textgreater{} Standard{-}errors: IID }
\CommentTok{\#\textgreater{}         Estimate Std. Error t value  Pr(\textgreater{}|t|)    }
\CommentTok{\#\textgreater{} funding 0.096365   0.004702 20.4926 \textless{} 2.2e{-}16 ***}
\CommentTok{\#\textgreater{} {-}{-}{-}}
\CommentTok{\#\textgreater{} Signif. codes:  0 \textquotesingle{}***\textquotesingle{} 0.001 \textquotesingle{}**\textquotesingle{} 0.01 \textquotesingle{}*\textquotesingle{} 0.05 \textquotesingle{}.\textquotesingle{} 0.1 \textquotesingle{} \textquotesingle{} 1}
\CommentTok{\#\textgreater{} RMSE: 4.09849     Adj. R2: 0.420879}
\CommentTok{\#\textgreater{}                 Within R2: 0.30388 }
\end{Highlighting}
\end{Shaded}

Observe that our estimate of the key \texttt{funding} coefficient
remains identical to our previous result of \texttt{0.096365}. Moreover,
because we specified \texttt{indiv} and \texttt{year} as fixed-effects,
the return object correctly recognizes these variables as controls
(``nuisance'' parameters) and avoids displaying the 100+ coefficients on
them. At the same time, it is important to note that the benefits extend
beyond aesthetics. This second estimation completes quicker than the
first one, exactly because the fixed-effects are handled by a
specialized and highly optimized routine. We demonstrate the speed
advantages in a benchmarking section later in the paper.

To interpret the coefficients, we need to know whether they are
statistically significant. Hence, we need to report their
standard-errors. By default \texttt{feols} assumes that the errors are
homoskedastic and non-correlated (labelled for convenience as ``IID'',
or ``independent and identically distributed'' in our return object
above).

But \pkg{fixest} makes it easy to specify a wide variety of alternate
standard-error types. For example, because our data is a panel (i.e.,
repeated cross-sections of the same individuals over time), we may be
worried that the errors are correlated within individuals. In
econometric parlance, this suggests the need to \emph{cluster} the
standard-errors at the individual level. We can implement this directly
at estimation time by using the argument
\texttt{vcov\ =\ \textasciitilde{}indiv}:

\begin{Shaded}
\begin{Highlighting}[]
\FunctionTok{feols}\NormalTok{(articles }\SpecialCharTok{\textasciitilde{}}\NormalTok{ funding }\SpecialCharTok{|}\NormalTok{ indiv }\SpecialCharTok{+}\NormalTok{ year, scipubs, }\AttributeTok{vcov =} \SpecialCharTok{\textasciitilde{}}\NormalTok{indiv)}
\CommentTok{\#\textgreater{} OLS estimation, Dep. Var.: articles}
\CommentTok{\#\textgreater{} Observations: 1,080}
\CommentTok{\#\textgreater{} Fixed{-}effects: indiv: 108,  year: 10}
\CommentTok{\#\textgreater{} Standard{-}errors: Clustered (indiv) }
\CommentTok{\#\textgreater{}         Estimate Std. Error t value  Pr(\textgreater{}|t|)    }
\CommentTok{\#\textgreater{} funding 0.096365   0.004542 21.2172 \textless{} 2.2e{-}16 ***}
\CommentTok{\#\textgreater{} {-}{-}{-}}
\CommentTok{\#\textgreater{} Signif. codes:  0 \textquotesingle{}***\textquotesingle{} 0.001 \textquotesingle{}**\textquotesingle{} 0.01 \textquotesingle{}*\textquotesingle{} 0.05 \textquotesingle{}.\textquotesingle{} 0.1 \textquotesingle{} \textquotesingle{} 1}
\CommentTok{\#\textgreater{} RMSE: 4.09849     Adj. R2: 0.420879}
\CommentTok{\#\textgreater{}                 Within R2: 0.30388 }
\end{Highlighting}
\end{Shaded}

Of course, other standard-error formulations are also possible. While
one may be tempted to re-run multiple versions of the same model with
slight variations to the standard-errors, this quickly becomes
inefficient. A core design principle of \pkg{fixest} is that computation
should be separate from inference. As such, \pkg{fixest} makes it easy
to adjust standard errors for existing models ``on the fly''. Almost all
of the package's post-estimation routines and methods allow for this
kind of \emph{post hoc} adjustment. Here we briefly demonstrate with a
simple side-by-side comparison of two standard types, using the
\texttt{etable} function to display the results in regression table
form:

\begin{Shaded}
\begin{Highlighting}[]
\FunctionTok{feols}\NormalTok{(articles }\SpecialCharTok{\textasciitilde{}}\NormalTok{ funding }\SpecialCharTok{|}\NormalTok{ indiv }\SpecialCharTok{+}\NormalTok{ year, scipubs) }\SpecialCharTok{|\textgreater{}}
  \FunctionTok{etable}\NormalTok{(}\AttributeTok{vcov =} \FunctionTok{list}\NormalTok{(}\StringTok{"iid"}\NormalTok{, }\SpecialCharTok{\textasciitilde{}}\NormalTok{indiv))}
\CommentTok{\#\textgreater{}                 feols(articles \textasciitilde{}.. feols(articles \textasciitilde{}...1}
\CommentTok{\#\textgreater{} Dependent Var.:           articles             articles}
\CommentTok{\#\textgreater{}                                                        }
\CommentTok{\#\textgreater{} funding         0.0964*** (0.0047)   0.0964*** (0.0045)}
\CommentTok{\#\textgreater{} Fixed{-}Effects:  {-}{-}{-}{-}{-}{-}{-}{-}{-}{-}{-}{-}{-}{-}{-}{-}{-}{-}   {-}{-}{-}{-}{-}{-}{-}{-}{-}{-}{-}{-}{-}{-}{-}{-}{-}{-}}
\CommentTok{\#\textgreater{} indiv                          Yes                  Yes}
\CommentTok{\#\textgreater{} year                           Yes                  Yes}
\CommentTok{\#\textgreater{} \_\_\_\_\_\_\_\_\_\_\_\_\_\_\_ \_\_\_\_\_\_\_\_\_\_\_\_\_\_\_\_\_\_   \_\_\_\_\_\_\_\_\_\_\_\_\_\_\_\_\_\_}
\CommentTok{\#\textgreater{} S.E. type                      IID            by: indiv}
\CommentTok{\#\textgreater{} Observations                 1,080                1,080}
\CommentTok{\#\textgreater{} R2                         0.48367              0.48367}
\CommentTok{\#\textgreater{} Within R2                  0.30388              0.30388}
\CommentTok{\#\textgreater{} {-}{-}{-}}
\CommentTok{\#\textgreater{} Signif. codes: 0 \textquotesingle{}***\textquotesingle{} 0.001 \textquotesingle{}**\textquotesingle{} 0.01 \textquotesingle{}*\textquotesingle{} 0.05 \textquotesingle{}.\textquotesingle{} 0.1 \textquotesingle{} \textquotesingle{} 1}
\end{Highlighting}
\end{Shaded}

The key argument above is
\texttt{vcov\ =\ list("iid",\ \textasciitilde{}indiv))}, which prompts
the display (and computation of) of both IID and clustered standard
errors alongside each other. Note that this type of functionality is
easily extended and that \pkg{fixest} supports many other standard-error
adjustments. We defer details until later in the paper and leave this as
a taster.

Let us return to our broader research question. Suppose we want to know
whether the effect of science funding is similar across the EU and US?
While there are multiple approaches to tackling this question, a
potential starting point is to estimate separate models for each region.
The \texttt{split} argument provides a convenient way to do this in a
single call. Again, we can pass the resulting object to \texttt{etable}
for convenient side-by-side display.\footnote{In truth, passing to
  \texttt{etable} is redundant since \pkg{fixest} multiple estimation
  objects are printed with \texttt{etable} by default. But we do so here
  to avoid assuming too much familiarity with the package at this
  introductory stage.}

\begin{Shaded}
\begin{Highlighting}[]
\FunctionTok{feols}\NormalTok{(articles }\SpecialCharTok{\textasciitilde{}}\NormalTok{ funding }\SpecialCharTok{|}\NormalTok{ indiv }\SpecialCharTok{+}\NormalTok{ year, scipubs, }\AttributeTok{split =} \SpecialCharTok{\textasciitilde{}}\NormalTok{eu\_us) }\SpecialCharTok{|\textgreater{}}
  \FunctionTok{etable}\NormalTok{()}
\CommentTok{\#\textgreater{}                 feols(articles ..1 feols(articles ..2}
\CommentTok{\#\textgreater{} Sample (eu\_us)                  EU                 US}
\CommentTok{\#\textgreater{} Dependent Var.:           articles           articles}
\CommentTok{\#\textgreater{}                                                      }
\CommentTok{\#\textgreater{} funding         0.0900*** (0.0061) 0.1035*** (0.0065)}
\CommentTok{\#\textgreater{} Fixed{-}Effects:  {-}{-}{-}{-}{-}{-}{-}{-}{-}{-}{-}{-}{-}{-}{-}{-}{-}{-} {-}{-}{-}{-}{-}{-}{-}{-}{-}{-}{-}{-}{-}{-}{-}{-}{-}{-}}
\CommentTok{\#\textgreater{} indiv                          Yes                Yes}
\CommentTok{\#\textgreater{} year                           Yes                Yes}
\CommentTok{\#\textgreater{} \_\_\_\_\_\_\_\_\_\_\_\_\_\_\_ \_\_\_\_\_\_\_\_\_\_\_\_\_\_\_\_\_\_ \_\_\_\_\_\_\_\_\_\_\_\_\_\_\_\_\_\_}
\CommentTok{\#\textgreater{} S.E. type                      IID                IID}
\CommentTok{\#\textgreater{} Observations                   550                530}
\CommentTok{\#\textgreater{} R2                         0.57106            0.44382}
\CommentTok{\#\textgreater{} Within R2                  0.30866            0.35002}
\CommentTok{\#\textgreater{} {-}{-}{-}}
\CommentTok{\#\textgreater{} Signif. codes: 0 \textquotesingle{}***\textquotesingle{} 0.001 \textquotesingle{}**\textquotesingle{} 0.01 \textquotesingle{}*\textquotesingle{} 0.05 \textquotesingle{}.\textquotesingle{} 0.1 \textquotesingle{} \textquotesingle{} 1}
\end{Highlighting}
\end{Shaded}

Thus far, we have made minimal assumptions about the causal direction of
our results. An obvious complication is the likely endogeneity of
scientific funding (e.g., better researchers with more publications get
more funding). We may therefore wish to \emph{instrument} funding with
an unexpected policy shock that changed funding availability in an
exogenous way. For our toy example here, we can use the \texttt{policy}
variable---corresponding to a dummy on certain years---as an instrument.
To perform an instrumental variable (IV) estimation in \pkg{fixest}, we
use the following syntax:

\[
\mathtt{dep\_var} \sim \mathtt{exogenous\_vars} \mathtt{\ |\ } \mathtt{fixed\_effects} \mathtt{\ |\ } \mathtt{endo\_vars} \sim \mathtt{instruments}
\]

Applying this syntax to our research productivity question, we have:

\begin{Shaded}
\begin{Highlighting}[]
\FunctionTok{feols}\NormalTok{(articles }\SpecialCharTok{\textasciitilde{}} \DecValTok{1} \SpecialCharTok{|}\NormalTok{ indiv }\SpecialCharTok{+}\NormalTok{ year }\SpecialCharTok{|}\NormalTok{ funding }\SpecialCharTok{\textasciitilde{}}\NormalTok{ policy, scipubs)}
\CommentTok{\#\textgreater{} TSLS estimation}
\CommentTok{\#\textgreater{} |{-} D.V.   : articles}
\CommentTok{\#\textgreater{} |{-} Endo.  : funding}
\CommentTok{\#\textgreater{} |{-} Instr. : policy}
\CommentTok{\#\textgreater{} |}
\CommentTok{\#\textgreater{} |=\textgreater{} Second Stage}
\CommentTok{\#\textgreater{} |   Dep. Var.: articles}
\CommentTok{\#\textgreater{} Observations: 1,080}
\CommentTok{\#\textgreater{} Fixed{-}effects: indiv: 108,  year: 10}
\CommentTok{\#\textgreater{} Standard{-}errors: IID }
\CommentTok{\#\textgreater{}              Estimate Std. Error   t value Pr(\textgreater{}|t|) }
\CommentTok{\#\textgreater{} fit\_funding {-}0.245182    1.63019 {-}0.150401  0.88048 }
\CommentTok{\#\textgreater{} {-}{-}{-}}
\CommentTok{\#\textgreater{} Signif. codes:  0 \textquotesingle{}***\textquotesingle{} 0.001 \textquotesingle{}**\textquotesingle{} 0.01 \textquotesingle{}*\textquotesingle{} 0.05 \textquotesingle{}.\textquotesingle{} 0.1 \textquotesingle{} \textquotesingle{} 1}
\CommentTok{\#\textgreater{} RMSE: 10.4     Adj. R2: 0.168161}
\CommentTok{\#\textgreater{}              Within R2: 1.061e{-}4}
\CommentTok{\#\textgreater{} F{-}test (1st stage), funding: stat = 0.05190, p = 0.819832, on 1 and 962 DoF.}
\CommentTok{\#\textgreater{}                  Wu{-}Hausman: stat = 0.28442, p = 0.593945, on 1 and 961 DoF.}
\end{Highlighting}
\end{Shaded}

The results of the second stage estimation are reported, as well as some
fit statistics, like the (weak instrument) F-test for the first stage
and the Wu-Hausman endogeneity test.

\subsection{Generalized linear model
estimation}\label{generalized-linear-model-estimation}

All of our estimations thus far have modeled \texttt{articles} as the
outcome of interest. But number of articles is a positive count
(integer). Hence it is arguably more natural to model the outcome as a
Poisson process. In \pkg{fixest} this is easily done via the
\texttt{fepois} function:

\begin{Shaded}
\begin{Highlighting}[]
\FunctionTok{fepois}\NormalTok{(articles }\SpecialCharTok{\textasciitilde{}}\NormalTok{ funding }\SpecialCharTok{|}\NormalTok{ indiv }\SpecialCharTok{+}\NormalTok{ year, scipubs, }\AttributeTok{vcov =} \SpecialCharTok{\textasciitilde{}}\NormalTok{indiv)}
\CommentTok{\#\textgreater{} Poisson estimation, Dep. Var.: articles}
\CommentTok{\#\textgreater{} Observations: 1,080}
\CommentTok{\#\textgreater{} Fixed{-}effects: indiv: 108,  year: 10}
\CommentTok{\#\textgreater{} Standard{-}errors: Clustered (indiv) }
\CommentTok{\#\textgreater{}         Estimate Std. Error z value  Pr(\textgreater{}|z|)    }
\CommentTok{\#\textgreater{} funding 0.005644   0.000292 19.3214 \textless{} 2.2e{-}16 ***}
\CommentTok{\#\textgreater{} {-}{-}{-}}
\CommentTok{\#\textgreater{} Signif. codes:  0 \textquotesingle{}***\textquotesingle{} 0.001 \textquotesingle{}**\textquotesingle{} 0.01 \textquotesingle{}*\textquotesingle{} 0.05 \textquotesingle{}.\textquotesingle{} 0.1 \textquotesingle{} \textquotesingle{} 1}
\CommentTok{\#\textgreater{} Log{-}Likelihood: {-}3,070.4   Adj. Pseudo R2: 0.105804}
\CommentTok{\#\textgreater{}            BIC:  6,964.9     Squared Cor.: 0.482025}
\end{Highlighting}
\end{Shaded}

As the reader will note, the syntax is identical to the \texttt{feols}
case; we simply switch out the main function call. The same is true for
a variety of other generalized model families that \pkg{fixest} supports
through its \texttt{feglm} and \texttt{femlm} functions. Again, however,
we leave this a taster and defer details until later in the paper.

\subsection{Presentation}\label{presentation}

Having iterated through various models, let us conclude this
introductory example by considering how we might present our findings.
Assume we have settled on the following variants: Poisson model with the
EU/US region split (and full sample), and the OLS-IV model.

The same \texttt{etable} function from before can be used to compile all
of our preferred results into a single \LaTeX table. We need simply
specify a valid \texttt{file} path to write \LaTeX output instead of
printing the table to screen. We also invoke some additional arguments:

\begin{itemize}
\tightlist
\item
  the \texttt{vcov} argument is used to adjust the standard-errors for
  all models on-the-fly.
\item
  the \texttt{stage} argument is used to display both the 1st and 2nd
  stage of the IV estimation.
\item
  the \texttt{dict} argument used to pass a dictionary for displaying
  nicer variable names.\\
\item
  the \texttt{caption} and \texttt{label} arguments pass helper macros
  to the saved \LaTeX table.
\end{itemize}

\begin{Shaded}
\begin{Highlighting}[]
\NormalTok{est\_split }\OtherTok{=} \FunctionTok{feols}\NormalTok{(articles }\SpecialCharTok{\textasciitilde{}}\NormalTok{ funding }\SpecialCharTok{|}\NormalTok{ indiv }\SpecialCharTok{+}\NormalTok{ year, scipubs,}
                  \AttributeTok{fsplit =} \SpecialCharTok{\textasciitilde{}}\NormalTok{ eu\_us)}
\NormalTok{est\_iv }\OtherTok{=} \FunctionTok{feols}\NormalTok{(articles }\SpecialCharTok{\textasciitilde{}} \DecValTok{1} \SpecialCharTok{|}\NormalTok{ indiv }\SpecialCharTok{+}\NormalTok{ year }\SpecialCharTok{|}\NormalTok{ funding }\SpecialCharTok{\textasciitilde{}}\NormalTok{ policy, scipubs)}
\FunctionTok{etable}\NormalTok{(est\_split, est\_iv,}
       \AttributeTok{vcov   =} \SpecialCharTok{\textasciitilde{}}\NormalTok{indiv,}
       \AttributeTok{stage  =} \DecValTok{1}\SpecialCharTok{:}\DecValTok{2}\NormalTok{,}
       \AttributeTok{dict   =} \FunctionTok{c}\NormalTok{(}\AttributeTok{articles =} \StringTok{"\# Articles"}\NormalTok{,}
                   \AttributeTok{funding =} \StringTok{"Funding (\textquotesingle{}000 $US)"}\NormalTok{,}
                    \AttributeTok{policy =} \StringTok{"Policy"}\NormalTok{,}
                    \AttributeTok{indiv =} \StringTok{"Researcher"}\NormalTok{, }
                     \AttributeTok{year =} \StringTok{"Year"}\NormalTok{,}
                    \AttributeTok{eu\_us =} \StringTok{"Region"}\NormalTok{),}
       \AttributeTok{file    =} \StringTok{"tables/first\_example.tex"}\NormalTok{,}
       \AttributeTok{caption =} \FunctionTok{c}\NormalTok{(}\StringTok{"Compiling the results from five estimations: "}\NormalTok{,}
                   \StringTok{"Default LaTeX rendering."}\NormalTok{),}
       \AttributeTok{label   =} \StringTok{"tab\_first\_ex"}\NormalTok{)}
\end{Highlighting}
\end{Shaded}

The resulting table is shown in Table \ref{tab_first_ex}.

\begin{table}[htbp]
   \caption{\label{tab_first_ex} Compiling the results from five estimations: Default LaTeX rendering.}
   \centering
   \begin{tabular}{lccccc}
      \tabularnewline \midrule \midrule
      Dependent Variables: & \multicolumn{3}{c}{\# Articles} & Funding ('000 \$US) & \# Articles\\
      Region & Full sample & EU & US & \multicolumn{2}{c}{} \\ 
      IV stages & \multicolumn{3}{c}{ } & First & Second \\ 
      Model:               & (1)            & (2)            & (3)            & (4)                  & (5)\\  
      \midrule
      \emph{Variables}\\
      Funding ('000 \$US)  & 0.0964$^{***}$ & 0.0900$^{***}$ & 0.1035$^{***}$ &                      & -0.2452\\   
                           & (0.0045)       & (0.0065)       & (0.0060)       &                      & (1.381)\\   
      Policy               &                &                &                & -1.710               &   \\   
                           &                &                &                & (9.013)              &   \\   
      \midrule
      \emph{Fixed-effects}\\
      Researcher           & Yes            & Yes            & Yes            & Yes                  & Yes\\  
      Year                 & Yes            & Yes            & Yes            & Yes                  & Yes\\  
      \midrule
      \emph{Fit statistics}\\
      Observations         & 1,080          & 550            & 530            & 1,080                & 1,080\\  
      R$^2$                & 0.48367        & 0.57106        & 0.44382        & 0.12131              & 0.25836\\  
      Within R$^2$         & 0.30388        & 0.30866        & 0.35002        & $5.4\times 10^{-5}$  & 0.00011\\  
      \midrule \midrule
      \multicolumn{6}{l}{\emph{Clustered (Researcher) standard-errors in parentheses}}\\
      \multicolumn{6}{l}{\emph{Signif. Codes: ***: 0.01, **: 0.05, *: 0.1}}\\
   \end{tabular}
\end{table}

\begin{figure}[htb]
  \centering
  \caption{\label{fig_first_example}Default \texttt{coefplot} graph for split sample estimation.}
  
  \includegraphics[width=1\textwidth]{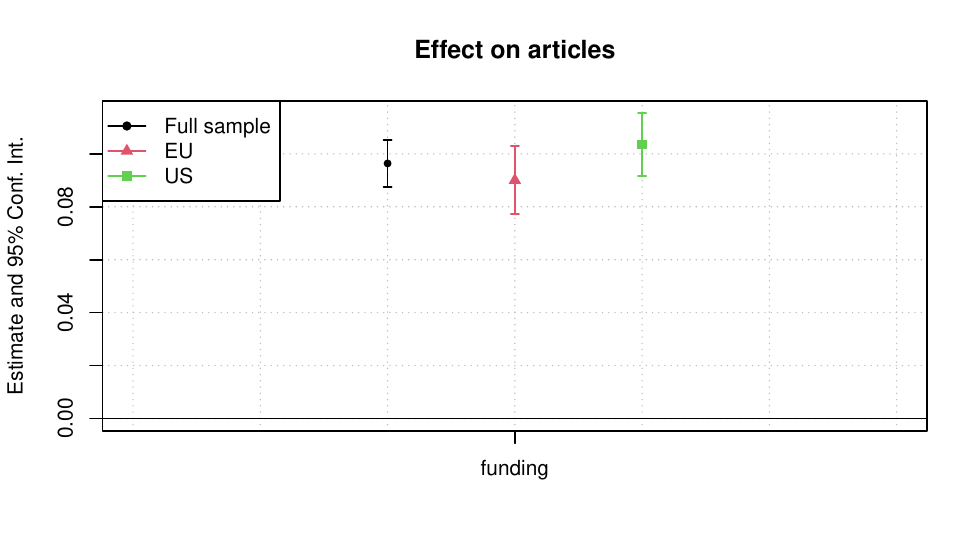}
\end{figure}

In addition to a regression table, we might also wish to present our
results in the form of a coefficient plot via the dedicated
\texttt{coefplot} function. Below we demonstrate with a simple example,
limited to the split-sample Poisson estimation. The resulting plot is
displayed in Figure \ref{fig_first_example}.

\begin{Shaded}
\begin{Highlighting}[]
\FunctionTok{coefplot}\NormalTok{(est\_split, }\AttributeTok{vcov =} \SpecialCharTok{\textasciitilde{}}\NormalTok{indiv)}
\FunctionTok{legend}\NormalTok{(}\StringTok{"topleft"}\NormalTok{, }\FunctionTok{c}\NormalTok{(}\StringTok{"Full sample"}\NormalTok{, }\StringTok{"EU"}\NormalTok{, }\StringTok{"US"}\NormalTok{), }
       \AttributeTok{col =} \DecValTok{1}\SpecialCharTok{:}\DecValTok{3}\NormalTok{, }\AttributeTok{lty =} \DecValTok{1}\NormalTok{, }\AttributeTok{pch =} \FunctionTok{c}\NormalTok{(}\DecValTok{16}\NormalTok{, }\DecValTok{17}\NormalTok{, }\DecValTok{15}\NormalTok{), }\AttributeTok{bg =} \StringTok{"white"}\NormalTok{)}
\end{Highlighting}
\end{Shaded}

Thus concludes our introductory example. Having established the core
intuition and syntax, we now examine the econometric theory and
computational algorithms that make \pkg{fixest} estimations fast and
reliable. We begin a theory review, before turning to a detailed review
of the package's main functions and features.

\section{Fixed-effects: Theory and the fixest
implementation}\label{fixed-effects-theory-and-the-fixest-implementation}

\label{sec_theory}

\pkg{fixest} originated as a package specifically optimized for tackling
\emph{fixed-effects} estimations. However, since the definition of the
term `fixed-effects' can differ greatly across fields, we first explain
the meaning of a fixed-effects estimation from our perspective. We also
show how it is tightly linked to variable demeaning. We then document
existing solutions to demeaning problems and detail \pkg{fixest}'s
specific approach.

We offer two expositional notes before proceeding. First, we focus on
OLS estimation. The extension to generalized linear models is discussed
in Section \ref{sec_glm_ml}. Second, we focus on a very simple setup;
namely, a balanced panel with only two fixed-effects dimensions
(individual and time). The logic readily extends to imbalanced panels
and any number of fixed-effects. However, this level of generality would
necessitate ugly notations that can obscure the key insights without
corresponding benefit. Hence, we refrain from attempting this here. See
\citet{Berge2018_FENmlm} for a treatment of the general case in the
context of maximum likelihood estimation.

\subsection{Fixed-effects and the FWL theorem: A demeaning
problem}\label{fixed-effects-and-the-fwl-theorem-a-demeaning-problem}

For illustration, consider the setup of a balanced panel in which we
follow individuals, indexed by \(i\), across years, indexed by \(t\).
There are \(N_I\) individuals and \(N_T\) periods, leading to
\(N = N_I\times N_T\) observations. Let \(y\) be the \(N\)-vector
representing the dependent variable and \(X\) a \(N \times K\) matrix of
explanatory variables. We are interested in the following data
generating process:

\begin{equation}\phantomsection\label{eq-model-panel}{
y_{it} = \alpha_i + \beta_t + X_{it} \gamma + \epsilon_{it},
}\end{equation}

with \(X_{it}\) the row-vector of explanatory variables for individual
\(i\) and time \(t\), and \(\gamma\) the \(K\)-vector of the
\emph{coefficients of interest}. In the context of the previous
equation, the coefficients \(\alpha_i\) capture any confounder
(i.e.~explanatory variables that are not in the equation) that is
individual-specific and time-invariant. Similarly, \(\beta_t\) captures
any time-specific confounder. Finally, \(\epsilon_{it}\) is a random
variable representing the error term.

It is important to stress that we are not directly interested in the
values of \(\alpha_i\) and \(\beta_t\). Rather, they are effectively
nuisance parameters, insofar as they provide a means of controlling for
a set of confounders. Indeed, \(\alpha_i\) and \(\beta_t\) here
represent the \emph{fixed effects} of the estimation, since they capture
invariant (fixed) confounders across the unit and time dimensions,
respectively. Fixed effects are often written as \(\alpha_i\), but this
is really short-hand notation for a set of \(N_I\) indicator variables
for each unit with a coefficient \(\alpha_i\) for each (equivalent to
\texttt{factor(unit)} in a regression). Extending this idea more
generally, we define ``fixed-effects estimation'' in this article simply
as an estimation that includes categorical variables as controls.

Now let us link fixed-effects with demeaning. Consider the estimation of
the coefficients of Equation \eqref{eq-model-panel} by OLS. With a hat
referring to the estimated coefficients and \(r_{it}\) to the residual,
we have:

\begin{equation}\phantomsection\label{eq-est-panel}{
y_{it} = \hat{\alpha}_i + \hat{\beta}_t + X_{it} \hat{\gamma} + r_{it}.
}\end{equation}

Let us define \(\tilde{y}\) as the residual of the OLS estimation of
\(y\) on the fixed-effects only:

\[
\tilde{y}_{it} \overset{\text{def}}{=} y_{it} - \hat{\alpha}_i^y - \hat{\beta}_t^y,
\]

with \(\hat{\alpha}_i^y\) the OLS estimator of the fixed-effect
coefficient.

In a similar fashion, let us define matrix \(\tilde{X}\) whose
\(k^{th}\) column is defined as the residual of the estimation of the
\(k^{th}\) column of \(X\) on the fixed-effects only. If we estimate
\(\tilde{y}\) on \(\tilde{X}\) by OLS, we have:

\begin{equation}\phantomsection\label{eq-est-reduced}{
\tilde{y}_{it} = \tilde{X}_{it} \hat{\gamma}_{bis} + r_{it}^{bis}.
}\end{equation}

The Frisch-Waugh-Lovell
\citep{frisch1933PartialTimeRegressions, lovell1963SeasonalAdjustmentEconomic}
theorem states that the coefficients \(\hat{\gamma}\) and the residuals
\(r_{it}\) of Equation \eqref{eq-est-panel} are the same as the ones in
Equation \eqref{eq-est-reduced}. That is
\(\hat{\gamma} = \hat{\gamma}_{bis}\) and \(r_{it} = r_{it}^{bis}\). It
follows that the variance-covariance matrices of the coefficients are
also similar:
\(V\left(\hat{\gamma}\right)=s\times V\left(\hat{\gamma}_{bis}\right)\),
with \(s \in \left[0;1\right]\) a scaling factor.\footnote{See
  \citet{ding2021FrischWaughLovell} for intuitive demonstrations.}

The equivalence between Equation \eqref{eq-est-panel} and Equation
\eqref{eq-est-reduced} means that instead of estimating
\(N_I + N_T + K\) coefficients, we can estimate only \(K\) coefficients.
Given that solving an OLS problem is of complexity
\(\mathcal{O}\left(n_{coef}^3\right)\) with \(n_{coef}\) the number of
coefficients to estimate, avoiding the estimation of \(N_I + N_T\)
coefficients greatly reduces the dimensionality of the
problem.\footnote{For large $N_I$, or $N_T$, the direct estimation of Equation \eqref{eq-est-panel} is simply intractable.}

However, to estimate Equation \eqref{eq-est-reduced}, one still needs to
compute \(\tilde{y}\) and \(\tilde{X}\) first. The good news is that
computing these values is generally fast, thanks the simple structure of
the fixed-effects. Specifically, the computing step amounts to a
demeaning exercise, which we now illustrate.

To compute \(\tilde{y}\), we need to find the coefficients
\(\hat{\alpha}_i^y\) and \(\hat{\beta}_t^y\) that solve the following
OLS minimization problem:

\begin{equation}\phantomsection\label{eq-opt-pblm}{
\left( \hat{\alpha}_i^y, \hat{\beta}_t^y \right)_{i,t} = 
  \arg \min_{\alpha_i,\beta_t} \sum_i \sum_t \left( y_i - \alpha_i - \beta_t\right)^2.
}\end{equation}

We cannot solve for \(\alpha_i\) and \(\beta_t\) directly, since we
would be confronted by the aforementioned
\(\mathcal{O}\left(n_{coef}^3\right)\) complexity of the OLS solution.
Instead, we can solve it iteratively by minimizing over \(\alpha_i\) and
\(\beta_t\) sequentially until we obtain the optimal coefficients. Let
\(\alpha_i^{(m)}\) be the coefficients minimizing Equation
\eqref{eq-opt-pblm} given \(\beta_t^{(m-1)}\) at iteration \(m\). We
define \(\beta_t^{(m)}\) similarly, with
\(\beta_t^{(0)} = 0, \forall t\). The first order conditions (FOC) of
Equation \eqref{eq-opt-pblm} lead to:

\begin{equation}\phantomsection\label{eq-FOC-solution}{
\begin{split}
\alpha_i^{(m)} &= \dfrac{\sum_t y_{it}-\beta_t^{(m-1)}}{N_T} \\
\beta_t^{(m)} &= \dfrac{\sum_i y_{it}-\alpha_i^{(m)}}{N_I}
\end{split}
}\end{equation}

As we can see, the coefficients satisfying the FOCs are simple group
means; hence our describing it as a demeaning problem. The complexity of
obtaining all \(N_I\) coefficients \(\alpha_i^{m}\) is
\(\mathcal{O}\left(n_{obs}\right)\). Let \(n_{iter}\) be the number of
iterations required for convergence, such that
\(|\alpha_i^{(n_{iter}+1)}-\alpha_i^{(n_{iter})}| \approx 0, \forall i\)
and
\(|\beta_t^{(n_{iter}+1)}-\beta_t^{(n_{iter})}| \approx 0, \forall t\).
Then for large \(N_I\), \(N_T\) and for small \(n_{iter}\), obtaining
\(\hat{\gamma}\) from Equation \eqref{eq-est-reduced} is much faster
than directly estimating Equation \eqref{eq-est-panel}.

To estimate \(\hat{\gamma}\) we end up with a total complexity of
\(\mathcal{O}\left(K \times n_{iter} \times n_{obs} \times n_{fixef}^2\right)\).
We can see that the number of iterations, \(n_{iter}\), plays a critical
role in reducing complexity. Defining an algorithmic routine that
ensures a low value of \(n_{iter}\) is thus essential to reducing
computing time and cost. Before detailing \pkg{fixest}'s novel approach,
we first describe existing solutions offered by other software.

\subsection{Fixed-effects solutions in other
software}\label{fixed-effects-solutions-in-other-software}

As we have shown, the key to a fast OLS estimation with fixed-effects is
finding a computationally efficient solution to the fixed-effects-only
regression described by Equation \eqref{eq-opt-pblm}. In fact, this
problem corresponds to a sparse linear regression, and various solutions
have been offered to solve it. The most widely known of which is the
method of alternating projects.

\subsubsection{Alternating projections}\label{alternating-projections}

The method of alternating projections \citep{gaure2013ols} is almost
equivalent to the demeaning problem defined in the previous section. The
only difference is that, instead of finding the solution coefficients,
only the residuals of the OLS regression is computed (\(\tilde{y}\)).

To illustrate this method, let us rewrite the problem with matrices. We
drop the double indexing \(it\) and use a single index \(o\) to refer to
the observations. Let \(D_I\) be the \(N \times N_I\) matrix such that
\((D_I)_{o,i}\) is equal to 1 if observation \(o\) refers to individual
\(i\) and 0 otherwise. We define \(D_T\) similarly.

Using these matrix notations, the regression of \(y\) on the individual
fixed-effects leads to the classic OLS solution
\(\hat\alpha = (D_I^\top D_I)^{-1}D_I^\top y\). Let the \(N \times N\)
matrix \(R_I\) be the \emph{residual maker}, defined as:

\[
R_I \overset{\text{def}}{=} I_N - D_I(D_I^\top D_I)^{-1}D_I^\top.
\]

The product \(R_I y\) gives the residual of \(y\) such that
\((R_I y)_{o} = y_o - \hat\alpha_i\), with observation \(o\) referring
to individual \(i\). You can see a direct equivalence between the value
of this residual and the right hand numerators in Equation
\eqref{eq-FOC-solution}.

With \(R_T\) defined similarly, we can reach the residual of the OLS
estimation on the \emph{two} fixed-effects by sequentially applying
\(R_I\) and \(R_T\), such that:

\[
\tilde{y} = R_T R_I ... R_T R_I y.
\]

One algorithm to obtain \(\tilde y\) is: set \(y^{(0)} = y\), then
compute iteratively \(y^{m} = R_T R_I y^{(m-1)}\) until
\(||y^{m} - y^{(m-1)}||^\infty \approx 0\). Each update of \(y^{(m)}\)
is known as an Halperin projection. This approach avoids to compute
explicitly the coefficients of the OLS estimations at each step and is
then slightly faster than the demeaning algorithm of the previous
section, although they have similar convergence properties.\footnote{The
  astute reader may wonder: why not simply applying a quick
  exponentiation to \((R_T R_I)\) before right multiplying by \(y\)? The
  problem is that while \(R_T\) and \(R_I\) are sparse matrices, their
  product is dense, which quickly becomes problematic for \(N \times N\)
  matrices.}

This method is used in the \proglang{R} packages \pkg{lfe}
\citep{gaure2013lfe} and \pkg{alpaca}
\citep{stammann2018FastFeasibleEstimation}, and in the \proglang{Python}
package \pkg{PyFixest} \citep{fischer2024Pyfixest}.

\subsubsection{Other methods}\label{other-methods}

In the language \proglang{Stata}, \pkg{reghdfe}
\citep{correia2016feasible} uses the method of
\citet{kelner2013SimpleCombinatorialAlgorithm} to solve the linear
system of equations given by Equation \eqref{eq-opt-pblm}.

In the language \proglang{Julia}, the software \pkg{FixedEffectModels}
\citep{gomez2024FixedEffectModelsJlFast} solves the linear system using
the LSMR method \citep{fong2011LSMRIterativeAlgorithm}.

\subsection{The fixest solution: A fixed-point problem with
acceleration}\label{the-fixest-solution-a-fixed-point-problem-with-acceleration}

The package \pkg{fixest} uses a demeaning algorithm to compute
\(\tilde{y}\) and \(\tilde{X}\). But it does not use the method of
alternating projections shown above. Instead it estimates the solution
coefficients directly. Depending on the structure of the fixed-effects,
how they are distributed across observations, the number of iterations
required to reach convergence can be arbitrarily high. To ensure good
convergence properties in all circumstances, \pkg{fixest} uses a
fixed-point acceleration algorithm whose implementation we now detail.

Before starting, we must say that the current example of a balanced
panel converges in exactly one iteration, making the demeaning algorithm
of Equation \eqref{eq-FOC-solution} the fastest available for this
situation and any acceleration unnecessary. Even though we proceed with
this balanced example for illustration purposes, it may be assumed that
our method is applied to more difficult contexts where convergence
requires a high number of iterations.\footnote{A canonical example of
  difficult convergence is the context of an estimation with employees
  and firms fixed-effects. Typically the majority of employees will be
  employed in only a handful of firms across their lifetime and the
  probability that two employees are employed in the same company is
  low. From a network perspective, where employees are connected to
  other employees through companies, this means that the average network
  distance between all employees is very high; making shocks very slow
  to propagate. This contrasts sharply against the balanced panel case,
  where the average network distance between each individual is exactly
  one, and shocks propagate instantly. In Section \ref{sec_benchmarks},
  we provide a simple benchmark example where changing the structure of
  the fixed-effects increases the computational time by several orders
  of magnitude.}

Let us rewrite the demeaning algorithm as a fixed-point problem. Having
the demeaning algorithm in this form proves useful as it unlocks the use
of fixed-point acceleration algorithms.

We define the function
\(f_I:\mathbb{R}^{N_T} \rightarrow \mathbb{R}^{N_I}\) such that
\(\alpha^{(m + 1)} = f_I (\beta^{(m)})\), with \(\alpha^{(m + 1)}\) and
\(\beta^{(m)}\) the vector versions of the values in Equation
\eqref{eq-FOC-solution}. Defining
\(f_T: \mathbb{R}^{N_I} \rightarrow \mathbb{R}^{N_T}\) similarly, we can
rewrite Equation \eqref{eq-FOC-solution} using these functions:

\[
\begin{split}
\alpha^{(m + 1)} &= f_I (\beta^{(m)}) \\
\beta^{(m + 1)}  &= f_T (\alpha^{(m + 1)})
\end{split}
\]

Replacing \(\alpha^{(m + 1)}\) by this definition leads to:

\[
\beta^{(m + 1)}  = f_T (f_I (\beta^{(m)})).
\]

We clearly see that \(\beta^{(m + 1)}\) is only a function of
\(\beta^{(m)}\) and does not depend on \(\alpha^{(m + 1)}\) in any way.
Defining \(F = f_T \circ f_I\),
\(F: \mathbb{R}^{N_T} \rightarrow \mathbb{R}^{N_T}\) our problem reduces
to finding \(\beta^*\) such that:

\[
\beta^* = F(\beta^*).
\]

This is a fixed-point problem with \(N_T\) parameters and, hence, a
canonical fixed-point algorithm can be leveraged to hasten the
convergence process \citep[e.g., see][]{ramiere2015iterative}. In the
\pkg{fixest} case, we use algorithm proposed by
\citet{irons1969version}.

Starting at \(\beta^0_t=0, \forall t\), the Irons and Tuck algorithm
replaces \(\beta^{(m + 1)} = F(\beta^{(m)})\) with the following
iteration:

\[
\beta^{(m+1)} = F(F(\beta^{(m)})) - \dfrac{\Delta F(\beta^{(m)}) \cdot \Delta^2 \beta^{(m)}}{||\Delta^2 \beta^{(m)}||^2}\Delta F(\beta^{(m)}),
\]

where \(\Delta \beta^{(m)} = F(\beta^{(m)}) - \beta^{(m)}\),
\(\Delta F(\beta^{(m)}) = F(F(\beta^{(m)})) - F(\beta^{(m)})\), and
\(\Delta^2 \beta^{(m)} = \Delta F(\beta^{(m)}) - \Delta \beta^{(m)}\).

Once the coefficients \(\beta^*\) are found, we can reconstruct the
other coefficients \(\alpha^* = f_I (\beta^*)\), compute the demeaned
variables \(\tilde{y}\) and \(\tilde{X}\), and obtain our coefficients
of interest \(\hat{\gamma}\) as the OLS coefficients of estimating
\(\tilde{y}\) on \(\tilde{X}\).

\subsection{Varying slopes}\label{varying-slopes}

\label{sec_theory_vvs}

On top of regular fixed-effects estimations, a unique strength of
\pkg{fixest} is that it efficiently handles variables with varying
slopes. The concept of varying slopes parallels that of random slopes in
the mixed-effects literature, where the coefficient on one variable is
modulated by the value of another (categorical) variable. For instance,
we might wish to model individual-specific time trends in our panel,
under the following data generating process:

\[
y_{it} = \alpha_{i} + t \times \beta_i + X_{it} \gamma + \epsilon_{it}.
\]

Each individual in this setting is modeled as having their own
intercept, \(\alpha_i\), as well as their own linear time trend
coefficient \(\beta_i\). Hence there would be \(N_I\) trends to
estimate. \pkg{fixest} is able to estimate the set of coefficients
\(\beta_i\) all at once. We briefly expose the idea of this algorithm.

Generalizing the above example, consider a model with individual
fixed-effects and a varying slopes variable \(z_{it}\):

\[
y_{it} = \alpha_{i} + z_{it} \times \beta_i + X_{it} \gamma + \epsilon_{it}.
\]

The first order conditions for the OLS problem leads to the following
optimal values for \(\alpha_i\) and \(\beta_i\):

\[
\alpha_{i} = \dfrac{1}{n} \sum_{t=0}^{N_T}\left( y_{it} - z_{it} \beta_i - X_{it} \gamma \right) \quad \beta_{i} = \dfrac{1}{\sum_{t=0}^{N_T} z_{it}^2} \sum_{t=0}^{N_T}\left( y_{it} - \alpha_i - X_{it} \gamma \right) 
\]

Taking all other values as given, we can solve for \(\alpha_i\) and
\(\beta_i\) at once, since this is a system of two equations with two
unknowns. There is a unique solution for them.\footnote{Internally, when
  \(\sum_{t=0}^{N_T} z_{it}^2 = 0\), we assume the absence of the
  coefficient \(\beta_i\).} Hence, to find all the coefficients
\(\alpha_i\) and \(\beta_i\), we need only solve \(N_I\) tiny systems of
equations; no iteration is needed. Internally, \pkg{fixest} uses
Gaussian elimination to solve these systems.

Finally, to obtain the coefficients \(\gamma\), we proceed exactly as in
the previous section. We regress \(y\) on the fixed-effects and the
variables with varying slopes to obtain the residual \(\tilde{y}\), and
obtain \(\tilde{X}\) similarly. Then we regress \(\tilde{y}\) on
\(\tilde{X}\). Using the technique mentioned above, the regression to
obtain \(\tilde{y}\) is direct and does not require any iteration.

Now let us generalize the previous example. Instead of having one
fixed-effect and one variable with varying slopes, we have a set of
\(L_I\) variables with varying slopes contained in the \(N \times L_I\)
matrix \(Z_I\), with the coefficients varying across the individual
dimension. For an individual \(i\), we have to estimate \(L_I\)
coefficients, i.e.~the vector \(\vec{\alpha}_i\). We can contrast this
against the simple fixed-effects case, where we had only had to estimate
one coefficient \(\alpha_i\). Similarly for the time dimension, we have
a set of \(L_T\) variables with varying slopes in the matrix \(Z_T\).
The model we estimate is:

\[
y_{it} = Z_I\vec{\alpha}_{i} + Z_T \vec{\beta}_t + X_{it} \gamma + \epsilon_{it}.
\]

Note that the individual fixed-effects, not attached to a variable, can
be represented by a column of 1s in the matrix \(Z_I\).

As always, to obtain the coefficient of interest \(\gamma\), we first
need to obtain the residual of the OLS regression of \(y\) on all the
coefficients with varying slopes (in \(Z_I\) and \(Z_T\)). To obtain the
\(L_I\) coefficients of \(\vec{\alpha}_{i}\) and the \(L_T\)
coefficients of \(\vec{\beta}_{t}\), we apply the same fixed-point
algorithm detailed in the previous section, where we iterate across
these two dimensions.

The fact that all coefficients in a given dimension (e.g.,
\(\vec{\alpha}_{i}\)) are found at once---irrespective of the number of
variable with varying slopes---greatly reduces the complexity of the
algorithm and increases numerical stability. To the best of our
knowledge, this feature is unique to \pkg{fixest}.

\subsection{Extending OLS to GLM and
MLE}\label{extending-ols-to-glm-and-mle}

\label{sec_glm_ml}

Our exposition thus far has been limited to the OLS framework. However,
\pkg{fixest} also handles fixed-effects efficiently in the context of
generalized linear models (GLM) estimations and maximum likelihood
estimation (MLE). Here we briefly outline \pkg{fixest}'s GLM and MLE
implementations, and discuss the pros and cons of each approach for
similar likelihood families.

In OLS, the conditional expectation of the dependent variable is a
linear combination of the predictor variables and the coefficients:
\(E\left(y \ |\ X\right) = X\gamma\). GLM/MLE offers more flexibility as
they model the conditional expectation of \(y\) as a function of the
linear combination of the predictors and their coefficients:
\(E\left(y \ |\ X\right) = f\left(X\gamma\right)\). Further, in both GLM
and MLE the dependent variable is modeled with a certain likelihood. The
objective of these algorithms is to maximize that likelihood with
respect to the coefficients \(\gamma\). For example, if \(y\) is a count
variable (i.e.~only zero or positive integers), one natural way to model
it is with a Poisson model, such that its expectation is modeled as:
\(E\left(y \ |\ X\right) = \exp\left(X\gamma\right)\).

In \pkg{fixest}, the maximization algorithms of GLM and MLE differ: GLM
uses the iteratively re-weighted least squares (IRLS), while MLE uses
the direct maximum likelihood estimation with a Newton-Raphson
algorithm. We describe the key differences between these two methods
below.

\subsubsection{fixest's GLM estimation}\label{fixests-glm-estimation}

Given that GLM's IRLS is a succession of OLS estimations, the OLS
algorithm detailed in the previous section can be directly leveraged.
Hence fixed-effects can be estimated directly, as well as variables with
varying slopes. Further, the method works for all likelihood families
that can be estimated via GLM (Gaussian, Poisson, binomial, Gamma, etc).

\subsubsection{fixest's MLE estimation}\label{fixests-mle-estimation}

\pkg{fixest} implements the MLE method for four likelihoods: Gaussian,
Poisson, Logit and Negative Binomial. Fixed-effects are implemented via
a concentration of the likelihood, now briefly detailed. See
\citet{Berge2018_FENmlm} for a full description.

Let \(\alpha\) represent all the fixed-effect coefficients and
\(\gamma\) the coefficients of interest. The log-likelihood can be
written as \(\ln L(y, X, \gamma, \alpha)\) and we want to find the
maximizing coefficients \(\hat\alpha\) and \(\hat\gamma\). Let
\(\ln \tilde L(y, X, \gamma)\) be the concentrated likelihood such that
\(\ln \tilde L(y, X, \gamma) = \max_{\alpha}\ln L(y, X, \gamma, \alpha)\).

In the case of the Gaussian likelihood, the coefficients to concentrate
the likelihood are obtained in a similar fashion as the OLS problem,
given that the first order conditions are similar. For the Poisson
likelihood, closed forms for the coefficients also exist while this is
not true for the other two likelihoods. For those last two, various
tricks are leveraged to hasten the computation of the concentrated
likelihood, detailed in \citet{Berge2018_FENmlm}.

\subsubsection{Differences between GLM and MLE procedures in
fixest}\label{differences-between-glm-and-mle-procedures-in-fixest}

Contrary to GLM, \pkg{fixest}'s MLE method suffers from the following
limitations: i) variables with varying slopes cannot be used, ii)
analytical weights cannot be used, iii) the families are limited to the
four hard coded ones, iv) the algorithm is more sensitive to starting
values.

However, contrary to GLM, the MLE estimation has two main advantages.
First, it includes the Negative Binomial. This likelihood family has an
ancillary parameter governing overdispersion, which cannot be directly
estimated via GLM. Second, the functional form of the right hand side is
not restricted to linear in parameter predictors. Indeed, the IRLS
algorithm of GLM estimations cannot be used directly when the predictors
are non-linearly related.

\section{Econometric estimations with
fixest}\label{econometric-estimations-with-fixest}

\label{sec_estimation}

\subsection{Core estimation functions}\label{core-estimation-functions}

Table \ref{tbl-est-funs} summarizes the core \pkg{fixest} functions for
performing econometric estimations:

\begin{table}[tbhp]
  \centering
  \caption{\label{tbl-est-funs}High-level estimation functions in \pkg{fixest}.}
  

\begin{tabular}{ p{ (\textwidth - 2\tabcolsep) * \real{0.1429} } p{ (\textwidth - 2\tabcolsep) * \real{0.82} } }
  \toprule
  Function & Description \\
  \midrule
  \texttt{feols} & for OLS and 2SLS estimations \\
  \texttt{feglm} & for GLM estimations \\
  \hspace{1em}\texttt{fepois} & \hspace{1em}dedicated GLM routine for Poisson estimation \\
  \texttt{femlm} & for MLE estimations (Gaussian, Poisson, Logit, or Negative Binomial) \\
  \hspace{1em}\texttt{fenegbin} & \hspace{1em}dedicated MLE routine for Negative Binomial estimation \\
  \hspace{1em}\texttt{feNmlm} & \hspace{1em}dedicated MLE routine for non-linear in parameters estimation \\
  
  \bottomrule
\end{tabular}

\end{table}

The user-facing API for these six core functions is virtually identical,
lowering the switching cost of moving between different estimation
routines. For example, the estimation functions share many of the same
arguments, of which the main ones are summarized in Table
\ref{tbl-main-args}. Among the listed arguments, \texttt{fml} and
\texttt{vcov} deserve special mention, since they govern much of
\pkg{fixest}'s special behavior. We cover these two arguments in
separate deep dives in Sections \ref{sec_fml} and \ref{sec_vcov},
respectively.

In addition to the high-level estimation functions, \pkg{fixest} also
offers low-level routines for OLS (\texttt{feols.fit}) and GLM
(\texttt{feglm.fit}), respectively. Similar to base
\texttt{stats::(g)lm.fit}, these low-level functions offer performance
boosts at the cost of placing greater strictures on the inputs;
i.e.~they require a matrix of predictors alongside the dependent
variable in vector form.

\makesavenoteenv{tabular}
\makesavenoteenv{table}
\begin{table}[h!]
  \centering
  \caption{\label{tbl-main-args}Main arguments for \pkg{fixest} estimation functions.}
  

\renewcommand{\arraystretch}{1.2}
\begin{tabular}{ p{ (\textwidth - 2\tabcolsep) * \real{0.14} } p{ (\textwidth - 2\tabcolsep) * \real{0.825} } }
  \toprule
  Argument & Description \\
  \midrule
  
  \texttt{fml} & 
  the formula to estimate. This formula plays a critical role in \pkg{fixest}: 
  it is where we specify the fixed-effects, where we add the IVs, where we can interpolate 
  the variables to estimate, where we specify multiple estimations, where we can lag the variables, etc.
  See Section \ref{sec_fml} for a deep dive on the formula. \\

  \texttt{data} & 
  a data.frame which contains the variables of the formula. Note that this argument 
  is required (as opposed to base \proglang{R}'s \texttt{stats::lm}).\footnotemark \\

  \texttt{vcov} & 
  specifies how to compute the variance-covariance matrix (VCOV) and, thus, standard errors. 
  \pkg{fixest} offers several built-in ways to adjust the VCOV (heteroskedasticity-robust, clustered, etc),
  with many options. Note that the VCOV can be adjusted post estimation, so this argument may be 
  provided only when displaying/exporting the results. See Section \ref{sec_vcov} for a
  deep dive on standard errors. \\

  \texttt{weights} & 
  a one sided-formula providing analytical weights for each observation.
  The variable specifying the weights should be in the data set. Alternatively, it can also be a vector. \\

  \texttt{offset} & 
  a one sided-formula providing an offset to the dependent variable. Alternatively, it can also be a vector. \\

  \texttt{subset} & 
  a one sided formula describing how the sample should be selected. Alternatively, it can also be a vector. \\

  \texttt{split} & 
  a one-sided formula representing a variable. If provided, the sample is split according 
  to this variable and one estimation is performed for each value of this variable. 
  The object returned is then a \texttt{fixest\_multi} object. \\

  \texttt{fsplit} & 
  same as \texttt{split}, but includes one extra estimation for the full sample. \\

  \texttt{panel.id} & 
  a one sided formula describing panel unit-time identifiers. If provided, the data set is considered to be a 
  panel and the formula accepts leads, lags and first differences for any variable.
  Using panel identifiers also facilitate the specification of panel-specific VCOVs. \\

  \texttt{collin.tol} & 
  a numeric scalar, default is \texttt{1e-10}. This is the collinearity threshold used in 
  the OLS estimation. This argument is important to know, sometimes it can be useful to set 
  it to lower/larger values. \\

  \texttt{nthreads} & 
  the number of threads to use in the computationally intensive parts of the estimation. 
  It defaults to half the available threads. \\

  \texttt{lean} & 
  logical, default is \texttt{FALSE}. If \texttt{TRUE}, any large object from the resulting 
  estimation is deleted, leading to a final object with tiny memory footprint. The drawback is that follow on
  methods may not work on this reduced object. \\

  \texttt{only.coef} & 
  logical, default is \texttt{FALSE}. If \texttt{TRUE}, then the vector of estimated coefficients is returned and
  the estimations is then faster. This is useful for simulations. \\
  
  \bottomrule
\end{tabular}
\footnotetext{The fact that the data is always required is a design choice. 
This ensures that all methods work down the road without any issue or unintentional mistake from the user-side.}

\end{table}

\subsection{The fixest object class and associated
methods}\label{the-fixest-object-class-and-associated-methods}

\label{sec_methods}

Consider a simple estimation on the base \proglang{R}
\texttt{airquality} dataset:

\begin{Shaded}
\begin{Highlighting}[]
\NormalTok{est }\OtherTok{=} \FunctionTok{feols}\NormalTok{(Ozone }\SpecialCharTok{\textasciitilde{}}\NormalTok{ Temp }\SpecialCharTok{|}\NormalTok{ Month, airquality)}
\CommentTok{\#\textgreater{} }\AlertTok{NOTE}\CommentTok{: 37 observations removed because of NA values (LHS: 37).}
\FunctionTok{class}\NormalTok{(est)}
\CommentTok{\#\textgreater{} [1] "fixest"}
\end{Highlighting}
\end{Shaded}

The core \pkg{fixest} estimation functions all return an object of class
\texttt{fixest}.\footnote{An exception is the case of multiple
  estimation, which returns an object of (sister) class
  \texttt{fixest\_multi}. However, the distinction is not important
  here. We revisit multiple estimation in Section \ref{sec_fml_multi}.}
This object is a list containing many elements that are useful for
post-estimation routines, including the model scores (first
derivatives), the unscaled VCOV, the coefficients, etc. While it is
possible to access these list elements directly---e.g,
\texttt{est\$nobs}---that is not the recommended approach. Rather,
\pkg{fixest} implements a wide array of dedicated methods and functions
that can be applied to \texttt{fixest} objects. Users should thus invoke
these methods---e.g., \texttt{nobs(est)}---since they provide additional
convenience facilities and safeguards.

\subsubsection{Methods from the stats
package}\label{methods-from-the-stats-package}

The base \proglang{R} \pkg{stats} package implements a variety of
generic statistical methods. Table \ref{tbl-stats-methods} lists the
\pkg{stats} methods supported by \texttt{fixest} objects, sometimes with
bespoke modifications.

\begin{table}[tbhp]
  \centering
  \caption{\label{tbl-stats-methods}Supported \pkg{stats} methods for \texttt{fixest} objects.}
  

\begin{tabular}{ p{ (\textwidth - 2\tabcolsep) * \real{0.18} } p{ (\textwidth - 2\tabcolsep) * \real{0.785} } }
  \toprule
  Method & Description \\
  \midrule
  
  \texttt{AIC} & Akaike's information criterion \\
  
  \texttt{BIC} & Bayesian information criterion \\
  
  \texttt{coef} & accesses the coefficients \\
  
  \texttt{confint} & confidence intervals for the estimated coefficients \\
  
  \texttt{deviance} & extracts the model deviance \\
  
  \texttt{df} & residual degrees-of-freedom \\
  
  \texttt{fitted} & extracts fitted values \\
  
  \texttt{formula} & extracts the formula \\
  
  \texttt{hatvalues} & extracts the leverage values for each observation \\
  
  \texttt{logLik} & extracts the log-likelihood \\
  
  \texttt{model.matrix} & accesses the design (model) matrix that was used
  in the estimation. Can retrieve either: the independent variables
  (default), the dependent variable, the IVs parts (endogenous variables,
  instruments), or the fixed-effects. \\
  
  \texttt{nobs} & extracts the number of observations in the estimation \\
  
  \texttt{predict} & predict method, where we can access to a prediction
  for input values different from the estimation sample \\
  
  \texttt{resid} & extracts the residuals \\
  
  \texttt{sigma} & residual standard deviation \\
  
  \texttt{summary} & computes a summary of the estimation. Bespoke
  \texttt{fixest} functionality includes on-the-fly adjustment of the
  VCOV. \\
  
  \texttt{terms} & extracts the terms \\
  
  \texttt{update} & updates a \pkg{fixest} estimation \\
  
  \texttt{vcov} & accesses the VCOV matrix as currently computed. Accepts
  arguments to change the way the VCOV is computed. \\
  
  \texttt{weights} & extracts the weights \\
  
  \bottomrule
\end{tabular}

\end{table}

Here is a simple example, where we use the \texttt{update.fixest} method
to add another explanatory variable (\texttt{Wind}) to our previous
model. We'll also save the resulting estimation for re-use:

\begin{Shaded}
\begin{Highlighting}[]
\NormalTok{(}\AttributeTok{est2 =} \FunctionTok{update}\NormalTok{(est, . }\SpecialCharTok{\textasciitilde{}}\NormalTok{ . }\SpecialCharTok{+}\NormalTok{ Wind))}
\CommentTok{\#\textgreater{} }\AlertTok{NOTE}\CommentTok{: 37 observations removed because of NA values (LHS: 37).}
\CommentTok{\#\textgreater{} OLS estimation, Dep. Var.: Ozone}
\CommentTok{\#\textgreater{} Observations: 116}
\CommentTok{\#\textgreater{} Fixed{-}effects: Month: 5}
\CommentTok{\#\textgreater{} Standard{-}errors: IID }
\CommentTok{\#\textgreater{}      Estimate Std. Error  t value   Pr(\textgreater{}|t|)    }
\CommentTok{\#\textgreater{} Temp  2.10485   0.330074  6.37692 4.4764e{-}09 ***}
\CommentTok{\#\textgreater{} Wind {-}2.78170   0.668766 {-}4.15945 6.3835e{-}05 ***}
\CommentTok{\#\textgreater{} {-}{-}{-}}
\CommentTok{\#\textgreater{} Signif. codes:  0 \textquotesingle{}***\textquotesingle{} 0.001 \textquotesingle{}**\textquotesingle{} 0.01 \textquotesingle{}*\textquotesingle{} 0.05 \textquotesingle{}.\textquotesingle{} 0.1 \textquotesingle{} \textquotesingle{} 1}
\CommentTok{\#\textgreater{} RMSE: 20.7     Adj. R2: 0.57962 }
\CommentTok{\#\textgreater{}              Within R2: 0.478995}
\end{Highlighting}
\end{Shaded}

As noted, \pkg{fixest} supports bespoke customizations of \pkg{stats}
methods. Here is an illustration where we apply the
\texttt{predict.fixest} method to the \texttt{est2} object that we just
created. The first two \texttt{predict} calls adopt standard arguments
to generate the same output as from, say, an equivalent
\texttt{update.lm} call. However, the third call demonstrates a bespoke
\pkg{fixest} customization, where we request predicted values for the
full original data sample (i.e, before any \texttt{NA} values have been
removed).

\begin{Shaded}
\begin{Highlighting}[]
\FunctionTok{head}\NormalTok{(}\FunctionTok{predict}\NormalTok{(est2))}
\CommentTok{\#\textgreater{} [1] 35.46936 44.32461 35.73850 13.54012 12.50175 27.92161}
\FunctionTok{predict}\NormalTok{(est2, }\AttributeTok{newdata =} \FunctionTok{data.frame}\NormalTok{(}\AttributeTok{Temp =} \DecValTok{72}\NormalTok{, }\AttributeTok{Wind =} \FloatTok{8.5}\NormalTok{, }\AttributeTok{Month =} \DecValTok{6}\NormalTok{))}
\CommentTok{\#\textgreater{} [1] 26.57805}
\FunctionTok{head}\NormalTok{(}\FunctionTok{predict}\NormalTok{(est2, }\AttributeTok{sample =} \StringTok{"original"}\NormalTok{))}
\CommentTok{\#\textgreater{} [1] 35.46936 44.32461 35.73850 13.54012       NA 12.50175}
\end{Highlighting}
\end{Shaded}

\subsubsection{fixest-specific methods and
functions}\label{fixest-specific-methods-and-functions}

In addition to the generic \pkg{stats} methods, \pkg{fixest} introduces
a variety of bespoke methods and functions for dedicated use with
\texttt{fixest} objects. These are summarized in Table
\ref{tbl-fixest-methods}, with a further set of worked examples provided
in the main text below.

\begin{table}[tbhp]
  \centering
  \caption{\label{tbl-fixest-methods}\pkg{fixest}-specific methods and functions.}
  

\begin{tabular}{ p{ (\textwidth - 2\tabcolsep) * \real{0.23} } p{ (\textwidth - 2\tabcolsep) * \real{0.735} } }
  \toprule
  Method & Description \\
  \midrule
  
  Coefficients & \\
  
  \hspace{1em}\texttt{coeftable} & reports the coefficients table, with
  standard-errors, t-stats and p-values \\
  
  \hspace{1em}\texttt{fixef} & extracts the fixed-effects coefficients
  from the estimation \\
  
  \hspace{1em}\texttt{se},\texttt{tstat},\texttt{pvalue} & these functions
  report the standard-errors, t-stats, p-values, of the estimated
  coefficients (not fixed-effects) \\
  
  & \\
  Data & \\
  
  \hspace{1em}\texttt{fixest\_data} & accesses the data set used to
  perform the estimation \\
  
  \hspace{1em}\texttt{obs} & accesses the indices of the used
  observations \\
  
  & \\
  Fit & \\
  
  \hspace{1em}\texttt{degrees\_freedom} & gets the degrees of freedom of a
  \pkg{fixest} estimation. This is similar to the \texttt{df} function but
  allows more arguments. \\
  
  \hspace{1em}\texttt{fitstat} & computes various fit statistics for
  \pkg{fixest} objects. This function is tightly coupled with other
  \pkg{fixest} functions that call it internally \\
  
  \hspace{1em}\texttt{r2} & R-squared of \pkg{fixest} models. Includes
  various definitions (e.g., adjusted, within, pseudo, etc.). Also
  available from \texttt{fitstat}. \\
  
  \hspace{1em}\texttt{wald} & Wald test of joint coefficient nullity. Also
  available from \texttt{fitstat}. \\
  
  & \\
  Presentation & \\
  
  \hspace{1em}\texttt{coefplot},\texttt{iplot} & draws a coefficient plot
  with confidence intervals. Supports multiple estimations in a single
  graph. \\
  
  \hspace{1em}\texttt{etable} & displays the results of several
  estimations in a table in the console or in \LaTeX \\
  
  \bottomrule
\end{tabular}

\end{table}

\paragraph{Coefficients}\label{coefficients}

The goal of the \pkg{fixest}-specific function and methods is to
supplement and, in many cases, expand upon the generic methods defined
by base \proglang{R}. Consider the difference between the (generic)
\texttt{coef} method and the more detailed \texttt{coeftable} output:

\begin{Shaded}
\begin{Highlighting}[]
\FunctionTok{coef}\NormalTok{(est2)}
\CommentTok{\#\textgreater{}      Temp      Wind }
\CommentTok{\#\textgreater{}  2.104854 {-}2.781701 }
\FunctionTok{coeftable}\NormalTok{(est2)}
\CommentTok{\#\textgreater{}       Estimate Std. Error   t value     Pr(\textgreater{}|t|)}
\CommentTok{\#\textgreater{} Temp  2.104854  0.3300739  6.376918 4.476401e{-}09}
\CommentTok{\#\textgreater{} Wind {-}2.781701  0.6687662 {-}4.159452 6.383503e{-}05}
\CommentTok{\#\textgreater{} attr(,"vcov\_type")}
\CommentTok{\#\textgreater{} [1] "IID"}
\end{Highlighting}
\end{Shaded}

Readers will note the special attribute \texttt{"vcov\_type"} associated
with the coefficient table. Like many \pkg{fixest} functions,
\texttt{coeftable} accepts a \texttt{vcov} argument for on-the-fly VCOV
adjustment. We return to this idea in the VCOV deep-dive in Section
\ref{sec_vcov}.

As detailed by our theory review in Section \ref{sec_theory},
\pkg{fixest} views the fixed-effects as a set of nuisance parameters
that need to be controlled for---as efficiently as possible---in order
to derive valid results for the main parameters of interest. However, it
can still be useful to interrogate the fixed-effects coefficients
themselves (e.g., to see if they accord with intuition). These
coefficients can be retrieved with the \texttt{fixef} method. Here we
show that \texttt{fixef} yields identical coefficients to an equivalent,
but less computationally efficient, estimation with dummy variables:

\begin{Shaded}
\begin{Highlighting}[]
\NormalTok{est2\_dummies }\OtherTok{=} \FunctionTok{feols}\NormalTok{(Ozone }\SpecialCharTok{\textasciitilde{}}\NormalTok{ Wind }\SpecialCharTok{+}\NormalTok{ Temp }\SpecialCharTok{+} \FunctionTok{as.factor}\NormalTok{(Month), airquality)}
\CommentTok{\#\textgreater{} }\AlertTok{NOTE}\CommentTok{: 37 observations removed because of NA values (LHS: 37).}
\FunctionTok{rbind}\NormalTok{(}
  \FunctionTok{coef}\NormalTok{(est2\_dummies, }\AttributeTok{keep =} \StringTok{"Month"}\NormalTok{),}
  \FunctionTok{fixef}\NormalTok{(est2)}\SpecialCharTok{$}\NormalTok{Month[}\SpecialCharTok{{-}}\DecValTok{1}\NormalTok{] }\SpecialCharTok{{-}} \FunctionTok{fixef}\NormalTok{(est2)}\SpecialCharTok{$}\NormalTok{Month[}\DecValTok{1}\NormalTok{]}
\NormalTok{)}
\CommentTok{\#\textgreater{}      as.factor(Month)6 as.factor(Month)7 as.factor(Month)8 as.factor(Month)9}
\CommentTok{\#\textgreater{} [1,]         {-}16.35571         {-}8.769567         {-}7.967637         {-}17.40844}
\CommentTok{\#\textgreater{} [2,]         {-}16.35571         {-}8.769567         {-}7.967637         {-}17.40844}
\end{Highlighting}
\end{Shaded}

\paragraph{Data accessors}\label{data-accessors}

The data-accessing methods of \pkg{fixest} deserve special attention
because they enable important internal functionality, as well as provide
building blocks for downstream packages that depend on \pkg{fixest}. Let
us consider some brief examples, continuing with our \texttt{est2}
estimation object.

The \texttt{airquality} dataset contains missing values and, as a
result, not all of the previous observations were used in the preceding
estimations. We can extract the indices of the used observations with
\texttt{obs}:

\begin{Shaded}
\begin{Highlighting}[]
\FunctionTok{head}\NormalTok{(}\FunctionTok{obs}\NormalTok{(est2))}
\CommentTok{\#\textgreater{} [1] 1 2 3 4 6 7}
\end{Highlighting}
\end{Shaded}

(Notice that the fifth observation is missing.)

Similarly, we can access the full data set with \texttt{fixest\_data},
or the left-hand-side and right-hand-sides with \texttt{model.matrix}:

\begin{Shaded}
\begin{Highlighting}[]
\FunctionTok{head}\NormalTok{(}\FunctionTok{fixest\_data}\NormalTok{(est2, }\AttributeTok{sample =} \StringTok{"estimation"}\NormalTok{))}
\CommentTok{\#\textgreater{}   Ozone Solar.R Wind Temp Month Day}
\CommentTok{\#\textgreater{} 1    41     190  7.4   67     5   1}
\CommentTok{\#\textgreater{} 2    36     118  8.0   72     5   2}
\CommentTok{\#\textgreater{} 3    12     149 12.6   74     5   3}
\CommentTok{\#\textgreater{} 4    18     313 11.5   62     5   4}
\CommentTok{\#\textgreater{} 6    28      NA 14.9   66     5   6}
\CommentTok{\#\textgreater{} 7    23     299  8.6   65     5   7}
\FunctionTok{head}\NormalTok{(}\FunctionTok{model.matrix}\NormalTok{(est2, }\AttributeTok{type =} \FunctionTok{c}\NormalTok{(}\StringTok{"lhs"}\NormalTok{, }\StringTok{"rhs"}\NormalTok{)))}
\CommentTok{\#\textgreater{}   Ozone Temp Wind}
\CommentTok{\#\textgreater{} 1    41   67  7.4}
\CommentTok{\#\textgreater{} 2    36   72  8.0}
\CommentTok{\#\textgreater{} 3    12   74 12.6}
\CommentTok{\#\textgreater{} 4    18   62 11.5}
\CommentTok{\#\textgreater{} 6    28   66 14.9}
\CommentTok{\#\textgreater{} 7    23   65  8.6}
\end{Highlighting}
\end{Shaded}

\paragraph{Fit statistics}\label{fit-statistics}

While \pkg{fixest} supports several functions for assessing model fit,
the single most important one to remember is \texttt{fitstat}. The
\texttt{fitstat} function provides a general interface for computing and
accessing fit statistics for \pkg{fixest} estimations. It is tightly
coupled with other \pkg{fixest} methods and functions like
\texttt{print.fixest}, where it serves as the internal engine for
retrieving relevant model fit information.

Consider an example where we want to report the following statistics
from our \texttt{est2} object: (a) the number of observations, (b) R2
(R-squared), (c) adjusted-R2, and (d) the Wald test of joint coefficient
nullity.

\begin{Shaded}
\begin{Highlighting}[]
\FunctionTok{fitstat}\NormalTok{(est2, }\FunctionTok{c}\NormalTok{(}\StringTok{"n"}\NormalTok{, }\StringTok{"r2"}\NormalTok{, }\StringTok{"ar2"}\NormalTok{, }\StringTok{"wald"}\NormalTok{))}
\CommentTok{\#\textgreater{}         Observations: 116}
\CommentTok{\#\textgreater{}                   R2: 0.601553}
\CommentTok{\#\textgreater{}              Adj. R2: 0.57962}
\CommentTok{\#\textgreater{} Wald (joint nullity): stat = 50.106, p = 3.697e{-}16, on 2 and 109 DoF, VCOV: IID.}
\end{Highlighting}
\end{Shaded}

\texttt{fitstat} returns an object of class \texttt{fixest\_fitstat}
which has a dedicated print method for pretty printing, as illustrated
above. Since this object is a list, each of the statistics can be easily
accessed programmatically. We can also pass the desired fit arguments as
a one-sided formula instead of a character vector. In the example below,
we again compute the R2 and Wald test, but this time assign it to an
object. This facilitates convenient post-computation access:

\begin{Shaded}
\begin{Highlighting}[]
\NormalTok{some\_stats }\OtherTok{=} \FunctionTok{fitstat}\NormalTok{(est2, }\SpecialCharTok{\textasciitilde{}}\NormalTok{r2 }\SpecialCharTok{+}\NormalTok{ wald)}
\NormalTok{some\_stats}\SpecialCharTok{$}\NormalTok{r2}
\CommentTok{\#\textgreater{}        r2 }
\CommentTok{\#\textgreater{} 0.6015532 }
\NormalTok{some\_stats}\SpecialCharTok{$}\NormalTok{wald}
\CommentTok{\#\textgreater{} $stat}
\CommentTok{\#\textgreater{} [1] 50.10555}
\CommentTok{\#\textgreater{} }
\CommentTok{\#\textgreater{} $p}
\CommentTok{\#\textgreater{} [1] 3.69724e{-}16}
\CommentTok{\#\textgreater{} }
\CommentTok{\#\textgreater{} $df1}
\CommentTok{\#\textgreater{} [1] 2}
\CommentTok{\#\textgreater{} }
\CommentTok{\#\textgreater{} $df2}
\CommentTok{\#\textgreater{} [1] 109}
\CommentTok{\#\textgreater{} }
\CommentTok{\#\textgreater{} $vcov}
\CommentTok{\#\textgreater{} [1] "IID"}
\CommentTok{\#\textgreater{} }
\end{Highlighting}
\end{Shaded}

As already mentioned, \texttt{fitstat} is tightly linked with other
\pkg{fixest} methods and functions like \texttt{print.fixest} and
\texttt{etable}. Indeed, many of these downstream functions accept a
\texttt{fitstat} argument for customizing their return output. Consider
a simple example, where we override the default \texttt{print.fixest}
return and instead retrieve the (raw) R2 and Wald test as the required
statistics:

\begin{Shaded}
\begin{Highlighting}[]
\FunctionTok{print}\NormalTok{(est2, }\AttributeTok{fitstat =} \SpecialCharTok{\textasciitilde{}}\NormalTok{r2 }\SpecialCharTok{+}\NormalTok{wald)}
\CommentTok{\#\textgreater{} OLS estimation, Dep. Var.: Ozone}
\CommentTok{\#\textgreater{} Observations: 116}
\CommentTok{\#\textgreater{} Fixed{-}effects: Month: 5}
\CommentTok{\#\textgreater{} Standard{-}errors: IID }
\CommentTok{\#\textgreater{}      Estimate Std. Error  t value   Pr(\textgreater{}|t|)    }
\CommentTok{\#\textgreater{} Temp  2.10485   0.330074  6.37692 4.4764e{-}09 ***}
\CommentTok{\#\textgreater{} Wind {-}2.78170   0.668766 {-}4.15945 6.3835e{-}05 ***}
\CommentTok{\#\textgreater{} {-}{-}{-}}
\CommentTok{\#\textgreater{} Signif. codes:  0 \textquotesingle{}***\textquotesingle{} 0.001 \textquotesingle{}**\textquotesingle{} 0.01 \textquotesingle{}*\textquotesingle{} 0.05 \textquotesingle{}.\textquotesingle{} 0.1 \textquotesingle{} \textquotesingle{} 1}
\CommentTok{\#\textgreater{}                   R2: 0.601553}
\CommentTok{\#\textgreater{} Wald (joint nullity): stat = 50.106, p = 3.697e{-}16, on 2 and 109 DoF, VCOV: IID.}
\end{Highlighting}
\end{Shaded}

Note that it is also possible to enable these customizations
\emph{globally} via the family of \texttt{setFixest\_*}
functions.\footnote{We have elided over them for brevity, but the
  \texttt{setFixest\_*} functions (and their \texttt{getFixest\_*}
  counterparts) provide convenient APIs for enabling and customizing
  many global settings. More information can be found in the package
  vignettes and relevant help files: \texttt{?setFixest\_estimation},
  \texttt{?setFixest\_etable}, etc.} For example with the following
command, the \texttt{print} methods always display the R2 and the Wald
test:

\begin{Shaded}
\begin{Highlighting}[]
\FunctionTok{setFixest\_print}\NormalTok{(}\AttributeTok{fitstat =} \SpecialCharTok{\textasciitilde{}}\NormalTok{r2 }\SpecialCharTok{+}\NormalTok{ wald)}
\end{Highlighting}
\end{Shaded}

This can be easily reverted to the factory default with:

\begin{Shaded}
\begin{Highlighting}[]
\FunctionTok{setFixest\_print}\NormalTok{(}\AttributeTok{fitstat =} \ConstantTok{NULL}\NormalTok{)}
\end{Highlighting}
\end{Shaded}

Although we do not show it here, the exact same idea can be applied to
\texttt{etable}, which also accepts a \texttt{fitstat} argument. We make
two additional observations, which apply to both \texttt{print.fixest}
and \texttt{etable}. First, when using a formula, you can add to the
default fit statistics---rather than override them---by using a point
(e.g., \texttt{\textasciitilde{}.\ +\ wald}). Second, when a particular
fit statistic contains sub-elements, like the Wald test, you can refer
to specific component directly (e.g, \texttt{wald.p} for the p-value,
\texttt{wald.stat} for the statistic, etc.)

Finally, please note that \emph{many} other fit statistics are available
beyond those that we have used here; see
\texttt{fitstat(show\_types~=~TRUE)} for a full list. It is also very
easy to add your own custom fit statistics with the function
\texttt{fitstat\_register}. Any custom statistics that are defined in
this way will automatically be available for use with downstream methods
and functions like \texttt{print} and \texttt{etable}. Again, we must
omit examples for brevity, but readers may consult the
\texttt{?fitstat\_register} documentation for more information.

\paragraph{Presentation}\label{presentation-1}

We refrain from a discussion of the \pkg{fixest} presentation methods
here, deferring instead to the deep-dive in Section
\ref{sec_presentation}.

\section{The fixest formula}\label{the-fixest-formula}

\label{sec_fml}

Any \pkg{fixest} estimation must start with a formula describing the
relationship between variables. This formula plays a critical role in
\pkg{fixest}. Indeed, many features are implemented directly within the
formula and cannot be otherwise accessed. In this section we describe
how to: i) add fixed-effects, ii) estimate instrumental variable models,
iii) implement multiple estimations, iv) interpolate variables, v)
creating categorical variables and interactions, and vi) use
\pkg{fixest}'s panel-data features. All these features are enabled via
the formula.

\subsection{Fixed-effects}\label{fixed-effects}

Table \ref{tab_fml_fe} summarizes the various ways by which
fixed-effects can be included in a formula. The table also provides
equivalences to non-\pkg{fixest} syntax. While we provide this reference
as a convenient ``cheat sheet'' guide, we expand on specific
fixed-effects use cases in more detail below.

\begin{table}[tbhp]
  \centering
  \caption{\label{tab_fml_fe}Cheat sheet for including fixed-effects in a \pkg{fixest} estimation.}
  

\small
\begin{tabular}{ m{3.3cm} m{5.5cm} m{5.7cm} }
  \toprule
  Formula & Comment & Equivalent to\\
  \midrule
  \texttt{y \textasciitilde x} & 
  regular estimation & 
  \texttt{y \textasciitilde x}\\
  
  \\
  
  \texttt{y \textasciitilde x | fe} & 
  estimation with fixed-effect \texttt{fe} & 
  \texttt{y \textasciitilde x + as.factor(fe)}\\
  
  \\
  
  \texttt{y \textasciitilde 1 | fe} & 
  estimation with only the fixed-effect \texttt{fe} & 
  \texttt{y \textasciitilde as.factor(fe)}\\
  
  \\
  
  \texttt{y \textasciitilde 1 | fe1 \textasciicircum fe2} & 
  the fixed-effects \texttt{fe1} and \texttt{fe2} are combined & 
  \texttt{y \textasciitilde as.factor(fe1):as.factor(fe2)}\\
  
  \\
  
  \texttt{y \textasciitilde x1 | fe[x2]} & 
  estimation with fixed-effect \texttt{fe} and variable with varying slope \texttt{x2} & 
  \texttt{y \textasciitilde x1 + as.factor(fe) * x2}\\
  
  \\
  
  \makecell[l]{\texttt{y \textasciitilde x1 | fe1[x2, x3]} \\ \texttt{~~~+ fe2} } & 
  estimation with fixed-effects \texttt{fe1} and \texttt{fe2}, and variables with varying slope \texttt{x2} and \texttt{x3} & 
  \makecell[l]{\texttt{y \textasciitilde x1 + as.factor(fe1)*(x2 + x3)} \\ \texttt{~~~+ as.factor(fe2)}}\\
  
  \bottomrule
\end{tabular}

\end{table}

\subsubsection{Individual fixed effects}\label{individual-fixed-effects}

To add fixed-effects in an estimation, you must add them after a pipe,
\texttt{\textbar{}}, immediately after the independent variables. The
basic syntax is:

\[
\mathtt{dep\_var} \sim \mathtt{indep\_vars} \mathtt{\ |\ } \mathtt{fixed\_effects}
\]

Let us consider an example from the \texttt{trade} dataset, which is
included in \pkg{fixest}.

\begin{Shaded}
\begin{Highlighting}[]
\FunctionTok{data}\NormalTok{(trade)}
\FunctionTok{head}\NormalTok{(trade)}
\CommentTok{\#\textgreater{}   Destination Origin Product Year  dist\_km    Euros}
\CommentTok{\#\textgreater{} 1          LU     BE       1 2007 139.5719  2966697}
\CommentTok{\#\textgreater{} 2          BE     LU       1 2007 139.5719  6755030}
\CommentTok{\#\textgreater{} 3          LU     BE       2 2007 139.5719 57078782}
\CommentTok{\#\textgreater{} 4          BE     LU       2 2007 139.5719  7117406}
\CommentTok{\#\textgreater{} 5          LU     BE       3 2007 139.5719 17379821}
\CommentTok{\#\textgreater{} 6          BE     LU       3 2007 139.5719  2622254}
\end{Highlighting}
\end{Shaded}

Below we estimate a so-called ``gravity'' trade model, where the trade
volume between two countries (in \texteuro) is modeled as a function of
their geographic distance. Alongside controlling for product and year
fixed-effects, note that we also include origin and destination
fixed-effects to control for bilateral trade relationships. As shown in
the third line of the output, there are 15 countries of origin and
destination, and 20 different products.

\begin{Shaded}
\begin{Highlighting}[]
\FunctionTok{feols}\NormalTok{(}\FunctionTok{log}\NormalTok{(Euros) }\SpecialCharTok{\textasciitilde{}} \FunctionTok{log}\NormalTok{(dist\_km) }\SpecialCharTok{|}\NormalTok{ Origin }\SpecialCharTok{+}\NormalTok{ Destination }\SpecialCharTok{+}\NormalTok{ Product }\SpecialCharTok{+}\NormalTok{ Year,}
      \AttributeTok{data =}\NormalTok{ trade)}
\CommentTok{\#\textgreater{} OLS estimation, Dep. Var.: log(Euros)}
\CommentTok{\#\textgreater{} Observations: 38,325}
\CommentTok{\#\textgreater{} Fixed{-}effects: Origin: 15,  Destination: 15,  Product: 20,  Year: 10}
\CommentTok{\#\textgreater{} Standard{-}errors: IID }
\CommentTok{\#\textgreater{}              Estimate Std. Error  t value  Pr(\textgreater{}|t|)    }
\CommentTok{\#\textgreater{} log(dist\_km) {-}2.16988   0.020928 {-}103.685 \textless{} 2.2e{-}16 ***}
\CommentTok{\#\textgreater{} {-}{-}{-}}
\CommentTok{\#\textgreater{} Signif. codes:  0 \textquotesingle{}***\textquotesingle{} 0.001 \textquotesingle{}**\textquotesingle{} 0.01 \textquotesingle{}*\textquotesingle{} 0.05 \textquotesingle{}.\textquotesingle{} 0.1 \textquotesingle{} \textquotesingle{} 1}
\CommentTok{\#\textgreater{} RMSE: 1.74337     Adj. R2: 0.705139}
\CommentTok{\#\textgreater{}                 Within R2: 0.219322}
\end{Highlighting}
\end{Shaded}

\subsubsection{Combined fixed-effects}\label{combined-fixed-effects}

Suppose that we believe the confounding effect of any particular product
on the gravity equation depends on the original and destination country.
\pkg{fixest} provides a special \texttt{\^{}} (caret) operator to
quickly combine fixest effects. Here we combine product with the origin
and destination country, respectively:

\begin{Shaded}
\begin{Highlighting}[]
\FunctionTok{feols}\NormalTok{(}\FunctionTok{log}\NormalTok{(Euros) }\SpecialCharTok{\textasciitilde{}} \FunctionTok{log}\NormalTok{(dist\_km) }\SpecialCharTok{|}\NormalTok{ Origin}\SpecialCharTok{\^{}}\NormalTok{Product }\SpecialCharTok{+}\NormalTok{ Destination}\SpecialCharTok{\^{}}\NormalTok{Product }\SpecialCharTok{+}\NormalTok{ Year,}
      \AttributeTok{data =}\NormalTok{ trade)}
\CommentTok{\#\textgreater{} OLS estimation, Dep. Var.: log(Euros)}
\CommentTok{\#\textgreater{} Observations: 38,325}
\CommentTok{\#\textgreater{} Fixed{-}effects: Origin\^{}Product: 300,  Destination\^{}Product: 300,  Year: 10}
\CommentTok{\#\textgreater{} Standard{-}errors: IID }
\CommentTok{\#\textgreater{}              Estimate Std. Error  t value  Pr(\textgreater{}|t|)    }
\CommentTok{\#\textgreater{} log(dist\_km) {-}2.19662   0.017087 {-}128.557 \textless{} 2.2e{-}16 ***}
\CommentTok{\#\textgreater{} {-}{-}{-}}
\CommentTok{\#\textgreater{} Signif. codes:  0 \textquotesingle{}***\textquotesingle{} 0.001 \textquotesingle{}**\textquotesingle{} 0.01 \textquotesingle{}*\textquotesingle{} 0.05 \textquotesingle{}.\textquotesingle{} 0.1 \textquotesingle{} \textquotesingle{} 1}
\CommentTok{\#\textgreater{} RMSE: 1.40871     Adj. R2: 0.804663}
\CommentTok{\#\textgreater{}                 Within R2: 0.304684}
\end{Highlighting}
\end{Shaded}

Now there are 300 country-product coefficients (the number of origin
countries times the number of products). The important increase in the
number of regressors lead to an improvement of the fit statistics, as
shown with the RMSE going down from 1.7 to 1.4.

\subsubsection{Varying slopes}\label{varying-slopes-1}

As described in Section \ref{sec_theory_vvs}, \pkg{fixest} is able to
handle varying slopes very efficiently. Again, we provide a special
formula syntax to denote variables with varying slopes. Say \texttt{fe}
is the fixed-effect and \texttt{z1} and \texttt{z2} are variables whose
slopes vary according to \texttt{fe}. Then we specify the model as:

\[
\mathtt{dep\_var} \sim \mathtt{indep\_vars} \mathtt{\ |\ } \mathtt{fe[z_1, z_2]}
\]

Continuing with our trade example, we can add a unique year trend for
each country of origin as follows:

\begin{Shaded}
\begin{Highlighting}[]
\FunctionTok{feols}\NormalTok{(}\FunctionTok{log}\NormalTok{(Euros) }\SpecialCharTok{\textasciitilde{}} \FunctionTok{log}\NormalTok{(dist\_km) }\SpecialCharTok{|}\NormalTok{ Origin[Year] }\SpecialCharTok{+}\NormalTok{ Destination, }\AttributeTok{data =}\NormalTok{ trade)}
\CommentTok{\#\textgreater{} OLS estimation, Dep. Var.: log(Euros)}
\CommentTok{\#\textgreater{} Observations: 38,325}
\CommentTok{\#\textgreater{} Fixed{-}effects: Origin: 15,  Destination: 15}
\CommentTok{\#\textgreater{} Varying slopes: Year (Origin): 15}
\CommentTok{\#\textgreater{} Standard{-}errors: IID }
\CommentTok{\#\textgreater{}              Estimate Std. Error  t value  Pr(\textgreater{}|t|)    }
\CommentTok{\#\textgreater{} log(dist\_km) {-}2.07276   0.027099 {-}76.4873 \textless{} 2.2e{-}16 ***}
\CommentTok{\#\textgreater{} {-}{-}{-}}
\CommentTok{\#\textgreater{} Signif. codes:  0 \textquotesingle{}***\textquotesingle{} 0.001 \textquotesingle{}**\textquotesingle{} 0.01 \textquotesingle{}*\textquotesingle{} 0.05 \textquotesingle{}.\textquotesingle{} 0.1 \textquotesingle{} \textquotesingle{} 1}
\CommentTok{\#\textgreater{} RMSE: 2.25895     Adj. R2: 0.505112}
\CommentTok{\#\textgreater{}                 Within R2: 0.132569}
\end{Highlighting}
\end{Shaded}

\[
\ln(Euros)_{ijt} = \alpha_i + \beta_j + t \times \theta_i + \gamma \ln(dist\_km) + \epsilon,
\]

with \(i\) the index of the country of origin, \(j\) the country of
destination, and \(t\) the year.\footnote{This estimation is analogous,
  albeit much faster and with an exact solution, to a mixed-effects
  model of form:
  \texttt{lme4::lmer(log(Euros)~\textasciitilde{}~log(dist\_km)~+~Year~+~(Year~\textbar{}~Origin)~+~(1~\textbar{}~Destination),\ trade)}.}

\subsection{Instrumental variables}\label{instrumental-variables}

\label{sec_fml_iv}

While it is not available for all model families, \pkg{fixest} supports
two stage least squares (2SLS) instrumental variables (IV) estimations
for linear models in \texttt{feols}. Table \ref{tab_fml_iv} provides a
summary cheat sheet for how to perform 2SLS estimations, including
without fixed-effects.

\begin{table}[tbh]
  \centering
  \caption{\label{tab_fml_iv}Cheat sheet for performing 2SLS/IV estimations in \texttt{feols}.}
  

\small
\begin{tabular}{ m{.7cm} m{7cm} m{6.6cm} }
  \toprule
  id & Formula & Comment \\
  \midrule

  \texttt{(1)} &
  \texttt{y \textasciitilde x | endo \textasciitilde inst} &
  one exogenous variable \texttt{x}, one endogenous variable \texttt{endo}, 
  instrumented with one variable \texttt{inst} \\
  
  \\
  
  \texttt{(2)} &
  \texttt{y \textasciitilde x | endo \textasciitilde inst1 + inst2} &
  same as \texttt{(1)}, with two instruments \texttt{inst1} and \texttt{inst2}\\
  
  \\
  
  \texttt{(3)} &
  \texttt{y \textasciitilde x | endo1 + endo2 \textasciitilde inst1 + inst2} &
  same as \texttt{(2)}, with two endogenous variables \texttt{endo1} and \texttt{endo2}\\
  
  \\
  
  \texttt{(4)} &
  \texttt{y \textasciitilde 1 | endo \textasciitilde inst} &
  same as \texttt{(1)}, but without any exogenous regressor\\
  
  \\
  
  \texttt{(5)} &
  \texttt{y \textasciitilde x | fe | endo \textasciitilde inst} &
  same as \texttt{(1)}, with (exogenous) fixed-effects \texttt{fe}\\
  
  \\
  
  \texttt{(6)} &
  \texttt{y \textasciitilde 1 | fe | endo \textasciitilde inst} &
  same as \texttt{(4)}, the exogenous regressors consist of only the fixed-effects \texttt{fe}\\
  
  \bottomrule
\end{tabular}

\end{table}

Similar to the fixed-effects component, the IV part of the formula is
specified in its own slot after a \texttt{\textbar{}}. The syntax is:

\[
\mathtt{dep\_var} \sim \mathtt{exogenous\_vars} \mathtt{\ |\ } \mathtt{fixed\_effects} \mathtt{\ |\ } \mathtt{endo\_vars} \sim \mathtt{instruments}
\]

Again, we emphasize the order importance of our multipart formula. The
fixed-effect part is optional (and comes second). The IV part always
comes always last. If there is no exogenous regressor, the part
containing the exogenous regressors should be replaced with a single
\texttt{1}.s

In 2SLS estimations, by default only the second stage is reported. To
access the first stage(s), use the argument stage of the
\texttt{summary} method (Section \ref{sec_methods}). Let us use data
from the Fulton Fish Market the introductory example from Section
\ref{sec_example}, where we instrument science funding with a policy
shock to estimate the causal effect on productivity. We display both the
second stage and the first stage:

\begin{Shaded}
\begin{Highlighting}[]
\NormalTok{est\_iv }\OtherTok{=} \FunctionTok{feols}\NormalTok{(articles }\SpecialCharTok{\textasciitilde{}} \DecValTok{1} \SpecialCharTok{|}\NormalTok{ indiv }\SpecialCharTok{+}\NormalTok{ year }\SpecialCharTok{|}\NormalTok{ funding }\SpecialCharTok{\textasciitilde{}}\NormalTok{ policy, scipubs)}
\FunctionTok{summary}\NormalTok{(est\_iv, }\AttributeTok{stage =} \DecValTok{1}\SpecialCharTok{:}\DecValTok{2}\NormalTok{)}
\CommentTok{\#\textgreater{} IV: First stage: funding}
\CommentTok{\#\textgreater{} TSLS estimation}
\CommentTok{\#\textgreater{} |{-} D.V.   : articles}
\CommentTok{\#\textgreater{} |{-} Endo.  : funding}
\CommentTok{\#\textgreater{} |{-} Instr. : policy}
\CommentTok{\#\textgreater{} |}
\CommentTok{\#\textgreater{} |=\textgreater{} First Stage}
\CommentTok{\#\textgreater{} |   Current Dep. Var.: funding}
\CommentTok{\#\textgreater{} Observations: 1,080}
\CommentTok{\#\textgreater{} Fixed{-}effects: indiv: 108,  year: 10}
\CommentTok{\#\textgreater{} Standard{-}errors: IID }
\CommentTok{\#\textgreater{}        Estimate Std. Error   t value Pr(\textgreater{}|t|) }
\CommentTok{\#\textgreater{} policy {-}1.70991    7.50542 {-}0.227824  0.81983 }
\CommentTok{\#\textgreater{} {-}{-}{-}}
\CommentTok{\#\textgreater{} Signif. codes:  0 \textquotesingle{}***\textquotesingle{} 0.001 \textquotesingle{}**\textquotesingle{} 0.01 \textquotesingle{}*\textquotesingle{} 0.05 \textquotesingle{}.\textquotesingle{} 0.1 \textquotesingle{} \textquotesingle{} 1}
\CommentTok{\#\textgreater{} RMSE: 28.1     Adj. R2: 0.014444}
\CommentTok{\#\textgreater{}              Within R2: 5.395e{-}5}
\CommentTok{\#\textgreater{} F{-}test (1st stage): stat = 0.05190, p = 0.819832, on 1 and 962 DoF.}
\CommentTok{\#\textgreater{} }
\CommentTok{\#\textgreater{} IV: Second stage}
\CommentTok{\#\textgreater{} TSLS estimation}
\CommentTok{\#\textgreater{} |{-} D.V.   : articles}
\CommentTok{\#\textgreater{} |{-} Endo.  : funding}
\CommentTok{\#\textgreater{} |{-} Instr. : policy}
\CommentTok{\#\textgreater{} |}
\CommentTok{\#\textgreater{} |=\textgreater{} Second Stage}
\CommentTok{\#\textgreater{} |   Dep. Var.: articles}
\CommentTok{\#\textgreater{} Observations: 1,080}
\CommentTok{\#\textgreater{} Fixed{-}effects: indiv: 108,  year: 10}
\CommentTok{\#\textgreater{} Standard{-}errors: IID }
\CommentTok{\#\textgreater{}              Estimate Std. Error   t value Pr(\textgreater{}|t|) }
\CommentTok{\#\textgreater{} fit\_funding {-}0.245182    1.63019 {-}0.150401  0.88048 }
\CommentTok{\#\textgreater{} {-}{-}{-}}
\CommentTok{\#\textgreater{} Signif. codes:  0 \textquotesingle{}***\textquotesingle{} 0.001 \textquotesingle{}**\textquotesingle{} 0.01 \textquotesingle{}*\textquotesingle{} 0.05 \textquotesingle{}.\textquotesingle{} 0.1 \textquotesingle{} \textquotesingle{} 1}
\CommentTok{\#\textgreater{} RMSE: 10.4     Adj. R2: 0.168161}
\CommentTok{\#\textgreater{}              Within R2: 1.061e{-}4}
\CommentTok{\#\textgreater{} F{-}test (1st stage), funding: stat = 0.05190, p = 0.819832, on 1 and 962 DoF.}
\CommentTok{\#\textgreater{}                  Wu{-}Hausman: stat = 0.28442, p = 0.593945, on 1 and 961 DoF.}
\end{Highlighting}
\end{Shaded}

The function \texttt{fitstat} can access the following fit statistics
when applied to instrumental variables estimations:

\begin{itemize}
\tightlist
\item
  \texttt{ivf}: the F-test of the first stage (weak instrument test)
\item
  \texttt{ivwald}: the Wald-test of the first stage (weak instrument
  test)
\item
  \texttt{cd}: the Cragg-Donald test for weak instruments
\item
  \texttt{wh}: the Wu-Hausman endogeneity test
\item
  \texttt{sargan}: the Sargan test of overidentifying restrictions
\end{itemize}

\subsection{Categorical variables and interactions:
i()}\label{categorical-variables-and-interactions-i}

\label{sec_i}

To tell an estimation procedure to handle numeric variables as
categorical variables, \texttt{as.factor} is the common way to do it in
\proglang{R}. \pkg{fixest} introduces a simple function \texttt{i}, to
handle categorical variables and provides easy access to: i) reordering,
ii) binning, iii) interactions.

Use the function \texttt{i} to treat any variable as a categorical
variable. This specific use can be seen as a replacement for
\texttt{as.factor}. When \texttt{i} is applied, the names of the
categories is as follows: \texttt{"var\_name::value"}, whereas when
using \texttt{as.factor} the new names are
\texttt{"as.factor(var\_name)value"}. Further, the function \texttt{i}
allows for richer integration with some \pkg{fixest} functions like
\texttt{etable} or \texttt{iplot}.

In the example below, let us treat the numeric variable \texttt{year} as
categorical, either using \texttt{as.factor}, either using \texttt{i}.
With \texttt{coef} we then show \texttt{i}'s specific formatting of the
coefficient names. We end by showing the interaction with
\texttt{etable} and how we end up with a display of greater clarity.

\begin{Shaded}
\begin{Highlighting}[]
\NormalTok{est\_factor }\OtherTok{=} \FunctionTok{feols}\NormalTok{(articles }\SpecialCharTok{\textasciitilde{}} \FunctionTok{as.factor}\NormalTok{(year), scipubs)}
\NormalTok{est\_i      }\OtherTok{=} \FunctionTok{feols}\NormalTok{(articles }\SpecialCharTok{\textasciitilde{}} \FunctionTok{i}\NormalTok{(year), scipubs)}
\FunctionTok{coef}\NormalTok{(est\_i)}
\CommentTok{\#\textgreater{} (Intercept)     year::2     year::3     year::4     year::5     year::6 }
\CommentTok{\#\textgreater{}  14.7777778   0.7500000   0.6203704   1.1388889   1.0740741   2.8611111 }
\CommentTok{\#\textgreater{}     year::7     year::8     year::9    year::10 }
\CommentTok{\#\textgreater{}   3.3703704   4.1296296   4.9444444   4.6018519 }
\FunctionTok{etable}\NormalTok{(est\_factor, est\_i, }\AttributeTok{drop.section =} \StringTok{"stat"}\NormalTok{)}
\CommentTok{\#\textgreater{}                          est\_factor             est\_i}
\CommentTok{\#\textgreater{} Dependent Var.:            articles          articles}
\CommentTok{\#\textgreater{}                                                      }
\CommentTok{\#\textgreater{} Constant          14.78*** (0.5250) 14.78*** (0.5250)}
\CommentTok{\#\textgreater{} as.factor(year)2    0.7500 (0.7424)                  }
\CommentTok{\#\textgreater{} as.factor(year)3    0.6204 (0.7424)                  }
\CommentTok{\#\textgreater{} as.factor(year)4     1.139 (0.7424)                  }
\CommentTok{\#\textgreater{} as.factor(year)5     1.074 (0.7424)                  }
\CommentTok{\#\textgreater{} as.factor(year)6  2.861*** (0.7424)                  }
\CommentTok{\#\textgreater{} as.factor(year)7  3.370*** (0.7424)                  }
\CommentTok{\#\textgreater{} as.factor(year)8  4.130*** (0.7424)                  }
\CommentTok{\#\textgreater{} as.factor(year)9  4.944*** (0.7424)                  }
\CommentTok{\#\textgreater{} as.factor(year)10 4.602*** (0.7424)                  }
\CommentTok{\#\textgreater{} year = 2                              0.7500 (0.7424)}
\CommentTok{\#\textgreater{} year = 3                              0.6204 (0.7424)}
\CommentTok{\#\textgreater{} year = 4                               1.139 (0.7424)}
\CommentTok{\#\textgreater{} year = 5                               1.074 (0.7424)}
\CommentTok{\#\textgreater{} year = 6                            2.861*** (0.7424)}
\CommentTok{\#\textgreater{} year = 7                            3.370*** (0.7424)}
\CommentTok{\#\textgreater{} year = 8                            4.130*** (0.7424)}
\CommentTok{\#\textgreater{} year = 9                            4.944*** (0.7424)}
\CommentTok{\#\textgreater{} year = 10                           4.602*** (0.7424)}
\CommentTok{\#\textgreater{} \_\_\_\_\_\_\_\_\_\_\_\_\_\_\_\_\_ \_\_\_\_\_\_\_\_\_\_\_\_\_\_\_\_\_ \_\_\_\_\_\_\_\_\_\_\_\_\_\_\_\_\_}
\CommentTok{\#\textgreater{} S.E. type                       IID               IID}
\CommentTok{\#\textgreater{} {-}{-}{-}}
\CommentTok{\#\textgreater{} Signif. codes: 0 \textquotesingle{}***\textquotesingle{} 0.001 \textquotesingle{}**\textquotesingle{} 0.01 \textquotesingle{}*\textquotesingle{} 0.05 \textquotesingle{}.\textquotesingle{} 0.1 \textquotesingle{} \textquotesingle{} 1}
\end{Highlighting}
\end{Shaded}

The function \texttt{i} offers two useful arguments to manipulate the
categorical variable it creates:

\begin{itemize}
\tightlist
\item
  \texttt{ref}: to define the reference of the categorical variable, it
  can accept several values
\item
  \texttt{bin}: to merge several categorical values into one
\end{itemize}

Using the example above, let us first take \texttt{Month\ =\ 9} as a
reference (instead of 5), then run the same initial estimation but
grouping the months 8 and 9, and finally display the results with
\texttt{etable}.

\begin{Shaded}
\begin{Highlighting}[]
\NormalTok{est\_ref }\OtherTok{=} \FunctionTok{feols}\NormalTok{(articles }\SpecialCharTok{\textasciitilde{}} \FunctionTok{i}\NormalTok{(year, }\AttributeTok{ref =} \DecValTok{9}\NormalTok{), scipubs)}
\NormalTok{est\_bin }\OtherTok{=} \FunctionTok{feols}\NormalTok{(articles }\SpecialCharTok{\textasciitilde{}} \FunctionTok{i}\NormalTok{(year, }\AttributeTok{bin =} \FunctionTok{list}\NormalTok{(}\StringTok{"8 {-} 9"} \OtherTok{=} \DecValTok{8}\SpecialCharTok{:}\DecValTok{9}\NormalTok{)), scipubs)}
\FunctionTok{etable}\NormalTok{(est\_i, est\_ref, est\_bin)}
\CommentTok{\#\textgreater{}                             est\_i            est\_ref           est\_bin}
\CommentTok{\#\textgreater{} Dependent Var.:          articles           articles          articles}
\CommentTok{\#\textgreater{}                                                                       }
\CommentTok{\#\textgreater{} Constant        14.78*** (0.5250)  19.72*** (0.5250) 14.78*** (0.5250)}
\CommentTok{\#\textgreater{} year = 2          0.7500 (0.7424) {-}4.194*** (0.7424)   0.7500 (0.7425)}
\CommentTok{\#\textgreater{} year = 3          0.6204 (0.7424) {-}4.324*** (0.7424)   0.6204 (0.7425)}
\CommentTok{\#\textgreater{} year = 4           1.139 (0.7424) {-}3.806*** (0.7424)    1.139 (0.7425)}
\CommentTok{\#\textgreater{} year = 5           1.074 (0.7424) {-}3.870*** (0.7424)    1.074 (0.7425)}
\CommentTok{\#\textgreater{} year = 6        2.861*** (0.7424)  {-}2.083** (0.7424) 2.861*** (0.7425)}
\CommentTok{\#\textgreater{} year = 7        3.370*** (0.7424)   {-}1.574* (0.7424) 3.370*** (0.7425)}
\CommentTok{\#\textgreater{} year = 8        4.130*** (0.7424)   {-}0.8148 (0.7424)                  }
\CommentTok{\#\textgreater{} year = 9        4.944*** (0.7424)                                     }
\CommentTok{\#\textgreater{} year = 10       4.602*** (0.7424)   {-}0.3426 (0.7424) 4.602*** (0.7425)}
\CommentTok{\#\textgreater{} year = 1                          {-}4.944*** (0.7424)                  }
\CommentTok{\#\textgreater{} year = 8 {-} 9                                         4.537*** (0.6430)}
\CommentTok{\#\textgreater{} \_\_\_\_\_\_\_\_\_\_\_\_\_\_\_ \_\_\_\_\_\_\_\_\_\_\_\_\_\_\_\_\_ \_\_\_\_\_\_\_\_\_\_\_\_\_\_\_\_\_\_ \_\_\_\_\_\_\_\_\_\_\_\_\_\_\_\_\_}
\CommentTok{\#\textgreater{} S.E. type                     IID                IID               IID}
\CommentTok{\#\textgreater{} Observations                1,080              1,080             1,080}
\CommentTok{\#\textgreater{} R2                        0.09357            0.09357           0.09255}
\CommentTok{\#\textgreater{} Adj. R2                   0.08594            0.08594           0.08577}
\CommentTok{\#\textgreater{} {-}{-}{-}}
\CommentTok{\#\textgreater{} Signif. codes: 0 \textquotesingle{}***\textquotesingle{} 0.001 \textquotesingle{}**\textquotesingle{} 0.01 \textquotesingle{}*\textquotesingle{} 0.05 \textquotesingle{}.\textquotesingle{} 0.1 \textquotesingle{} \textquotesingle{} 1}
\end{Highlighting}
\end{Shaded}

As we can see, the first two estimations are the same, only the
reference of the categorical variable has changed, leading to different
coefficient estimates (but same R2, also note that
\(76.90 - 11.35 = 65.55\)). In the last estimation where we grouped the
coefficients for the two last months, we can see that the estimates for
the other coefficients stay the same.

Now say we want to estimate yearly time trends for the EU and the US
separately, we can do so with \texttt{i(cat,\ numerical)}:

\begin{Shaded}
\begin{Highlighting}[]
\FunctionTok{feols}\NormalTok{(articles }\SpecialCharTok{\textasciitilde{}} \FunctionTok{i}\NormalTok{(eu\_us) }\SpecialCharTok{+} \FunctionTok{i}\NormalTok{(eu\_us, year), scipubs)}
\CommentTok{\#\textgreater{} OLS estimation, Dep. Var.: articles}
\CommentTok{\#\textgreater{} Observations: 1,080}
\CommentTok{\#\textgreater{} Standard{-}errors: IID }
\CommentTok{\#\textgreater{}                 Estimate Std. Error  t value   Pr(\textgreater{}|t|)    }
\CommentTok{\#\textgreater{} (Intercept)    12.593939   0.471652 26.70174  \textless{} 2.2e{-}16 ***}
\CommentTok{\#\textgreater{} eu\_us::US       2.661407   0.673280  3.95289 8.2252e{-}05 ***}
\CommentTok{\#\textgreater{} eu\_us::EU:year  1.055978   0.076014 13.89195  \textless{} 2.2e{-}16 ***}
\CommentTok{\#\textgreater{} eu\_us::US:year  0.099714   0.077435  1.28772 1.9812e{-}01    }
\CommentTok{\#\textgreater{} {-}{-}{-}}
\CommentTok{\#\textgreater{} Signif. codes:  0 \textquotesingle{}***\textquotesingle{} 0.001 \textquotesingle{}**\textquotesingle{} 0.01 \textquotesingle{}*\textquotesingle{} 0.05 \textquotesingle{}.\textquotesingle{} 0.1 \textquotesingle{} \textquotesingle{} 1}
\CommentTok{\#\textgreater{} RMSE: 5.11086   Adj. R2: 0.194856}
\end{Highlighting}
\end{Shaded}

If we wanted to include indicators for region by year fixed effects with
\texttt{i(cat,\ i.cat2)}. The \texttt{i.} specifies that the second
variable is intended to be treated as a categorical as well (borrowing
the syntax from Stata). We could alternatively specify this in the
fixed-effect portion with \texttt{\textbar{}\ eu\_us\^{}year}.

\begin{Shaded}
\begin{Highlighting}[]
\FunctionTok{feols}\NormalTok{(articles }\SpecialCharTok{\textasciitilde{}} \FunctionTok{i}\NormalTok{(eu\_us) }\SpecialCharTok{+} \FunctionTok{i}\NormalTok{(eu\_us, i.year), scipubs)}
\CommentTok{\#\textgreater{} The variable \textquotesingle{}eu\_us::US:year::10\textquotesingle{} has been removed because of collinearity (see}
\CommentTok{\#\textgreater{} $collin.var).}
\CommentTok{\#\textgreater{} OLS estimation, Dep. Var.: articles}
\CommentTok{\#\textgreater{} Observations: 1,080}
\CommentTok{\#\textgreater{} Standard{-}errors: IID }
\CommentTok{\#\textgreater{}                    Estimate Std. Error   t value   Pr(\textgreater{}|t|)    }
\CommentTok{\#\textgreater{} (Intercept)       14.254545   0.690330 20.648878  \textless{} 2.2e{-}16 ***}
\CommentTok{\#\textgreater{} eu\_us::US          1.179417   0.985442  1.196841 2.3164e{-}01    }
\CommentTok{\#\textgreater{} eu\_us::EU:year::2  0.872727   0.976274  0.893936 3.7156e{-}01    }
\CommentTok{\#\textgreater{} eu\_us::EU:year::3  1.836364   0.976274  1.880991 6.0247e{-}02 .  }
\CommentTok{\#\textgreater{} eu\_us::EU:year::4  1.654545   0.976274  1.694754 9.0416e{-}02 .  }
\CommentTok{\#\textgreater{} eu\_us::EU:year::5  2.090909   0.976274  2.141723 3.2443e{-}02 *  }
\CommentTok{\#\textgreater{} eu\_us::EU:year::6  4.436364   0.976274  4.544177 6.1487e{-}06 ***}
\CommentTok{\#\textgreater{} eu\_us::EU:year::7  6.454545   0.976274  6.611405 6.0220e{-}11 ***}
\CommentTok{\#\textgreater{} ... 12 coefficients remaining (display them with summary() or use}
\CommentTok{\#\textgreater{} argument n)}
\CommentTok{\#\textgreater{} ... 1 variable was removed because of collinearity (eu\_us::US:year::10)}
\CommentTok{\#\textgreater{} {-}{-}{-}}
\CommentTok{\#\textgreater{} Signif. codes:  0 \textquotesingle{}***\textquotesingle{} 0.001 \textquotesingle{}**\textquotesingle{} 0.01 \textquotesingle{}*\textquotesingle{} 0.05 \textquotesingle{}.\textquotesingle{} 0.1 \textquotesingle{} \textquotesingle{} 1}
\CommentTok{\#\textgreater{} RMSE: 5.072   Adj. R2: 0.195084}
\end{Highlighting}
\end{Shaded}

\subsection{Multiple estimations}\label{multiple-estimations}

\label{sec_fml_multi}

A core design feature of \pkg{fixest} is that it should be easy to
estimate multiple models (or functional forms) all at once in a single
function call. In this section we focus on: i) multiple dependent
variables and ii) stepwise regressions. Again, we provide a cheat sheet
for \pkg{fixest}'s multiple estimation syntax in Table
\ref{tab_fml_multiple}.

\begin{table}[tbh!]
  \centering
  \caption{\label{tab_fml_multiple}Cheat sheet for multiple estimation in \pkg{fixest}.}
  

\small
\begin{tabular}{ p{4.4cm} p{5.8cm} p{4cm} }
  \toprule
  Formula & Estimates & Comment \\
  \midrule

  \makecell[tl]{\texttt{c(y1, y2) \textasciitilde x}} &
  \vspace{-0.4cm}
  \begin{itemize}
    \item \texttt{y1 \textasciitilde x}
    \item \texttt{y2 \textasciitilde x}
  \end{itemize} &
  multiple left hand sides\\
  
  \\
  
  \texttt{y \textasciitilde sw(x1, x2)} &
  \vspace{-0.4cm}
  \begin{itemize}
    \item \texttt{y \textasciitilde x1}
    \item \texttt{y \textasciitilde x2}
  \end{itemize} &
  \textbf{S}tep\textbf{W}ise\\
  
  \\
  
  \texttt{y \textasciitilde x | sw0(fe)} &
  \vspace{-0.4cm}
  \begin{itemize}
    \item \texttt{y \textasciitilde x}
    \item \texttt{y \textasciitilde x | fe}
  \end{itemize} &
  stepwise, starting with the empty element\\
  
  \\
  
  \makecell[tl]{
    \texttt{y \textasciitilde x1 + csw0(x2) | csw(fe1,} \\
    \texttt{~~~~~~~~~~~~~~~~~~~~~~~fe2)}
  } &
  \vspace{-0.4cm}
  \begin{itemize}
    \item \texttt{y \textasciitilde x1 | fe1}
    \item \texttt{y \textasciitilde x1 | fe1 + fe2}
    \item \texttt{y \textasciitilde x1 + x2 | fe1}
    \item \texttt{y \textasciitilde x1 + x2 | fe1 + fe2}
  \end{itemize} &
  \textbf{C}umulative \textbf{S}tep\textbf{W}ise\\
  
  \\
  
  \texttt{y \textasciitilde mvsw(x1, x2, x3)} &
  \vspace{-0.4cm}
  \begin{itemize}
    \item \texttt{y \textasciitilde 1}
    \item \texttt{y \textasciitilde x1}
    \item \texttt{y \textasciitilde x2}
    \item \texttt{y \textasciitilde x3}
    \item \texttt{y \textasciitilde x1 + x2}
    \item \texttt{y \textasciitilde x1 + x3}
    \item \texttt{y \textasciitilde x2 + x3}
    \item \texttt{y \textasciitilde x1 + x2 + x3}
  \end{itemize} &
  \textbf{M}ulti \textbf{V}erse  \textbf{S}tep\textbf{W}ise: includes all the combinations of variables\\
  
  \bottomrule
\end{tabular}

\end{table}

\subsubsection{Multiple dependent
variables}\label{multiple-dependent-variables}

To estimate a model for multiple dependent variables, you can nest them
within the \texttt{c()} function on the formula LHS:

\[
\mathtt{c(y_1, y_2)} \sim \mathtt{indep\_vars} \mathtt{\ |\ } \mathtt{fixed\_effects}.
\]

This will run two regressions: one with \(\mathtt{y_1}\) as dependent
variable and one with \(\mathtt{y_2}\). Here is a simple example using
the built-in \texttt{airquality} dataset. Note that the default print
method automatically reverts to \texttt{etable}, conveniently displaying
both regressions alongside each other.

\begin{Shaded}
\begin{Highlighting}[]
\FunctionTok{feols}\NormalTok{(}\FunctionTok{c}\NormalTok{(Ozone, Solar.R) }\SpecialCharTok{\textasciitilde{}}\NormalTok{ Wind }\SpecialCharTok{+}\NormalTok{ Temp, airquality)}
\CommentTok{\#\textgreater{}                                x.1               x.2}
\CommentTok{\#\textgreater{} Dependent Var.:              Ozone           Solar.R}
\CommentTok{\#\textgreater{}                                                     }
\CommentTok{\#\textgreater{} Constant          {-}71.03** (23.58)    {-}76.36 (82.00)}
\CommentTok{\#\textgreater{} Wind            {-}3.055*** (0.6633)     2.211 (2.308)}
\CommentTok{\#\textgreater{} Temp             1.840*** (0.2500) 3.075*** (0.8778)}
\CommentTok{\#\textgreater{} \_\_\_\_\_\_\_\_\_\_\_\_\_\_\_ \_\_\_\_\_\_\_\_\_\_\_\_\_\_\_\_\_\_ \_\_\_\_\_\_\_\_\_\_\_\_\_\_\_\_\_}
\CommentTok{\#\textgreater{} S.E. type                      IID               IID}
\CommentTok{\#\textgreater{} Observations                   116               146}
\CommentTok{\#\textgreater{} R2                         0.56871           0.08198}
\CommentTok{\#\textgreater{} Adj. R2                    0.56108           0.06914}
\CommentTok{\#\textgreater{} {-}{-}{-}}
\CommentTok{\#\textgreater{} Signif. codes: 0 \textquotesingle{}***\textquotesingle{} 0.001 \textquotesingle{}**\textquotesingle{} 0.01 \textquotesingle{}*\textquotesingle{} 0.05 \textquotesingle{}.\textquotesingle{} 0.1 \textquotesingle{} \textquotesingle{} 1}
\end{Highlighting}
\end{Shaded}

Beyond syntactic convenience, estimating multiple dependent outcomes in
this way can yield dramatic performance benefits. The bulk of the
computing time in a fixed-effects estimation is spent dealing with the
fixed-effects themselves. In a multiple estimation, \pkg{fixest} is able
to pool this computationally-expensive step across the various models,
thus avoiding the effort duplication that would arise from estimating
the models separately. The practical upside is that estimating, say, ten
models will typically take the same amount of time as estimating a
single model on its own.\footnote{An important caveat is that this
  pooling acceleration shortcut only applies to OLS. For GLM models, the
  more complex model structure means that the same shortcut cannot
  readily be used.} We benchmark these gains with a large dataset in
Section \ref{sec_benchmarks}.

\subsubsection{Stepwise regressions}\label{stepwise-regressions}

Multiple estimation is also supported on the formula RHS. For example,
all \pkg{fixest} estimations support a family of special functions for
stepwise regressions: \texttt{sw} (stepwise), \texttt{sw0} (stepwise,
starting from the empty element), \texttt{csw} (cumulative stepwise),
\texttt{csw0} (cumulative stepwise starting with the empty element), and
\texttt{mvsw} (multiverse stepwise, i.e.~crossing). These functions
accept any number of variables as argument, and can be used directly in
a formula as regular variables. Moreover, they can be used in both the
regular part of the formula RHS (containing the explanatory variables),
or in the fixed-effects part. However, note that only stepwise function
is allowed per part.

Here is an example where we run six estimations in a single
\texttt{feols} call. Notice how expressive the syntax is, despite its
compactness. Once you know that \texttt{sw} stands for
\textbf{s}tep\textbf{w}ise and \texttt{csw} for \textbf{c}umulative
\textbf{s}tep\textbf{w}ise, and that a \texttt{0} suffix enforces an
empty starting element, the code is very easy to understand.

\begin{Shaded}
\begin{Highlighting}[]
\FunctionTok{feols}\NormalTok{(Ozone }\SpecialCharTok{\textasciitilde{}} \FunctionTok{sw}\NormalTok{(Wind, Temp) }\SpecialCharTok{|} \FunctionTok{csw0}\NormalTok{(Month, Day), airquality)}
\CommentTok{\#\textgreater{} }\AlertTok{NOTE}\CommentTok{: 37 observations removed because of NA values (LHS: 37).}
\CommentTok{\#\textgreater{}       |{-}\textgreater{} this msg only concerns the variables common to all estimations}
\CommentTok{\#\textgreater{}                                x.1               x.2                x.3}
\CommentTok{\#\textgreater{} Dependent Var.:              Ozone             Ozone              Ozone}
\CommentTok{\#\textgreater{}                                                                        }
\CommentTok{\#\textgreater{} Constant          96.87*** (7.239) {-}147.0*** (18.29)                   }
\CommentTok{\#\textgreater{} Wind            {-}5.551*** (0.6904)                   {-}4.643*** (0.7019)}
\CommentTok{\#\textgreater{} Temp                               2.429*** (0.2331)                   }
\CommentTok{\#\textgreater{} Fixed{-}Effects:  {-}{-}{-}{-}{-}{-}{-}{-}{-}{-}{-}{-}{-}{-}{-}{-}{-}{-} {-}{-}{-}{-}{-}{-}{-}{-}{-}{-}{-}{-}{-}{-}{-}{-}{-} {-}{-}{-}{-}{-}{-}{-}{-}{-}{-}{-}{-}{-}{-}{-}{-}{-}{-}}
\CommentTok{\#\textgreater{} Month                           No                No                Yes}
\CommentTok{\#\textgreater{} Day                             No                No                 No}
\CommentTok{\#\textgreater{} \_\_\_\_\_\_\_\_\_\_\_\_\_\_\_ \_\_\_\_\_\_\_\_\_\_\_\_\_\_\_\_\_\_ \_\_\_\_\_\_\_\_\_\_\_\_\_\_\_\_\_ \_\_\_\_\_\_\_\_\_\_\_\_\_\_\_\_\_\_}
\CommentTok{\#\textgreater{} S.E. type                      IID               IID                IID}
\CommentTok{\#\textgreater{} Observations                   116               116                116}
\CommentTok{\#\textgreater{} R2                         0.36186           0.48771            0.45290}
\CommentTok{\#\textgreater{} Within R2                       {-}{-}                {-}{-}            0.28462}
\CommentTok{\#\textgreater{} }
\CommentTok{\#\textgreater{}                               x.4                x.5               x.6}
\CommentTok{\#\textgreater{} Dependent Var.:             Ozone              Ozone             Ozone}
\CommentTok{\#\textgreater{}                                                                       }
\CommentTok{\#\textgreater{} Constant                                                              }
\CommentTok{\#\textgreater{} Wind                              {-}4.454*** (0.7554)                  }
\CommentTok{\#\textgreater{} Temp            2.704*** (0.3182)                    2.952*** (0.3579)}
\CommentTok{\#\textgreater{} Fixed{-}Effects:  {-}{-}{-}{-}{-}{-}{-}{-}{-}{-}{-}{-}{-}{-}{-}{-}{-} {-}{-}{-}{-}{-}{-}{-}{-}{-}{-}{-}{-}{-}{-}{-}{-}{-}{-} {-}{-}{-}{-}{-}{-}{-}{-}{-}{-}{-}{-}{-}{-}{-}{-}{-}}
\CommentTok{\#\textgreater{} Month                         Yes                Yes               Yes}
\CommentTok{\#\textgreater{} Day                            No                Yes               Yes}
\CommentTok{\#\textgreater{} \_\_\_\_\_\_\_\_\_\_\_\_\_\_\_ \_\_\_\_\_\_\_\_\_\_\_\_\_\_\_\_\_ \_\_\_\_\_\_\_\_\_\_\_\_\_\_\_\_\_\_ \_\_\_\_\_\_\_\_\_\_\_\_\_\_\_\_\_}
\CommentTok{\#\textgreater{} S.E. type                     IID                IID               IID}
\CommentTok{\#\textgreater{} Observations                  116                115               115}
\CommentTok{\#\textgreater{} R2                        0.53831            0.66733           0.74210}
\CommentTok{\#\textgreater{} Within R2                 0.39630            0.30290           0.45957}
\CommentTok{\#\textgreater{} {-}{-}{-}}
\CommentTok{\#\textgreater{} Signif. codes: 0 \textquotesingle{}***\textquotesingle{} 0.001 \textquotesingle{}**\textquotesingle{} 0.01 \textquotesingle{}*\textquotesingle{} 0.05 \textquotesingle{}.\textquotesingle{} 0.1 \textquotesingle{} \textquotesingle{} 1}
\end{Highlighting}
\end{Shaded}

\subsection{Formula interpolation}\label{formula-interpolation}

\label{sec_fml_interpol}

Another key design feature of \pkg{fixest} formulas is the ease of
interpolating variables. Programmatic formula manipulation in
\proglang{R} is a long-standing issue, for which users often resort to
\emph{ad hoc} solutions.\footnote{For example,
  \texttt{as.formula(paste(...))}. While \texttt{stats::reformulate}
  provides a mechanism for programmatic formula construction, it is
  limited to simple (single part) formulas and does not support
  interpolation.} In contrast, \pkg{fixest} support three primary ways
to interpolate variables directly in the formula itself: i) with formula
macros, ii) with regular expressions, and iii) with the
dot-square-bracket operator.

A cheat sheet of all possible interpolations in Table
\ref{tab_fml_interpol}. We also work through some explicit examples
below, and explain the role of the \texttt{xpd} function, which powers
all of \pkg{fixest}'s interpolation features under the hood.

\begin{table}[tbh!]
  \centering
  \caption{\label{tab_fml_interpol}Cheat sheet for variable interpolation in \texttt{fixest} estimations.}
  

\small
\begin{tabular}{ m{4.3cm} m{3.9cm} m{5.4cm} }
  \toprule
  Preceding code & Formula & Interpolates as\\
  \midrule
  
  \multicolumn{3}{l}{
    \textit{all examples with the \texttt{airquality} data set, with variables:} 
    \texttt{Ozone}, \texttt{Solar.R}, \texttt{Wind}, \texttt{Temp}, \texttt{Month}, \texttt{Day}
  }\\
  
  \\
  
  \multicolumn{3}{l}{
    \textsc{interpolation with} \texttt{.[]}
  }\\
  
  \\
  
  \texttt{x = c("Temp", "Month")} &
  \texttt{Ozone \textasciitilde .[x]} &
  \texttt{Ozone \textasciitilde Temp + Month}\\
  
  \\
  
  \texttt{x = \textasciitilde Temp + Month} &
  \texttt{Ozone \textasciitilde .[x]} &
  \texttt{Ozone \textasciitilde Temp + Month}\\
  
  \\
  
  \texttt{x = character(0)} &
  \texttt{Ozone \textasciitilde .[x]} &
  \texttt{Ozone \textasciitilde 1}\\
  
  \\
  
  \texttt{x = \textasciitilde Temp | Month} &
  \texttt{Ozone \textasciitilde .[x]} &
  \texttt{Ozone \textasciitilde Temp | Month}\\
  
  \\
  
  \texttt{y = c("Ozone", "Wind")} &
  \texttt{.[y] \textasciitilde Temp + Month} &
  \texttt{c(Ozone, Wind) \textasciitilde Temp + Month}\\
  
  \multicolumn{3}{l}{
    \footnotesize \textit{note:} passing several variables in the left-hand-side nests them in \texttt{c()} 
    to trigger multiple estimations
  }\\
  
  \\
  
  \texttt{x = c("Temp", "Month")} &
  \texttt{Ozone \textasciitilde sw(.[, x])} &
  \texttt{Ozone \textasciitilde sw(Temp, Month)}\\
  
  \\
  
  \texttt{n = 1:3} &
  \texttt{y \textasciitilde x.[n]} &
  \texttt{y \textasciitilde x1 + x2 + x3}\\
  
  \\
  
  \midrule
  
  \multicolumn{3}{l}{
    \textsc{macro variables starting with} \texttt{..}
  }\\
  
  \\
  
  \makecell[l]{\texttt{setFixest\_fml(} \\ \texttt{~~..x = \textasciitilde Temp + Month} \\ \texttt{)} } &
  \texttt{Ozone \textasciitilde ..x} &
  \texttt{Ozone \textasciitilde Temp + Month}\\
  
  \\
  
  \makecell[l]{\texttt{setFixest\_fml(} \\ \texttt{~~..y = \textasciitilde c(Ozone, Wind)} \\ \texttt{)} } &
  \texttt{..y \textasciitilde Temp + Month} &
  \texttt{c(Ozone, Wind) \textasciitilde Temp + Month}\\ 
  
  \\
  
  \midrule
  
  \multicolumn{3}{l}{
    \textsc{interpolations tied to a data set}
  }\\
  
  \multicolumn{3}{l}{
    \hspace{1em} \textit{using} \texttt{..("reg.-expr.")} \textit{or} \texttt{regex("reg.-expr.")}
  }\\
      
  \\
  
   &
  \texttt{Ozone \textasciitilde ..("Te|Mo")} &
  \texttt{Ozone \textasciitilde Temp + Month} \\
  
  \\
  
   &
  \texttt{Ozone \textasciitilde regex("Te|Mo")} &
  \texttt{Ozone \textasciitilde Temp + Month} \\
  
  \\
  
  
  \multicolumn{3}{l}{
    \hspace{1em} \textit{using name completion with} \texttt{name..}
  }\\
  
  \\
  
   &
  \texttt{Ozone \textasciitilde T.. + M..} &
  \texttt{Ozone \textasciitilde Temp + Month} \\
  
  \bottomrule
\end{tabular}

\end{table}

\subsubsection{Formula macros}\label{formula-macros}

Users can define (global) formula macros in \pkg{fixest} with the
function \texttt{setFixest\_fml}. While the function accepts any number
of arguments, each argument name must start with two dots and the
arguments themselves must be either a character vector or a one-sided
formula.\footnote{The \texttt{..\textless{}macro\textgreater{}} naming
  restriction ensures that macros can easily be distinguished from
  regular variables.} For example
\texttt{setFixest\_fml(..controls\ =\ c("Wind",\ "Temp"))} creates a
macro variable named \texttt{..controls}. This variable can now be used
in any \pkg{fixest} estimation as shown below:

\begin{Shaded}
\begin{Highlighting}[]
\FunctionTok{setFixest\_fml}\NormalTok{(}\AttributeTok{..controls =} \SpecialCharTok{\textasciitilde{}}\NormalTok{Wind }\SpecialCharTok{+}\NormalTok{ Temp)}
\FunctionTok{formula}\NormalTok{(}\FunctionTok{feols}\NormalTok{(Ozone }\SpecialCharTok{\textasciitilde{}}\NormalTok{ ..controls, airquality, }\AttributeTok{notes =} \ConstantTok{FALSE}\NormalTok{))}
\CommentTok{\#\textgreater{} Ozone \textasciitilde{} Wind + Temp}
\CommentTok{\#\textgreater{} \textless{}environment: 0x000002372212fea8\textgreater{}}
\end{Highlighting}
\end{Shaded}

Users may create as many macros as they wish through
\texttt{setFixest\_fml}. Macros can also contain fixed-effects if they
are created with a formula,
e.g.~\texttt{..controls\ =\ \textasciitilde{}Wind\ \textbar{}\ Month}.
Macros can prove especially useful in analysis scripts, with clear
definitions of recurring sets of variables. One caveat, though is that
we don't recommend using macros in loops. For this use case,
interpolation with the dot-square-bracket operator, later described, is
better suited.

\subsubsection{Regex interpolation}\label{regex-interpolation}

Another way to interpolate variables in formulas is through regular
expressions. Use \texttt{..("regex")} or \texttt{regex("regex")} to
include in the formula any variable in the data set that matches the
regular expression. When a variable name in the formula ends with two
dots, any variable in the data set starting with the same letters are
included. Here is an example of these two cases:

\begin{Shaded}
\begin{Highlighting}[]
\FunctionTok{formula}\NormalTok{(}\FunctionTok{feols}\NormalTok{(Ozone }\SpecialCharTok{\textasciitilde{}} \FunctionTok{..}\NormalTok{(}\StringTok{"Wi|Tem"}\NormalTok{) }\SpecialCharTok{|}\NormalTok{ Mo.., airquality, }\AttributeTok{notes =} \ConstantTok{FALSE}\NormalTok{))}
\CommentTok{\#\textgreater{} Ozone \textasciitilde{} Wind + Temp | Month}
\end{Highlighting}
\end{Shaded}

\subsubsection{Dot-square-bracket}\label{dot-square-bracket}

Finally, a powerful way to interpolate variables is with the
\emph{dot-square-bracket} operator. When expressions of the form
\texttt{.{[}x{]}} are found in a formula, the value of the variable
\texttt{x} is directly inserted in the formula. When \texttt{x} is a
vector, the values are combined with a \texttt{+}.

For example, let us reimplement our formula macro via the
dot-square-bracket operator:

\begin{Shaded}
\begin{Highlighting}[]
\NormalTok{x }\OtherTok{=} \FunctionTok{c}\NormalTok{(}\StringTok{"Wind"}\NormalTok{, }\StringTok{"Temp"}\NormalTok{)}
\FunctionTok{formula}\NormalTok{(}\FunctionTok{feols}\NormalTok{(Ozone }\SpecialCharTok{\textasciitilde{}}\NormalTok{ .[x], airquality, }\AttributeTok{notes =} \ConstantTok{FALSE}\NormalTok{))}
\CommentTok{\#\textgreater{} Ozone \textasciitilde{} Wind + Temp}
\CommentTok{\#\textgreater{} \textless{}environment: 0x000002371dac77a0\textgreater{}}
\end{Highlighting}
\end{Shaded}

This interpolation makes it very easy to work with formulas
programmatically within loops.

As illustrated above, when the interpolation is on several variables,
they are combined with a plus (i.e.~if \texttt{x\ =\ c("a",\ "b")}, then
\texttt{.{[}x{]}} becomes \texttt{a\ +\ b} in the formula). However,
sometimes we may want to combine them with a comma, especially when
inserting the variables as several arguments of a function. You can
override the default behavior but adding a comma right after the opening
bracket, as in \texttt{.{[},x{]}}. Let us estimate multiple models that
way and access the coefficients:

\begin{Shaded}
\begin{Highlighting}[]
\NormalTok{x }\OtherTok{=} \FunctionTok{c}\NormalTok{(}\StringTok{"Wind"}\NormalTok{, }\StringTok{"Temp"}\NormalTok{)}
\FunctionTok{feols}\NormalTok{(Ozone }\SpecialCharTok{\textasciitilde{}} \FunctionTok{sw}\NormalTok{(.[,x]), airquality, }\AttributeTok{notes =} \ConstantTok{FALSE}\NormalTok{) }\SpecialCharTok{|\textgreater{}}
  \FunctionTok{coef}\NormalTok{()}
\CommentTok{\#\textgreater{}   id  rhs (Intercept)      Wind     Temp}
\CommentTok{\#\textgreater{} 1  1 Wind    96.87289 {-}5.550923       NA}
\CommentTok{\#\textgreater{} 2  2 Temp  {-}146.99549        NA 2.428703}
\end{Highlighting}
\end{Shaded}

\subsubsection{xpd: generalized formula interpolation and
expansion}\label{xpd-generalized-formula-interpolation-and-expansion}

Under the hood, all interpolation features in \pkg{fixest} are powered
by the function \texttt{xpd}. This function ensures that interpolated
formulas are expanded to their correct full form. For example, we could
have used \texttt{xpd} to verify that the expanded formula in our
previous example, prior to running the actual estimation:

\begin{Shaded}
\begin{Highlighting}[]
\FunctionTok{xpd}\NormalTok{(Ozone }\SpecialCharTok{\textasciitilde{}} \FunctionTok{sw}\NormalTok{(.[,x]))}
\CommentTok{\#\textgreater{} Ozone \textasciitilde{} sw(Wind, Temp)}
\CommentTok{\#\textgreater{} \textless{}environment: 0x000002371fd59ac8\textgreater{}}
\end{Highlighting}
\end{Shaded}

As the reader may have noticed, this example also demonstrates that
\texttt{xpd} can be used on ``regular'' formulas outside of the main
\pkg{fixest} estimation functions. The only difference is that these
native \pkg{fixest} functions automatically use \texttt{xpd} to expand
their formulas internally and thus do not need an explicit \texttt{xpd}
call.

The \texttt{?xpd} documentation covers a variety of use-cases and
features, which we do not recapitulate here. Instead, we content
ourselves with a final example where the desired outcome is
interpolating over variables with numerical suffixes. In this particular
case, assume that we wish include the variables \texttt{x1}--\texttt{x5}
in a regression. When the dot-square-bracket operator is attached to a
variable, this variable can be used as the interpolation prefix, as
follows:

\begin{Shaded}
\begin{Highlighting}[]
\FunctionTok{xpd}\NormalTok{(y }\SpecialCharTok{\textasciitilde{}}\NormalTok{ x.[}\DecValTok{1}\SpecialCharTok{:}\DecValTok{5}\NormalTok{])}
\CommentTok{\#\textgreater{} y \textasciitilde{} x1 + x2 + x3 + x4 + x5}
\CommentTok{\#\textgreater{} \textless{}environment: 0x000002371fdf3568\textgreater{}}
\end{Highlighting}
\end{Shaded}

\subsection{Formula features for panel
data}\label{formula-features-for-panel-data}

The last features we will highlight in this section are functions that
are designed to work with time-series or panel data. In panel data
settings, where you have unit-time identifiers, it is often useful to
quickly compute the leads, lags, and/or differences of variables. The
\pkg{fixest} formula provides a family of convenience functions for
computing these transformations on the fly: \texttt{l}, \texttt{f} and
\texttt{d}. However, to access these functions \pkg{fixest} must first
know what the unit and time identifiers are. This information can be
passed either via the run-time \texttt{panel.id} argument, or by
explicitly setting the dataset as a panel via the \texttt{panel}
function.

Let us demonstrate with a quick example of the former approach. Here we
specify the panel identifiers as formula to \texttt{panel.id}, which in
turn allows us to concisely estimate the lead of the dependent variable
(\texttt{f(y)}) on the dependent variable and its three first lags
(\texttt{l(x1,\ 0:3)}).

\begin{Shaded}
\begin{Highlighting}[]
\FunctionTok{data}\NormalTok{(base\_did)}
\FunctionTok{feols}\NormalTok{(}\FunctionTok{f}\NormalTok{(y) }\SpecialCharTok{\textasciitilde{}} \FunctionTok{l}\NormalTok{(x1, }\DecValTok{0}\SpecialCharTok{:}\DecValTok{3}\NormalTok{), base\_did, }\AttributeTok{panel.id =} \SpecialCharTok{\textasciitilde{}}\NormalTok{id }\SpecialCharTok{+}\NormalTok{ period)}
\CommentTok{\#\textgreater{} }\AlertTok{NOTE}\CommentTok{: 432 observations removed because of NA values (LHS: 108, RHS: 324).}
\CommentTok{\#\textgreater{} OLS estimation, Dep. Var.: f(y)}
\CommentTok{\#\textgreater{} Observations: 648}
\CommentTok{\#\textgreater{} Standard{-}errors: IID }
\CommentTok{\#\textgreater{}             Estimate Std. Error   t value  Pr(\textgreater{}|t|)    }
\CommentTok{\#\textgreater{} (Intercept) 3.138156   0.228711 13.721072 \textless{} 2.2e{-}16 ***}
\CommentTok{\#\textgreater{} l(x1, 0)    0.024668   0.080943  0.304762  0.760646    }
\CommentTok{\#\textgreater{} l(x1, 1)    0.095704   0.080963  1.182069  0.237615    }
\CommentTok{\#\textgreater{} l(x1, 2)    0.069380   0.078556  0.883194  0.377462    }
\CommentTok{\#\textgreater{} l(x1, 3)    0.162121   0.077900  2.081127  0.037817 *  }
\CommentTok{\#\textgreater{} {-}{-}{-}}
\CommentTok{\#\textgreater{} Signif. codes:  0 \textquotesingle{}***\textquotesingle{} 0.001 \textquotesingle{}**\textquotesingle{} 0.01 \textquotesingle{}*\textquotesingle{} 0.05 \textquotesingle{}.\textquotesingle{} 0.1 \textquotesingle{} \textquotesingle{} 1}
\CommentTok{\#\textgreater{} RMSE: 5.79176   Adj. R2: 0.004676}
\end{Highlighting}
\end{Shaded}

It should be emphasized that \pkg{fixest}'s panel functionality offers
rich support for handling ambiguous and irregular timesteps, as well as
missing observations. For example, users can pass a \texttt{time.step}
argument to the \texttt{panel} function to delineate between daily data
that excludes weekends. Again, we refer the interested reader to the
package documentation (e.g., \texttt{?panel}) for more examples.

\subsubsection{Differences-in-differences}\label{differences-in-differences}

\pkg{fixest} also has a set of tools for estimating
difference-in-differences (DiD) models, which are a particularly popular
research design in applied econometrics settings
\citep{huntington2021effect}. Using the \texttt{base\_stagg} data set
provided in \pkg{fixest}, let us illustrate a DiD analysis. In this
example we interact the variable \texttt{period}, used as categorical,
with the variable \texttt{treat} and set the reference to the period 5,
the period right before the treatment takes place. We also include unit
and time fixed-effects.

\begin{Shaded}
\begin{Highlighting}[]
\FunctionTok{data}\NormalTok{(base\_stagg, }\AttributeTok{package =} \StringTok{"fixest"}\NormalTok{)}
\NormalTok{est\_did }\OtherTok{=} \FunctionTok{feols}\NormalTok{(y }\SpecialCharTok{\textasciitilde{}} \FunctionTok{i}\NormalTok{(time\_to\_treatment, }\AttributeTok{ref =} \FunctionTok{c}\NormalTok{(}\SpecialCharTok{{-}}\DecValTok{1000}\NormalTok{, }\SpecialCharTok{{-}}\DecValTok{1}\NormalTok{)) }\SpecialCharTok{|}\NormalTok{ id }\SpecialCharTok{+}\NormalTok{ year, }
\NormalTok{                base\_stagg)}
\NormalTok{est\_did}
\CommentTok{\#\textgreater{} OLS estimation, Dep. Var.: y}
\CommentTok{\#\textgreater{} Observations: 950}
\CommentTok{\#\textgreater{} Fixed{-}effects: id: 95,  year: 10}
\CommentTok{\#\textgreater{} Standard{-}errors: IID }
\CommentTok{\#\textgreater{}                       Estimate Std. Error  t value  Pr(\textgreater{}|t|)    }
\CommentTok{\#\textgreater{} time\_to\_treatment::{-}9 3.146558   1.190534 2.642982 0.0083731 ** }
\CommentTok{\#\textgreater{} time\_to\_treatment::{-}8 0.664518   0.882894 0.752659 0.4518683    }
\CommentTok{\#\textgreater{} time\_to\_treatment::{-}7 1.049788   0.745073 1.408973 0.1592180    }
\CommentTok{\#\textgreater{} time\_to\_treatment::{-}6 0.815505   0.661099 1.233560 0.2177166    }
\CommentTok{\#\textgreater{} time\_to\_treatment::{-}5 0.349854   0.602687 0.580490 0.5617420    }
\CommentTok{\#\textgreater{} time\_to\_treatment::{-}4 0.870255   0.559299 1.555975 0.1200957    }
\CommentTok{\#\textgreater{} time\_to\_treatment::{-}3 0.384655   0.526132 0.731100 0.4649248    }
\CommentTok{\#\textgreater{} time\_to\_treatment::{-}2 0.552281   0.500778 1.102846 0.2704143    }
\CommentTok{\#\textgreater{} ... 9 coefficients remaining (display them with summary() or use}
\CommentTok{\#\textgreater{} argument n)}
\CommentTok{\#\textgreater{} {-}{-}{-}}
\CommentTok{\#\textgreater{} Signif. codes:  0 \textquotesingle{}***\textquotesingle{} 0.001 \textquotesingle{}**\textquotesingle{} 0.01 \textquotesingle{}*\textquotesingle{} 0.05 \textquotesingle{}.\textquotesingle{} 0.1 \textquotesingle{} \textquotesingle{} 1}
\CommentTok{\#\textgreater{} RMSE: 2.13618     Adj. R2: 0.445625}
\CommentTok{\#\textgreater{}                 Within R2: 0.33628 }
\end{Highlighting}
\end{Shaded}

Alternatively, there has been new proposals for robust DID estimators
that are valid when treatment-timing is staggered. The \texttt{sunab()}
function implements the method from \citet{sun2021estimating}.

\begin{Shaded}
\begin{Highlighting}[]
\NormalTok{est\_did\_sunab }\OtherTok{=} \FunctionTok{feols}\NormalTok{(y }\SpecialCharTok{\textasciitilde{}} \FunctionTok{sunab}\NormalTok{(year\_treated, year) }\SpecialCharTok{|}\NormalTok{ id }\SpecialCharTok{+}\NormalTok{ year, base\_stagg)}
\end{Highlighting}
\end{Shaded}

It is very easy to plot both of these estimates thanks to the
\texttt{iplot} function.

\begin{Shaded}
\begin{Highlighting}[]
\FunctionTok{iplot}\NormalTok{(est\_did, est\_did\_sunab, }\AttributeTok{drop =} \StringTok{"1000"}\NormalTok{)}
\FunctionTok{legend}\NormalTok{(}\StringTok{"topleft"}\NormalTok{, }\FunctionTok{c}\NormalTok{(}\StringTok{"TWFE"}\NormalTok{, }\StringTok{"Sun and Abraham"}\NormalTok{), }\AttributeTok{col =} \DecValTok{1}\SpecialCharTok{:}\DecValTok{2}\NormalTok{, }\AttributeTok{lty =} \DecValTok{1}\NormalTok{)}
\end{Highlighting}
\end{Shaded}

\begin{figure}[htbp]
  \centering
  \caption{\label{fig_inter_iplot}Default \texttt{iplot} graph for DiD estimations.}
  
  \includegraphics[width=1\textwidth]{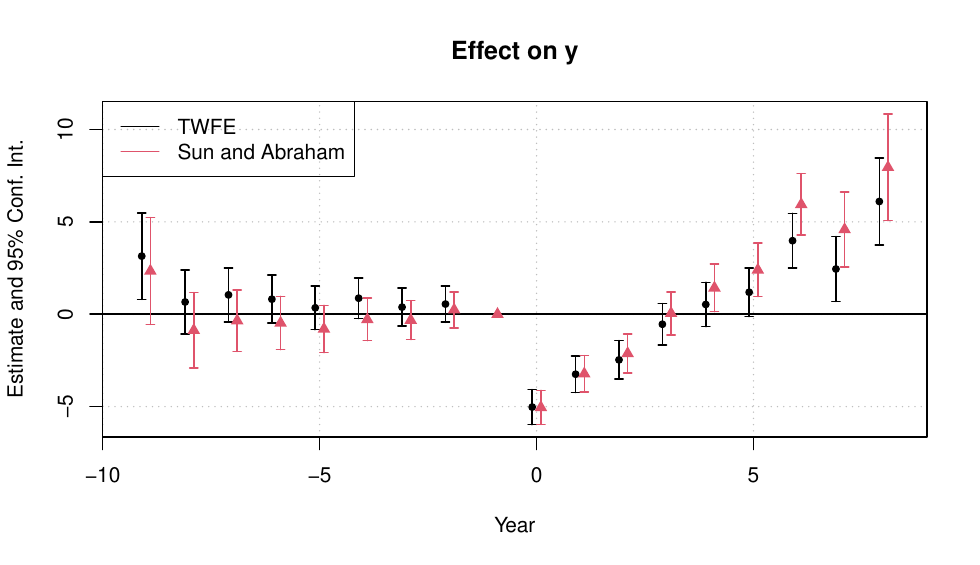}
\end{figure}

The graph is represented in Figure \ref{fig_inter_iplot}.

\section{Adjusting the
standard-errors}\label{adjusting-the-standard-errors}

\label{sec_vcov}

Next to the coefficients themselves, the variance-covariance matrix
(VCOV) is arguably the most important element of any estimation. This is
the arbiter of statistical significance and the determinant for any
inferential claims about our results. Given its prominence, \pkg{fixest}
aims to streamline the way VCOVs are selected through a simple and
intuitive interface. Specifically, all of the core estimating functions,
as well as many of the methods, accept a named \texttt{vcov} argument.
This interface is tightly integrated with the rest of package and
enables many synergies, such as accessing multiple VCOVs at estimation
time or when exporting the results. Moreover, a core design principle of
\pkg{fixest} is that it should be easy to change a model's VCOV
\emph{post} estimation, which in turn enables on-the-fly inference
adjustments to standard-errors.

In this section, we first present \pkg{fixest}'s built-in VCOVs and how
to parameterize them. We then show how to integrate the package with
other software to compute additional VCOVs.

\subsection{Built-in VCOVs}\label{built-in-vcovs}

\pkg{fixest} provides native support for a variety of VCOVs. These
built-in VCOVs cover some of the most common adjustments that are
required in applied econometrics research and also offer best-in-class
performance. The full set of these built-in VCOVs are documented in
Table \ref{tab_vcov_builtin}, including specification options and
required arguments. For the sake of brevity, we list the keyword
shortcuts for these built-in VCOVs below:

\begin{itemize}
\tightlist
\item
  \texttt{"iid"}: the default VCOV, without any adjustment, where it is
  assumed that the errors are homoskedastic and non-correlated
\item
  \texttt{"hc(1-3)"}: various heteroskedasticity-robust VCOVs, where
  each error has its own variance \emph{a la}
  \citet{white1980heteroskedasticity} and
  \citet{mackinnon1985heteroskedasticity}; note that \texttt{"hetero"}
  is an alias for \texttt{"hc1"}
\item
  \texttt{"cluster"}: clustered VCOV, where arbitrary correlation within
  groups (clusters) is allowed
\item
  \texttt{"twoway"}: double clustering of the VCOV
\item
  \texttt{"conley"}: VCOV that accounts for spatial correlation across
  errors, \emph{a la} \citet{conley1999gmm}
\item
  \texttt{"NW"}: in a time-series or panel time-series context, the
  Newey-West VCOV accounts for serial correlation within units and
  heteroskedasticity \citep{newey1987simple}
\item
  \texttt{"DK"}: in a time-series or panel time-series context, the
  Driscoll-Kraay VCOV accounts for serial correlation within units and
  arbitrary cross-sections; correlation across units within a time
  period \citep{driscoll1998consistent}
\end{itemize}

To invoke any of these built-in VCOVs, users need simply pass the
relevant keyword to the \texttt{vcov} argument. As we have tried to
emphasize, this can be done either at estimation time or as part of a
post-estimation method. Many \pkg{fixest} functions accept the
\texttt{vcov} argument and in the below example we demonstrate using the
\texttt{se} function, which retrieves standard errors. First, we specify
heteroskedastic-robust errors at estimation, before reverting to
spherical errors for comparison in a subsequent call.

\begin{Shaded}
\begin{Highlighting}[]
\NormalTok{est\_vcov }\OtherTok{=} \FunctionTok{feols}\NormalTok{(y }\SpecialCharTok{\textasciitilde{}}\NormalTok{ x1, base\_did, }\AttributeTok{vcov =} \StringTok{"hc1"}\NormalTok{)}
\FunctionTok{rbind}\NormalTok{(}\AttributeTok{hc1 =} \FunctionTok{se}\NormalTok{(est\_vcov),}
      \AttributeTok{iid =} \FunctionTok{se}\NormalTok{(est\_vcov, }\AttributeTok{vcov =} \StringTok{"iid"}\NormalTok{))}
\CommentTok{\#\textgreater{}     (Intercept)         x1}
\CommentTok{\#\textgreater{} hc1   0.1490257 0.05101605}
\CommentTok{\#\textgreater{} iid   0.1491554 0.05011910}
\end{Highlighting}
\end{Shaded}

While this example is very simple, it demonstrates the ease with which
\pkg{fixest} allows users to adjust VCOVs and, thus, standard errors.
Moreover, the \emph{on-the-fly} adjustment underscores a key
\pkg{fixest} design principle: estimation is separate from inference.
This principle enables user-level flexibility in addition to
facilitating computational efficiency, since we only need to recompute
the model scores instead of re-estimating the entire model.

Beyond keyword strings, the \texttt{vcov} argument also accepts various
other inputs. Again, these are detailed in full in Table
\ref{tab_vcov_builtin}. To quickly highlight one particularly common
use-case in panel-based econometrics, users can specify clustered
standard errors by passing a one-sided formula denoting which
variable(s) to cluster on. For example:

\begin{Shaded}
\begin{Highlighting}[]
\FunctionTok{feols}\NormalTok{(y }\SpecialCharTok{\textasciitilde{}}\NormalTok{ x1, base\_did, }\AttributeTok{vcov =} \SpecialCharTok{\textasciitilde{}}\NormalTok{id)}
\CommentTok{\#\textgreater{} OLS estimation, Dep. Var.: y}
\CommentTok{\#\textgreater{} Observations: 1,080}
\CommentTok{\#\textgreater{} Standard{-}errors: Clustered (id) }
\CommentTok{\#\textgreater{}             Estimate Std. Error t value  Pr(\textgreater{}|t|)    }
\CommentTok{\#\textgreater{} (Intercept) 1.988753   0.194352 10.2327 \textless{} 2.2e{-}16 ***}
\CommentTok{\#\textgreater{} x1          0.983110   0.046789 21.0115 \textless{} 2.2e{-}16 ***}
\CommentTok{\#\textgreater{} {-}{-}{-}}
\CommentTok{\#\textgreater{} Signif. codes:  0 \textquotesingle{}***\textquotesingle{} 0.001 \textquotesingle{}**\textquotesingle{} 0.01 \textquotesingle{}*\textquotesingle{} 0.05 \textquotesingle{}.\textquotesingle{} 0.1 \textquotesingle{} \textquotesingle{} 1}
\CommentTok{\#\textgreater{} RMSE: 4.89686   Adj. R2: 0.262357}
\end{Highlighting}
\end{Shaded}

Continuing with non-keyword inputs to \texttt{vcov}, note that some
VCOVs may require additional information. For example, Driscoll-Kraay
VCOVs require unit-time identifiers, while Conley VCOVs require
geographic information. In general, \pkg{fixest} tries to deduce any
additional VCOV arguments using simple but robust rules of
thumb.\footnote{For example,
  \texttt{feols(depth\ \textasciitilde{}\ mag,\ quakes,\ vcov\ =\ "conley")}
  specifies Conley standard errors in the base R \texttt{quakes} dataset
  to allow for arbitrary spatial auto correlation. Normally this
  requires explicit latitude and longitude information. But \pkg{fixest}
  automatically deduces that this information is furnished by the
  \texttt{lat} and \texttt{long} columns of the dataset.} But users can
also pass additional arguments explicitly in one of two ways. First, by
specifying a special helper function, whose name is the VCOV keyword
prefixed with \texttt{vcov\_}. For example, \texttt{vcov\_cluster} has
the argument \texttt{cluster} accepting a one-sided formula or a
character scalar specifying the cluster variable. Second, by invoking a
formula with the following syntax:

\[
\mathtt{vcov} = \mathtt{VCOV\_name} \sim \mathtt{variables}.
\]

Here, \texttt{VCOV\_name} is a function that takes additional arguments.
For example, \texttt{NW} and \texttt{DK} accept the argument lag, and
\texttt{conley} accept the arguments \texttt{cutoff}, \texttt{pixel} and
\texttt{distance}.

Re-using the same estimation from above, let us report the Newey-West
and clustered standard-errors using these two approaches. This time,
we'll pass the key \texttt{vcov} argument a list of VCOVs from an
\texttt{etable} call.

\begin{Shaded}
\begin{Highlighting}[]
\FunctionTok{etable}\NormalTok{(est\_vcov, }\AttributeTok{vcov =} \FunctionTok{list}\NormalTok{(}\StringTok{"hc1"}\NormalTok{, NW }\SpecialCharTok{\textasciitilde{}}\NormalTok{ id }\SpecialCharTok{+}\NormalTok{ period, }\FunctionTok{vcov\_cluster}\NormalTok{(}\SpecialCharTok{\textasciitilde{}}\NormalTok{ id)))}
\CommentTok{\#\textgreater{}                           est\_vcov         est\_vcov.1         est\_vcov.2}
\CommentTok{\#\textgreater{} Dependent Var.:                  y                  y                  y}
\CommentTok{\#\textgreater{}                                                                         }
\CommentTok{\#\textgreater{} Constant         1.989*** (0.1490)  1.989*** (0.1741)  1.989*** (0.1944)}
\CommentTok{\#\textgreater{} x1              0.9831*** (0.0510) 0.9831*** (0.0527) 0.9831*** (0.0468)}
\CommentTok{\#\textgreater{} \_\_\_\_\_\_\_\_\_\_\_\_\_\_\_ \_\_\_\_\_\_\_\_\_\_\_\_\_\_\_\_\_\_ \_\_\_\_\_\_\_\_\_\_\_\_\_\_\_\_\_\_ \_\_\_\_\_\_\_\_\_\_\_\_\_\_\_\_\_\_}
\CommentTok{\#\textgreater{} S.E. type       Heteroskedas.{-}rob.   Newey{-}West (L=1)             by: id}
\CommentTok{\#\textgreater{} Observations                 1,080              1,080              1,080}
\CommentTok{\#\textgreater{} R2                         0.26304            0.26304            0.26304}
\CommentTok{\#\textgreater{} Adj. R2                    0.26236            0.26236            0.26236}
\CommentTok{\#\textgreater{} {-}{-}{-}}
\CommentTok{\#\textgreater{} Signif. codes: 0 \textquotesingle{}***\textquotesingle{} 0.001 \textquotesingle{}**\textquotesingle{} 0.01 \textquotesingle{}*\textquotesingle{} 0.05 \textquotesingle{}.\textquotesingle{} 0.1 \textquotesingle{} \textquotesingle{} 1}
\end{Highlighting}
\end{Shaded}

As a final remark about the built-in VCOVs, \pkg{fixest} applies a small
sample correction of \(N / (N - K)\), where \(N\) is the number of
observations and \(K\) the number of variables. However, this small
sample correction can be disabled or adjusted in various ways via the
\texttt{ssc} function. Small sample corrections can play a surprisingly
prominent role in lining up results from different econometrics
software, particularly in the context of clustered VCOVs (where multiple
corrections are potentially valid from a theory perspective). We omit
further discussion for the sake of brevity here. But the interested
reader is referred to the dedicated
``\href{https://lrberge.github.io/fixest/articles/standard_errors.html}{On
standard errors}'' vignette for an exhaustive discussion and comparison
between \pkg{fixest} and other popular panel econometrics packages.

\begin{table}[htpb!]
  \centering
  \caption{\label{tab_vcov_builtin}Description of \texttt{fixest} built-in VCOVs, and how to write the \texttt{vcov} argument included in many functions of this package.}
  

\footnotesize

\begin{tabular}{ p{2.4cm} p{6.5cm} p{5.4cm} }
  \toprule
  Keyword & 
  Requirement & 
  \makecell[tl]{
    Provide requirement with \\
    \texttt{vcov = function} or \texttt{formula} 
  }
  \\
  \midrule
  
  \multicolumn{3}{l}{\textit{Spherical errors}}\\
  
  \texttt{"iid"} &
  \textit{No requirement} &
  \\
  
  \\
  
  \multicolumn{3}{l}{\textit{Heteroskedastic errors}}\\
  
  \makecell[tl]{
    \texttt{"hetero"}, \\
    \texttt{"hc1"}, \texttt{"hc2"} \\ 
    \texttt{"hc3"}
  }&
  \textit{No requirement} &
  \\
  
  \\
  
  \multicolumn{3}{l}{\textit{Clustered, within group correlation}}\\
  
  \texttt{"cluster"} &
  \makecell[tl]{
    req. 1: group identifier \\
    ~~\textit{default (if available, in that order):} \\
    \tabitem the unit identifier from the panel \\
    \tabitem the first fixed-effect
  } &
  \makecell[tl]{
    \texttt{vcov\_cluster("id")} \\
    \texttt{cluster \textasciitilde \, id} \\
    \texttt{\textasciitilde id} (i.e., no  formula lhs)
  }\\
  
  \\
  
  \multicolumn{3}{l}{\textit{Clustered, within groups correlations}}\\
  
  \texttt{"twoway"} &
  \makecell[tl]{
    req. 1: two group identifiers \\
    ~~\textit{default (if available, in that order):} \\
    \tabitem the unit and time identifiers \\ ~~~~~~ from the panel \\
    \tabitem the first two fixed-effects
  }&
  \makecell[tl]{
    \texttt{vcov\_cluster(c("id1", "id2"))} \\
    \texttt{cluster \textasciitilde \, id1 + id2} \\
    \texttt{\textasciitilde id1 + id2} (i.e., no  formula lhs)
  }\\
  
  \\
  
  \multicolumn{3}{l}{\textit{Newey West, serial correlation}}\\
  
  \texttt{"NW"} &
  \makecell[tl]{
    req. 1: unit and time identifiers \\
    ~~\textit{default (if available):} \\
    \tabitem the panel unit and time identifiers
  }&
  \makecell[tl]{
    \texttt{vcov\_NW(unit = "id", time = "year")} \\
    \texttt{NW \textasciitilde \,id + year}
  }\\
  
   &
  \makecell[tl]{
    req. 2: lag \\
    ~~\textit{default:} \\
    \tabitem default lag provided by \\ ~~~~~~ \texttt{sandwich::bwNeweyWest} 
  }&
  \makecell[tl]{
    \texttt{vcov\_NW(lag = 2, "id", "year")} \\
    \texttt{NW(lag = 2) \textasciitilde \,id + year}
  }\\
  
  \\
  
  \multicolumn{3}{l}{\textit{Driscoll Kraay, serial correlation}}\\
  
  \texttt{"DK"} &
  \makecell[tl]{
    req. 1: time identifier \\
    ~~\textit{default (if available):} \\
    \tabitem the time identifier from the panel
  }&
  \makecell[tl]{
    \texttt{vcov\_DK(time = "year")} \\
    \texttt{DK \textasciitilde \,year}
  }\\
  
   &
  \makecell[tl]{
    req. 2: lag \\
    ~~\textit{default:} \\
    \tabitem default lag equal to $N_T^{0.25}$
  }&
  \makecell[tl]{
    \texttt{vcov\_DK("year", lag = 4)} \\
    \texttt{DK(lag = 4) \textasciitilde \,year}
  }\\
  
  \\
  
  \multicolumn{3}{l}{\textit{Conley, spatial corelation}}\\
  
  \texttt{"conley"} &
  \makecell[tl]{
    req. 1: latitude and longitude \\
    ~~\textit{default (if available):} \\
    \tabitem lat. and long. variables from the data set
  } &
  \makecell[tl]{
    \texttt{vcov\_conley("lat", "lng")} \\
    \texttt{conley \textasciitilde \,lat + lng}
  }\\
  
   &
  \makecell[tl]{
    req. 2: distance cutoff \\
    ~~\textit{default:} \\
    \tabitem the default cutoff is based on an \\ ~~~~~~ internal algorithm using the median \\ ~~~~~~ distance across a sample of units
  }&
  \makecell[tl]{
    \texttt{vcov\_conley("lat", "lng", "100mi")} \\
    \texttt{conley(cutoff = "100mi") \textasciitilde \,lat + lng}
  }\\
  
  \bottomrule
\end{tabular}

\end{table}

\subsection{Access to external VCOVs}\label{access-to-external-vcovs}

Beyond the built-in VCOVs, the \texttt{vcov} argument also accepts
custom functions and VCOVs from packages. For example, \pkg{fixest} has
a relatively good compatibility with the well-known \pkg{sandwich}
package
\citep{zeileis2006ObjectorientedComputationSandwich, zeileis2020VariousVersatileVariances},
which specializes in computing the variance-covariance matrices.
However, there are a few important caveats. Some functions from
\pkg{sandwich} do not work well with \pkg{fixest} estimations when
fixed-effects are present, due to the way that these fixed-effects are
handled. Similarly, while most \pkg{sandwich} functions have a
\texttt{fixest} method, there is not a drop-in replacement for
\texttt{vcovBS()}, \texttt{vcovJK()}, and \texttt{vcovOPG()}. For these
reasons, we recommend that users rely on native \pkg{fixest} VCOVs where
possible.

Users can also define their custom VCOVs and pass them to
\texttt{summary}, \texttt{etable}, \texttt{coeftable}, etc. In this next
example, we compute a bootstrapped VCOV and display the associated
results.\footnote{As an aside, \pkg{fixest} provides a dedicated
  estimation environment that, among other things, allows for extremely
  efficient Bayesian bootstrap computation. See the \texttt{?est\_env}
  documentation for an example.} Note that we pass our custom VCOV
through a named list for pretty printing and display.

\begin{Shaded}
\begin{Highlighting}[]
\FunctionTok{data}\NormalTok{(iris)}
\NormalTok{est }\OtherTok{=} \FunctionTok{feols}\NormalTok{(Petal.Length }\SpecialCharTok{\textasciitilde{}}\NormalTok{ Sepal.Length, iris)}

\FunctionTok{set.seed}\NormalTok{(}\DecValTok{1}\NormalTok{)}
\NormalTok{n }\OtherTok{=} \FunctionTok{nrow}\NormalTok{(iris)}
\NormalTok{all\_coefs }\OtherTok{=} \FunctionTok{lapply}\NormalTok{(}\DecValTok{1}\SpecialCharTok{:}\DecValTok{99}\NormalTok{, }\ControlFlowTok{function}\NormalTok{(i) \{}
  \FunctionTok{feols}\NormalTok{(Petal.Length }\SpecialCharTok{\textasciitilde{}}\NormalTok{ Sepal.Length, iris[}\FunctionTok{sample}\NormalTok{(}\DecValTok{1}\SpecialCharTok{:}\NormalTok{n, n, }\AttributeTok{replace =} \ConstantTok{TRUE}\NormalTok{), ], }
        \AttributeTok{only.coef =} \ConstantTok{TRUE}\NormalTok{)}
\NormalTok{\})}
\NormalTok{all\_coefs }\OtherTok{=} \FunctionTok{do.call}\NormalTok{(rbind, all\_coefs)}
\NormalTok{boot\_vcov }\OtherTok{=} \FunctionTok{var}\NormalTok{(all\_coefs)}

\NormalTok{est\_boot }\OtherTok{=} \FunctionTok{summary}\NormalTok{(est, }\AttributeTok{vcov =} \FunctionTok{list}\NormalTok{(}\StringTok{"My Bootstrapped SEs"} \OtherTok{=}\NormalTok{ boot\_vcov))}
\FunctionTok{print}\NormalTok{(est\_boot)}
\CommentTok{\#\textgreater{} OLS estimation, Dep. Var.: Petal.Length}
\CommentTok{\#\textgreater{} Observations: 150}
\CommentTok{\#\textgreater{} Standard{-}errors: My Bootstrapped SEs }
\CommentTok{\#\textgreater{}              Estimate Std. Error  t value  Pr(\textgreater{}|t|)    }
\CommentTok{\#\textgreater{} (Intercept)  {-}7.10144   0.414480 {-}17.1334 \textless{} 2.2e{-}16 ***}
\CommentTok{\#\textgreater{} Sepal.Length  1.85843   0.067392  27.5764 \textless{} 2.2e{-}16 ***}
\CommentTok{\#\textgreater{} {-}{-}{-}}
\CommentTok{\#\textgreater{} Signif. codes:  0 \textquotesingle{}***\textquotesingle{} 0.001 \textquotesingle{}**\textquotesingle{} 0.01 \textquotesingle{}*\textquotesingle{} 0.05 \textquotesingle{}.\textquotesingle{} 0.1 \textquotesingle{} \textquotesingle{} 1}
\CommentTok{\#\textgreater{} RMSE: 0.86201   Adj. R2: 0.758333}
\end{Highlighting}
\end{Shaded}

\section{Presentation}\label{presentation-2}

\label{sec_presentation}

In this section, we discuss in turn how to display the results in a
table or in a graph.

\subsection{Formatting the results into a
table}\label{formatting-the-results-into-a-table}

\label{sec_etable}

As we have already seen from the preceding examples, the \texttt{etable}
function allows us to export the results from (multiple) \pkg{fixest}
estimations to a nicely formatted regression table. The table can either
be displayed directly in the R console or exported as a \LaTeX~table.

It has the following main arguments:

\begin{itemize}
\tightlist
\item
  \texttt{...}: this is where we can pass any number of estimations
  (\pkg{fixest} of \texttt{fixest\_multi} objects)
\item
  \texttt{vcov}: to define the way the VCOV is computed for all
  estimations, see Section \ref{sec_vcov} for details. It is possible to
  provide a list of VCOVs in which case the models are replicated for
  each VCOV.
\item
  \texttt{dict}: a dictionary in the form of a named vector whose names
  are the variable names and values are the labels to be displayed
\item
  \texttt{keep}, \texttt{drop}, \texttt{order}: vectors of regular
  expressions to keep, drop or order the coefficients in the table. It
  is possible to negate the regular expression by starting it with an
  exclamation mark.
\item
  \texttt{tex}: logical. If \texttt{FALSE}, the function returns a
  \texttt{data.frame} of class \texttt{etable\_df} with a custom print
  method. If \texttt{TRUE}, the function returns a character vector
  containing a \LaTeX table. By default, it is equal to \texttt{FALSE}
  if the argument \texttt{file} is missing, and true otherwise.
\item
  \texttt{file}: path to a file where to write the table
\item
  \texttt{style.df}: an object returned by \texttt{style.df()}. How to
  style the table displayed in the console.
\item
  \texttt{style.tex}: an object returned by \texttt{style.tex()}. How to
  style the \LaTeX, table.
\end{itemize}

The factory defaults of \texttt{etable} are usually good enough for
displaying results in the R console and there is little need for further
customization. However, the situation is more complicated when creating
\LaTeX~publication table, since these often require fine-grained control
and detailing. The default \LaTeX~table produced by \texttt{etable} is
quite plain, yet contains all of the necessary information to uniquely
identify the context of an estimation, as well as interpret it.
Specifically, we report: the sample (if the sample was split), the type
of standard errors, the family when several estimations have different
families (e.g., OLS or Poisson), convergence state when one model did
not converge, etc. We can therefore think of the default \texttt{etable}
output as appropriate for an exploratory table, where we want as much
information as possible without worrying too much about aesthetics.
Table \ref{tab_first_ex}, from our motivating example back in Section
\ref{sec_example}, provides an example of such a table.

For published tables, the ability to style, as well as precisely add
pieces of information, take on heightened importance. \texttt{etable}
caters to the needs of publication-focused users by offering over 60
arguments to customize their table. Striking the balance for usability
here requires nuance, since the number of functions can be overwhelming.
However, the basic philosophy is to simply set your table preferences
once and then forget about them. Thanks to the function
\texttt{setFixest\_etable()} it is possible to set the styling arguments
globally. Further, the function \texttt{setFixest\_dict()} sets the
labels of the arguments globally.

To illustrate, let us revisit our earlier Table \ref{tab_first_ex}
example. We re-use the saved \texttt{est\_split} and \texttt{est\_iv}
estimations from before and our \texttt{etable} call also remains
unchanged (we pass the exact same arguments). Our only changes are i)
adding labels to the variables with \texttt{setFixest\_dict}, and ii)
changing the default \LaTeX~style with \texttt{setFixest\_etable}. The
resulting table is displayed in Table \ref{tab_first_ex_options} and
reveal a number of aesthetic improvements.

\begin{Shaded}
\begin{Highlighting}[]
\FunctionTok{setFixest\_dict}\NormalTok{(}\FunctionTok{c}\NormalTok{(}\AttributeTok{articles =} \StringTok{"\# Articles"}\NormalTok{, }\AttributeTok{funding =} \StringTok{"Funding (\textquotesingle{}000 $US)"}\NormalTok{, }
                 \AttributeTok{year =} \StringTok{"Year"}\NormalTok{, }\AttributeTok{eu\_us =} \StringTok{"Region"}\NormalTok{, }\AttributeTok{policy =} \StringTok{"Policy"}\NormalTok{, }
                 \AttributeTok{indiv =} \StringTok{"Researcher"}\NormalTok{))}
\FunctionTok{setFixest\_etable}\NormalTok{(}\AttributeTok{style.tex =} \FunctionTok{style.tex}\NormalTok{(}\StringTok{"aer"}\NormalTok{, }\AttributeTok{signif.code =} \ConstantTok{NA}\NormalTok{), }
                 \AttributeTok{fitstat =} \SpecialCharTok{\textasciitilde{}}\NormalTok{n }\SpecialCharTok{+}\NormalTok{ my)}
\FunctionTok{etable}\NormalTok{(est\_split, est\_iv, }
       \AttributeTok{vcov    =} \SpecialCharTok{\textasciitilde{}}\NormalTok{indiv,}
       \AttributeTok{stage   =} \DecValTok{1}\SpecialCharTok{:}\DecValTok{2}\NormalTok{, }
       \AttributeTok{caption =} \FunctionTok{c}\NormalTok{(}\StringTok{"Compiling the results from five estimations: "}\NormalTok{,}
                   \StringTok{"LaTeX rendering with options."}\NormalTok{),}
       \AttributeTok{label   =} \StringTok{"tab\_first\_ex\_options"}\NormalTok{, }
       \AttributeTok{file    =} \StringTok{"tables/first\_example\_options.tex"}\NormalTok{)}
\end{Highlighting}
\end{Shaded}

\begin{table}[htbp]
   \caption{\label{tab_first_ex_options} Compiling the results from five estimations: LaTeX rendering with options.}
   \bigskip
   \centering
   \begin{tabular}{lccccc}
      \toprule
       & \multicolumn{3}{c}{\# Articles} & Funding ('000 \$US) & \# Articles\\
      Region & Full sample & EU & US & \multicolumn{2}{c}{} \\ 
      IV stages & \multicolumn{3}{c}{ } & First & Second \\ 
                               & (1)           & (2)           & (3)           & (4)           & (5)\\  
      \midrule 
      Funding ('000 \$US)      & 0.0964        & 0.0900        & 0.1035        &               & -0.2452\\   
                               & (0.0045)      & (0.0065)      & (0.0060)      &               & (1.381)\\   
      Policy                   &               &               &               & -1.710        &   \\   
                               &               &               &               & (9.013)       &   \\   
       \\
      Observations             & 1,080         & 550           & 530           & 1,080         & 1,080\\  
      Dependent variable mean  & 17.127        & 18.402        & 15.804        & 106.19        & 17.127\\  
       \\
      Researcher fixed effects & $\checkmark$  & $\checkmark$  & $\checkmark$  & $\checkmark$  & $\checkmark$\\   
      Year fixed effects       & $\checkmark$  & $\checkmark$  & $\checkmark$  & $\checkmark$  & $\checkmark$\\   
      \bottomrule
   \end{tabular}
\end{table}

There are many \texttt{etable} options beyond the ones presented here,
including special integrations for R Markdown and Quarto documents, as
well as support for coefficient highlighting, three-part tables, etc.
The \pkg{fixest} website provides further details on how to customize
regression tables in the form of standalone documentation
(e.g.~\href{https://lrberge.github.io/fixest/reference/etable.html}{\texttt{?etable}},
\href{https://lrberge.github.io/fixest/reference/style.tex.html}{\texttt{?style.tex}})
and two vignettes
(``\href{https://lrberge.github.io/fixest/articles/exporting_tables.html}{Exporting
estimation tables}'' and
``\href{https://lrberge.github.io/fixest/articles/etable_new_features.html}{\texttt{etable}:
new features in \texttt{fixest} 0.10.2}'').

\subsection{Plotting the coefficients}\label{plotting-the-coefficients}

\label{sec_coefplot}

Alongside tabling support, \pkg{fixest} also provides two dedicated
plotting functions for plotting the coefficients and confidence
intervals: \texttt{coefplot} and \texttt{iplot}. \texttt{coefplot} plots
the point estimates and confidence intervals of coefficients from one or
several \pkg{fixest} estimations. \texttt{iplot} does the same, but
limited to categorical variables or interaction terms created with the
function \texttt{i()}; see Section \ref{sec_i}.

Some of the main arguments are:

\begin{itemize}
\tightlist
\item
  \texttt{...}: it receives \pkg{fixest} estimations or a list of
  \pkg{fixest} estimations
\item
  \texttt{vcov}: how to compute the VCOV for the estimations in input,
  it can be a list of different ways to compute the VCOVs
\item
  \texttt{ci\_level}: the level of the confidence interval, the default
  is 0.95
\item
  \texttt{keep}, \texttt{drop}, \texttt{order}: vectors of regular
  expressions to keep, drop or order the coefficients in the table. It
  is possible to negate the regular expression by starting it with an
  exclamation mark
\item
  \texttt{pt.col}, \texttt{pt.cex}, \texttt{pt.lwd}, \texttt{pt.pch}:
  color, size, width and point type of the coefficient estimate
\item
  \texttt{ci.col}, \texttt{ci.lty}, \texttt{ci.lwd}: color, line type
  and line width of the edge of the confidence interval
\item
  \texttt{pt.join}, \texttt{pt.join.par}: whether the point estimated
  should be joined with a line and, in that case, the parameters of the
  line (a list of \texttt{lwd}, \texttt{col}, and \texttt{lty})
\item
  \texttt{ci.join}, \texttt{ci.join.par}: same as above, for the edges
  of the confidence intervals
\item
  \texttt{ci.fill}, \texttt{ci.fill.par}: whether the space between the
  confidence intervals should be filled with a color, and the parameters
  of such filling (\texttt{col} and \texttt{alpha})
\item
  \texttt{group}: a list containing variables to appear grouped, by
  default it groups the factors
\end{itemize}

Paralleling \texttt{etable}, users can set any of these parameters
globally via the \texttt{setFixest\_coefplot()} function, and thus
define a preferred aesthetic for all of their plots.

In Figure \ref{fig_coefplot} below, we combine several examples to
demonstrate \pkg{fixest}'s plotting facilities. We start by performing a
single estimation (with \texttt{"iid"} errors) and then obtaining a
second version with clustered standard errors.

\begin{Shaded}
\begin{Highlighting}[]
\NormalTok{est\_airq }\OtherTok{=} \FunctionTok{feols}\NormalTok{(Ozone }\SpecialCharTok{\textasciitilde{}}\NormalTok{ Temp }\SpecialCharTok{+}\NormalTok{ Wind }\SpecialCharTok{+} \FunctionTok{i}\NormalTok{(Month), airquality)}
\NormalTok{est\_airq\_clu }\OtherTok{=} \FunctionTok{summary}\NormalTok{(est\_airq, }\AttributeTok{vcov =} \SpecialCharTok{\textasciitilde{}}\NormalTok{Day)}
\end{Highlighting}
\end{Shaded}

Our first graph (top left) is a simple \texttt{coefplot} call,
demonstrating the default behavior and the ease with which multiple
estimations are passed.

\begin{Shaded}
\begin{Highlighting}[]
\FunctionTok{coefplot}\NormalTok{(est\_airq, est\_airq\_clu, }\AttributeTok{drop =} \StringTok{"Constant"}\NormalTok{)}
\end{Highlighting}
\end{Shaded}

The second graph (top right) is a tweaked version of the first, flipping
the plot axes and passing various other customizations arguments.

\begin{Shaded}
\begin{Highlighting}[]
\FunctionTok{coefplot}\NormalTok{(est\_airq, est\_airq\_clu, }\AttributeTok{drop =} \StringTok{"Constant"}\NormalTok{, }\AttributeTok{horiz =} \ConstantTok{TRUE}\NormalTok{, }
         \AttributeTok{lwd =} \DecValTok{2}\NormalTok{, }\AttributeTok{pt.pch =} \FunctionTok{c}\NormalTok{(}\DecValTok{20}\SpecialCharTok{:}\DecValTok{21}\NormalTok{), }\AttributeTok{ci.lty =} \DecValTok{1}\SpecialCharTok{:}\DecValTok{2}\NormalTok{,}
         \AttributeTok{zero.par =} \FunctionTok{list}\NormalTok{(}\AttributeTok{lty =} \DecValTok{3}\NormalTok{, }\AttributeTok{col =} \StringTok{"darkgrey"}\NormalTok{, }\AttributeTok{lwd =} \DecValTok{3}\NormalTok{))}
\end{Highlighting}
\end{Shaded}

The third graph (bottom left) is the same data plotted with
\texttt{iplot} and its default values, which then focuses only on the
categorical variable that was included with \texttt{i()}. Moreover,
instead of having two estimation objects with different standard-errors,
here we demonstrate that users can also have only one estimation object
and pass several VCOVs using the argument \texttt{vcov}:

\begin{Shaded}
\begin{Highlighting}[]
\FunctionTok{iplot}\NormalTok{(est\_airq, }\AttributeTok{vcov =} \FunctionTok{list}\NormalTok{(}\StringTok{"iid"}\NormalTok{, }\SpecialCharTok{\textasciitilde{}}\NormalTok{Day))}
\end{Highlighting}
\end{Shaded}

The last graph (bottom right) reports the coefficients of an event study
plotted with \texttt{iplot} and a custom style displaying a ribbon:

\begin{Shaded}
\begin{Highlighting}[]
\NormalTok{est\_did }\OtherTok{=} \FunctionTok{feols}\NormalTok{(y }\SpecialCharTok{\textasciitilde{}}\NormalTok{ x1 }\SpecialCharTok{+} \FunctionTok{i}\NormalTok{(period, treat, }\DecValTok{5}\NormalTok{) }\SpecialCharTok{|}\NormalTok{ id }\SpecialCharTok{+}\NormalTok{ period, base\_did)}
\FunctionTok{iplot}\NormalTok{(est\_did, }\AttributeTok{ci.fill =} \ConstantTok{TRUE}\NormalTok{, }\AttributeTok{ci.fill.par =} \FunctionTok{list}\NormalTok{(}\AttributeTok{col =} \StringTok{"lightblue"}\NormalTok{), }
      \AttributeTok{ci.lwd =} \DecValTok{0}\NormalTok{, }\AttributeTok{pt.join =} \ConstantTok{TRUE}\NormalTok{)}
\end{Highlighting}
\end{Shaded}

\begin{figure}[thbp]
  \centering
  \caption{\label{fig_coefplot}Example of graphs created with \texttt{coefplot} (above) and \texttt{iplot} (below), with default (left) and customized (right) layout.}
  
  \includegraphics[width=1\textwidth]{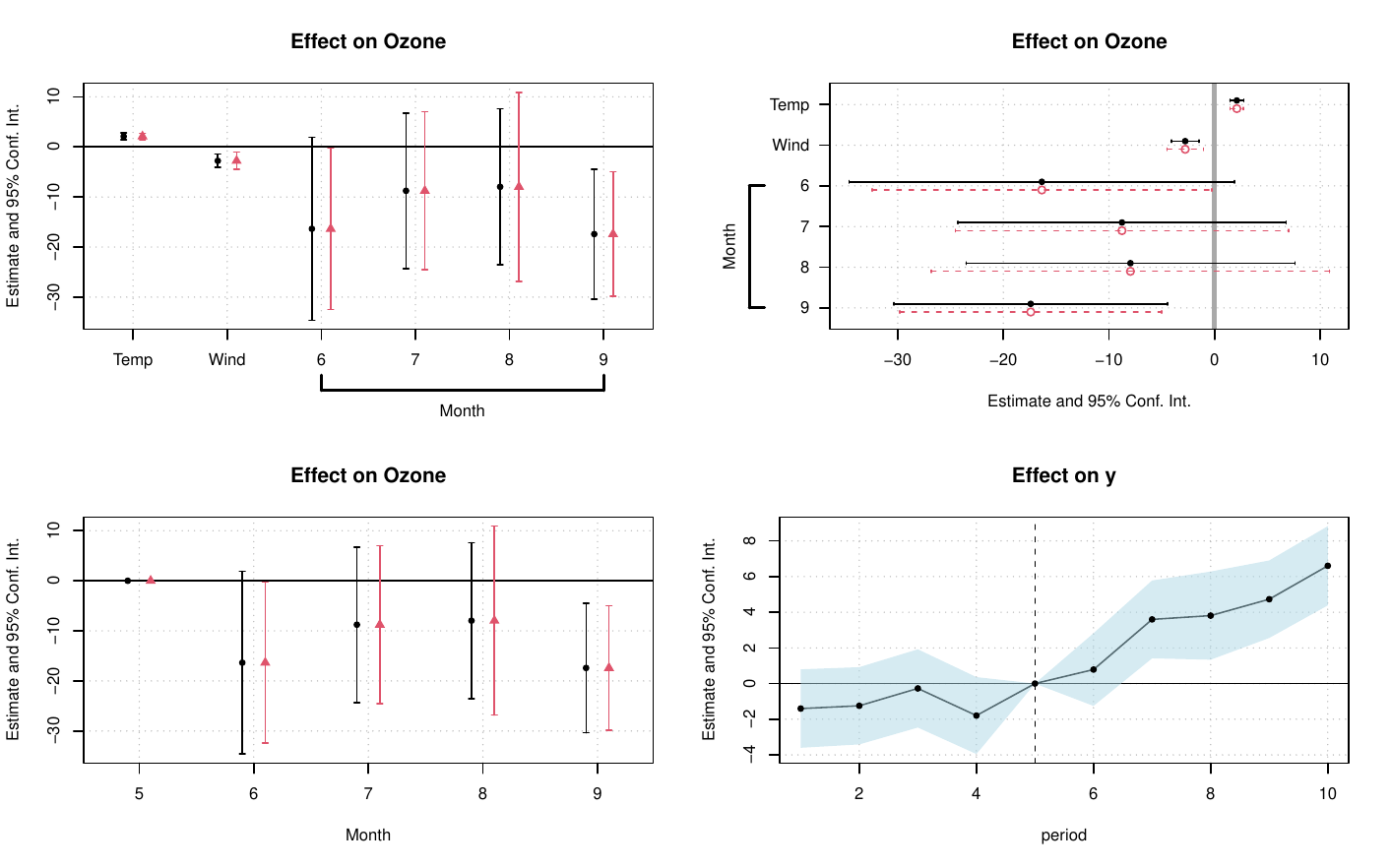}
\end{figure}

Even though \texttt{coefplot}/\texttt{iplot} accommodate many options to
customize the graphs, it may not be enough to create very precise
publication graphs. In that case, the user can access the dataset used
to build the graph (without plotting) by invoking the
\texttt{only.params\ =\ TRUE} argument. Similarly, these \pkg{fixest}
plotting functions yield base R graphics, which may or may not be to the
user's taste. The \pkg{ggfixest} package \citep{ggfixest} offers
equivalent functionality in the form of dedicated \pkg{ggplot2}
\citep{wickham2016Ggplot2ElegantGraphics} methods for \pkg{fixest}
objects, as well as a set of ancillary plotting features.

\section{Benchmarks}\label{benchmarks}

\label{sec_benchmarks}

In this section, we present a set of benchmarks that demonstrate
\pkg{fixest}'s performance against alternative state of the art
software, across a broad range of settings. For the sake of parsimony
and reproducibility, we restrict our comparisons to free and open-source
software (FOSS) which specialize in fixed-effects
estimations.\footnote{The alternatives that we do not benchmark here,
  including base \proglang{R} routines such as \texttt{(g)lm} and
  proprietary software such as \proglang{Stata}, are generally slower
  than our comparison FOSS group for this set of tasks.} Specifically,
we benchmark \pkg{fixest} against \pkg{alpaca}
\citep{stammann2018FastFeasibleEstimation} and \pkg{lfe}
\citep{gaure2013lfe} in \proglang{R}, \pkg{FixedEffectModels}
\citep{gomez2024FixedEffectModelsJlFast} and \pkg{GLFixedEffectModels}
\citep{boehm2025GLFixedEffectModelsJl} in \proglang{Julia}, and
\pkg{PyFixest} \citep{fischer2024Pyfixest} in
\proglang{Python}.\footnote{For the sake of clarification, please note
  that \pkg{PyFixest} is developed by independent authors. As the name
  suggests, it is a port of \pkg{fixest}'s syntax and features into
  \proglang{Python}, although some of the core internal routines are
  different; as we documented in Section \ref{sec_theory}. The
  \pkg{PyFixest} team recently began work to replace their existing
  alternating projections routine with \pkg{fixest}'s fixed-point
  algorithm. This means that the performance of these two libraries
  should become closer in the near future, as they will both rely on the
  same method and implementation.} We use the latest available versions
of these software, as of 15 January 2026. The benchmarks presented here
are run on a Windows PC with 128GB of RAM and an
Intel\textsuperscript{\tiny\textregistered} Core\textsuperscript{TM}
i7-10850H CPU @ 2.70GHz processor. Separate benchmarks run on an Apple
Mac M4 Pro laptop with 48 GB of RAM yield outcomes that are
proportionally very similar.

Before continuing, we issue the standard disclaimer that benchmarks
should always be interpreted with a degree of caution, given the
inherent sensitivity to arbitrary data and estimation choices.
Nonetheless, we believe that we provide a fair and representative
overview, which seeks to minimize bias by running multiple benchmarks
over many use-cases, using both simulated and real-life data. In
particular, we present two broad sets of benchmarks. The first uses
simulated datasets with different convergence properties, providing
insight into how the performance of our different software scales with
the number of observations. The second evaluates performance on a
demanding real-life dataset, where we vary the type of model estimation
and focus on specific advantages provided by \pkg{fixest}'s feature set.
All benchmarking code is provided with the replication packet for this
paper.

\subsection{Simulated data}\label{simulated-data}

\begin{figure}[t]
  \centering
  \caption{\label{fig_ols_bench}OLS benchmark results on simulated data.}
  
  \includegraphics[width=1\textwidth]{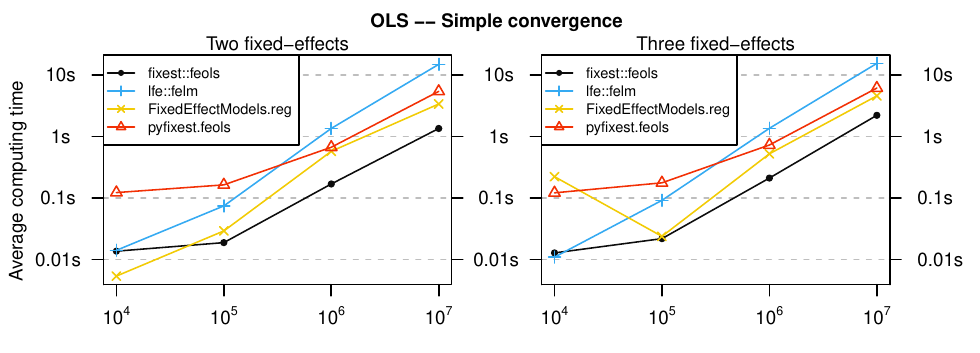}
  \includegraphics[width=1\textwidth]{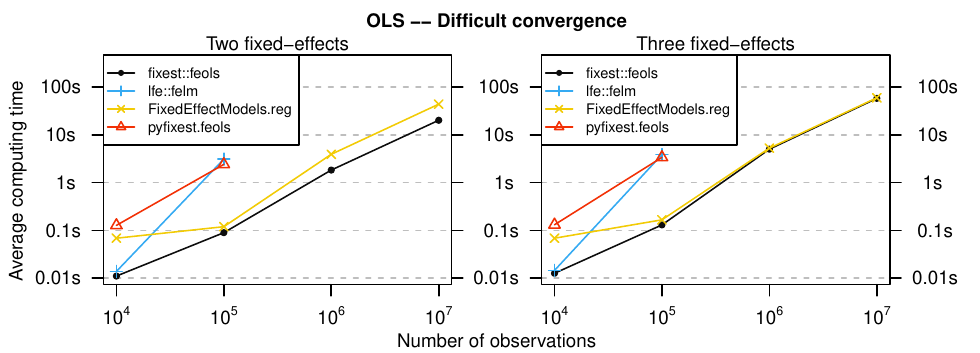}
  
\end{figure}

\begin{figure}[t]
  \centering
  \caption{\label{fig_poisson_bench}Poisson benchmark results on simulated data.}
  
  \includegraphics[width=1\textwidth]{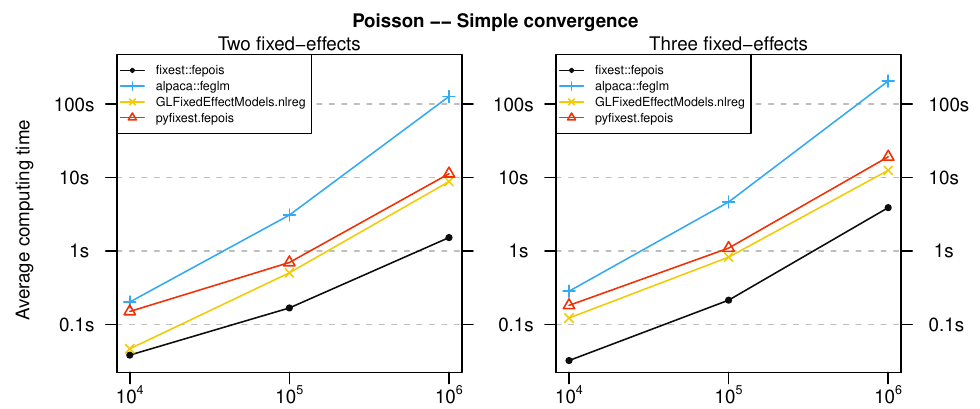}
  \includegraphics[width=1\textwidth]{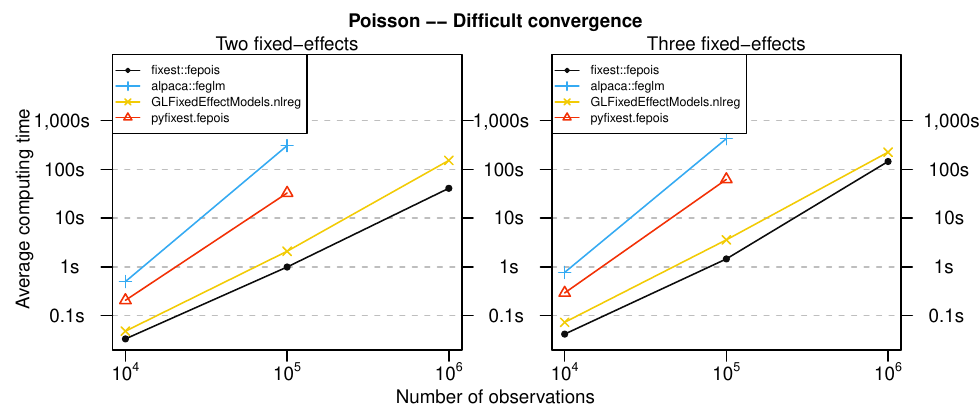}
  
\end{figure}

Our simulated dataset draws inspiration from labor economics by
generating a panel of employees and firms over time. Our specific
regression task is estimating wages as a function of covariates (like
training), whilst controlling for individual, firm and year
fixed-effects. Importantly, we consider two data generating processes
(DGPs) for our simulated data: one for which the convergence of the
estimated fixed-effects is ``simple'' and another for which it is
``difficult''. The next two paragraphs provide intuition for why
convergence can be simple (fast) or difficult (slow). Readers
uninterested in the intuition can safely skip these paragraphs and jump
straight to the results in Figures \ref{fig_ols_bench} and
\ref{fig_poisson_bench}.

Consider a balanced panel of employees. The assignation of firms to
employees is the key determinant of convergence speed, which in turn
elevates the importance of algorithmic routines for estimating the
individual and firm fixed-effects. In the ``simple'' case, we assign
firms randomly to employees, resulting in each firm being connected to
many different employees. From a network perspective, this creates a
dense network where firms share common employees, which in turn means
that any changes to a single firm coefficient will propagate quickly
through the system, enabling fast convergence. In contrast, for the
``difficult'' case, we assign firms sequentially to employees, resulting
in firms being connected to very few employees. Now our network is
extremely sparse with few connections between firms, which means that
changes to firm coefficients are slow to propagate, yielding slow
convergence.

Our DGP is as follows. For \(N\) observations, we generate a panel of
\(N_I = N/N_T\) individuals over \(N_T = 10\) years. These employees
work across \(N_F = N_I / 23\) companies. In the simple case, we assign
companies randomly to employees. In the difficult case, we assign them
sequentially. All the fixed-effect coefficients are generated with a
standard normal distribution. There are two covariates \(x_1\),
generated from a standard normal distribution, and \(x_2 = x_1^2\).
Finally the outcome, \(y\) is obtained as follows: \[
  y_{ift} = x_{1, ift} + 0.05 x_{2, ift} + \mu_i + \lambda_t + \nu_f + \varepsilon_{ift},
\]

where \(\varepsilon_{ift}\) is drawn from a standard normal
distribution. We report below the code to generate the data:

\begin{Shaded}
\begin{Highlighting}[]
\NormalTok{base\_dgp }\OtherTok{=} \ControlFlowTok{function}\NormalTok{(}\AttributeTok{n =} \DecValTok{1000}\NormalTok{) \{}
  
\NormalTok{  nb\_year }\OtherTok{=} \DecValTok{10}
\NormalTok{  nb\_indiv\_per\_firm }\OtherTok{=} \DecValTok{23}
  
\NormalTok{  nb\_indiv }\OtherTok{=} \FunctionTok{round}\NormalTok{(n }\SpecialCharTok{/}\NormalTok{ nb\_year)}
\NormalTok{  nb\_firm }\OtherTok{=} \FunctionTok{round}\NormalTok{(nb\_indiv }\SpecialCharTok{/}\NormalTok{ nb\_indiv\_per\_firm)}
\NormalTok{  indiv\_id }\OtherTok{=} \FunctionTok{rep}\NormalTok{(}\DecValTok{1}\SpecialCharTok{:}\NormalTok{nb\_indiv, }\AttributeTok{each =}\NormalTok{ nb\_year)}
\NormalTok{  year }\OtherTok{=} \FunctionTok{rep}\NormalTok{(}\DecValTok{1}\SpecialCharTok{:}\NormalTok{nb\_year, }\AttributeTok{times =}\NormalTok{ nb\_indiv)}

\NormalTok{  firm\_id\_simple }\OtherTok{=} \FunctionTok{sample}\NormalTok{(}\DecValTok{1}\SpecialCharTok{:}\NormalTok{nb\_firm, n, }\AttributeTok{replace =} \ConstantTok{TRUE}\NormalTok{)}
\NormalTok{  firm\_id\_difficult }\OtherTok{=} \FunctionTok{rep}\NormalTok{(}\DecValTok{1}\SpecialCharTok{:}\NormalTok{nb\_firm, }\AttributeTok{length.out =}\NormalTok{ n)}

\NormalTok{  unit\_fe }\OtherTok{=} \FunctionTok{rnorm}\NormalTok{(nb\_indiv)[indiv\_id]}
\NormalTok{  year\_fe }\OtherTok{=} \FunctionTok{rnorm}\NormalTok{(nb\_year)[year]}
\NormalTok{  firm\_fe }\OtherTok{=} \FunctionTok{rnorm}\NormalTok{(nb\_firm)[firm\_id\_simple]}
  
\NormalTok{  x1 }\OtherTok{=} \FunctionTok{rnorm}\NormalTok{(n)}
\NormalTok{  y }\OtherTok{=} \DecValTok{1} \SpecialCharTok{*}\NormalTok{ x1 }\SpecialCharTok{+} \FloatTok{0.05} \SpecialCharTok{*}\NormalTok{ x1}\SpecialCharTok{\^{}}\DecValTok{2} \SpecialCharTok{+}\NormalTok{ firm\_fe }\SpecialCharTok{+}\NormalTok{ unit\_fe }\SpecialCharTok{+}\NormalTok{ year\_fe }\SpecialCharTok{+} \FunctionTok{rnorm}\NormalTok{(n)}
\NormalTok{  df }\OtherTok{=} \FunctionTok{data.frame}\NormalTok{(}
    \AttributeTok{indiv\_id          =}\NormalTok{ indiv\_id, }
    \AttributeTok{year              =}\NormalTok{ year,}
    \AttributeTok{firm\_id           =}\NormalTok{ firm\_id\_simple,}
    \AttributeTok{firm\_id\_difficult =}\NormalTok{ firm\_id\_difficult, }
    \AttributeTok{x1                =}\NormalTok{ x1,}
    \AttributeTok{x2                =}\NormalTok{ x1}\SpecialCharTok{\^{}}\DecValTok{2}\NormalTok{,}
    \AttributeTok{y                 =}\NormalTok{ y}
\NormalTok{  )}
  
  \FunctionTok{return}\NormalTok{(df)}
\NormalTok{\}}
\end{Highlighting}
\end{Shaded}

Our simulated benchmarks evaluate both the OLS and Poisson GLM
estimators. Specifically, the benchmarks estimate a set of models where:
i) the dependent variable is \(y\) for OLS and \(exp(y)\) for Poisson,
ii) the explanatory variables are \(x_1\) and \(x_2\), iii) there are
two or three sets of fixed-effects, and iv) the convergence of the
fixed-effects can be ``simple'' or ``difficult''. Two fixed-effects
correspond to individual and firms, while three fixed-effects adds the
year. Finally we also vary the number of observations from 10,000 to
10M.

The results of the OLS benchmark are represented in Figure
\ref{fig_ols_bench}. It represents the average computing time across
three replications, in logarithmic scale. The top row represents the
results for the ``simple'' case. There we see that
\texttt{fixest::feols} is consistently faster than the alternative
software, with a computing time of about 1s for 10M observations and two
fixed-effects, while on the other end of the scale \texttt{lfe::felm}
takes an order of magnitude longer at 10s.

An interesting observation about this ``simple'' case is that there are
effectively no scaling effects; the relative differences between
software do not meaningfully change with the number of observations.
However, this is not true for our second, ``difficult'' set of results,
which we report in the bottom row of Figure \ref{fig_ols_bench}. Both
\pkg{fixest} and \proglang{Julia}'s \pkg{FixedEffectModels} continue to
scale almost linearly as the dataset grows. Somewhat remarkably, these
two implementations are effectively identical for our most difficult
case of three-fixed effects and 10M observations at just over 60s. In
contrast, the convergence times of \pkg{lfe} and \pkg{PyFixest} explode
in this more difficult setting. We do not report benchmark times for
these two libraries above 100,000 observations, since neither could fit
the model within our constraints.\footnote{Either erroring due to
  absence of convergence, or exceeding our 1hr timeout.}

Our Poisson benchmarks are reported in Figure \ref{fig_poisson_bench}
and follow much the same pattern as the OLS results. Again, we observe
that \pkg{fixest} times are consistently faster across the board, with
good scaling properties for both the ``simple'' (top) and ``difficult''
(bottom) DGP cases. \pkg{GLFixedEffectModels} is nearest in performance
and also scales well across our different settings. In contrast,
\pkg{alpaca} and \pkg{PyFixest} struggle in the difficult convergence
case, and were unable to complete within our 1hr timeout constraint for
1M or more observations.

\subsection{NYC taxi data}\label{nyc-taxi-data}

Having evaluated the impact that different convergence properties can
have on simulated data, we now illustrate \pkg{fixest}'s performance on
a demanding real-life example. Specifically, we use the well-known NYC
Taxi and Limousine Commission dataset.\footnote{See:
  \url{https://www.nyc.gov/site/tlc/about/tlc-trip-record-data.page}.}
Our sample covers three months of taxi trip data from Jan--Mar 2012 and
is rather large at over 42M observations.

The results, which are the average timings across three runs, are
reported in Table \ref{tab_bench_taxi}, where we run four basic
specifications that are intended to highlight \pkg{fixest}-specific
features:

\begin{enumerate}
\def\labelenumi{\arabic{enumi}.}
\tightlist
\item
  \emph{Baseline}. Standard OLS with two variables and three
  fixed-effects
\item
  \emph{Varying slopes}. Same as \#1 but adding varying slopes to one
  explanatory variable
\item
  \emph{Multiple outcomes}. Same as \#1 but with two outcome variables
\item
  \emph{Multiple VCOVs}. Same as \#1 but with with both standard and
  clustered VCOVs
\end{enumerate}

\begin{table}[thb]
  \centering
  \caption{\label{tab_bench_taxi}OLS benchmarks using NYC taxi data ($\approx$ 42M rows).}
  

\small
\begin{tabular}{ m{7.1cm} m{4.3cm} m{2.1cm} }
  \toprule
  Estimated model & Method & Time\\
  \midrule
  
  \textbf{Baseline} & & \\
  \texttt{y \textasciitilde~ x1 + x2 | fe1 + fe2 + fe3} & 
  \texttt{fixest::feols} &
  9.9s
  \\
  
  & \texttt{FixedEffectModels.reg} &
  18.1s
  \\
  
  & \texttt{pyfixest.feols} &
  23.6s
  \\
  
  \textbf{Varying Slopes} & & \\
  \texttt{y \textasciitilde~ x1 | fe1[x2] + fe2 + fe3} & 
  \texttt{fixest::feols} &
  8.9s
  \\
  
  & \texttt{FixedEffectModels.reg} &
  15.3s
  \\
  
  & \texttt{pyfixest.feols} &
  not available 
  \\
  
  \textbf{Multiple Outcomes} & & \\
  \texttt{c(y1, y2) \textasciitilde~ x1 + x2 | fe1 + fe2 + fe3} & 
  \texttt{fixest::feols} &
  13.5s
  \\
  
  & \texttt{FixedEffectModels.reg} &
  35.2s
  \\
  
  & \texttt{pyfixest.feols} &
  43.2s
  \\
  
  \textbf{Multiple VCOVs} & & \\
  \texttt{y \textasciitilde~ x1 + x2 | fe1 + fe2 + fe3} & 
  \texttt{fixest::feols} &
  10.7s
  \\
  
  \texttt{vcov = "iid"} & 
  \texttt{FixedEffectModels.reg} &
  36.6s
  \\
  
  \texttt{vcov = vcov\_cluster("fe1")} & 
  \texttt{pyfixest.feols} &
  37.8s
  \\
  
  \bottomrule
\end{tabular}

\end{table}

Starting with the baseline estimation in the top row of Table
\ref{tab_bench_taxi}, \pkg{fixest} is fastest at only 9.9s, followed by
\pkg{FixedEffectModels} at 18s and then \pkg{PyFixest} at 23s.

Adding varying slopes to one explanatory variable reduces the estimation
time for \pkg{fixest} and \pkg{FixedEffectModels} to 8.9s and 15s
respectively, while this feature is unavailable in \pkg{PyFixest}.

Estimating two outcomes simultaneously adds less than 4s in
\pkg{fixest}, thanks to its highly efficient handling of multiple
estimations. Since this feature in unavailable in
\pkg{FixedEffectModels}, the estimation time is effectively doubled to
35s. \pkg{PyFixest} also handles multiple estimations, although their
computation time increases by 80\% to 43s.

Finally, as we have emphasized, \pkg{fixest}'s design separates
estimation from inference. This means that estimating two models with
different VCOVs (standard-errors) incurs virtually no overhead.
Conversely, for \pkg{FixedEffectModels} the time is doubled.
\pkg{PyFixest} also separates inference from estimation, albeit not as
efficiently, thus its timing increases by 60\%.

\section{Conclusion and discussion}\label{conclusion-and-discussion}

\label{sec_conclu}

The \pkg{fixest} package represents a mature econometric toolkit that we
hope serves diverse user communities effectively. While this article has
focused on the core user-facing features, we also wish to highlight two
additional user groups: package developers and educators.

For package developers, \pkg{fixest} provides a robust foundation with
numerous convenience features. Many econometric methods are implemented
via a regression model and, by relying on \pkg{fixest} for estimation,
package developers automatically gain access to all of these features.
This allows developers to focus on methodological innovations, rather
than worrying about speed or re-implementing standard features like
fixed-effects handling, robust standard errors, plotting, tabling, and
basic method support. Moreover, the \pkg{fixest} ecosystem extends
seamlessly through integrations with \pkg{broom}
\citep{robinson2025BroomConvertStatistical}, \pkg{parameters}
\citep{parameters}, \pkg{marginaleffects}
\citep{arel-bundock2024marginaleffects}, \pkg{gtsummary}
\citep{gtsummary}, and \pkg{modelsummary}
\citep{arel-bundock2022modelsummary}, providing even more functionality
that package developers can leverage.

Second, we believe that \pkg{fixest} can prove useful for educators. The
range of \pkg{fixest} features are intended to cover large swathes of
the applied econometrics toolkit. More to the point, incorporating all
of these features into a single package with a unified syntax allows
instructors to focus on methodology, rather than software proliferation.
As a leading example, consider \emph{The Effect} textbook
\citep{huntington2021effect}, which aims to teach causal inference at
the undergraduate level. Code covering basic regression models, robust
inference, instrumental variables, fixed-effects, DiD, and regression
discontinuity all are written using the single function \texttt{feols}.

We would like to conclude with a few words on \pkg{fixest}'s development
philosophy. This software is built on three cornerstones: speed,
robustness, and stability.

On speed. All data intensive operations in \pkg{fixest} have been
implemented with as many optimizations as possible, involving
methodological innovations when needed. This is highlighted in the
benchmarks for the core estimating functions, but the same philosophy
extends to many other functions.

On robustness. Error-handling is central to \pkg{fixest} since it is
critical to a good user experience. We have tried our best to anticipate
any potential problems from the user side, even unlikely ones, and the
\pkg{fixest} codebase contains hundreds, possibly thousands, of custom
error messages and exceptions. For the remaining bugs that are
inevitably left over, we could count on the numerous feedback from a
fantastic community that has helped iron them out. \pkg{fixest} is by
now a mature package, forged by battle-hardened testing over several
years and across an array of contexts.

On Stability. As core developers and users of \pkg{fixest}, we value
stability very highly. Any design choices made in consideration of the
codebase and user-facing API are done with that desideratum in mind. We
would make three observations in support of this claim. First,
\pkg{fixest} has no hard dependencies that are at risk themselves. The
core codebase relies only on three non-base \proglang{R} packages,
namely \pkg{Rcpp} \citep{rcpp2024software}, \pkg{stringmagic}
\citep{berge2025stringmagic} and \pkg{dreamerr}
\citep{berge2025dreamerr}. The former already acts as the solid
foundation of thousands of \proglang{R} packages, while the latter two
are spinoffs from \pkg{fixest} itself. Second, only a handful of
breaking changes to non-core \pkg{fixest} functions have been introduced
over its seven years of existence. This, despite the introduction of
many new features and considerable growth of the codebase. Third, we
strive to avoid making rushed design decisions. Rather, we introspect
carefully and then seek feedback to ensure that any new features are as
stable and future proof as they can.

By continuing to build on these foundations, we hope that \pkg{fixest}
will prove to be a valuable and reliable companion for empirical work in
the years to come.

\section*{Acknowledgments}

Laurent Bergé wishes to thank Grant McDermott whose early publicity and
genuine enthusiasm led to unforeseen levels of development, Dirk
Eddelbuettel for creating and maintaining \pkg{Rcpp} (without this key
technology \pkg{fixest} would have never existed), the community which
made this software infinitely more robust thanks to bug reports, and the
\proglang{R} authors and the CRAN team for making such an unbelievably
reliable, and fun, programming language.

\renewcommand\refname{References}
\bibliography{fixest.bib}

\end{document}